\documentclass[
 reprint,
nofootinbib,
 amsmath,amssymb,
 aps, pra,
]{revtex4-1}
\usepackage{xcolor}
\usepackage{dsfont}
\usepackage{graphicx}
\usepackage{dcolumn}
\usepackage{bm}

\usepackage{verbatim}
\usepackage{listings}
\usepackage[english]{babel}
\usepackage[utf8]{inputenc}
\usepackage{braket}

\newcommand{\eye}{\mathds{1}}
\newcommand{\ten}{\otimes}
\newcommand{\sig}{\mathcal{\sigma}}

\newcommand{\Tr}{\text{Tr}}

\newcommand{\s}{$\mathcal{S}$ }

\newcommand{\Ac}{{\mathcal{A}_{\rm C}}}
\newcommand{\Rc}{\hat{\bold{R}}_{\rm{C}}}

\newcommand{\Gc}{\Gamma_{\rm C}}

\begin{document}

\title{Non-Markov Enhancement of Maximum Power for  Quantum Thermal Machines}

\author{Paolo Abiuso}
 \affiliation{Scuola Normale Superiore, I-56126 Pisa, Italy}
 \affiliation{ICFO – Institut de Ciències Fotòniques, The Barcelona Institute of Science and Technology,08860 Castelldefels (Barcelona), Spain}
\email{paolo.abiuso@icfo.eu}
\author{Vittorio Giovannetti}
\affiliation{NEST, Scuola Normale Superiore and Istituto Nanoscienze-CNR, I-56126 Pisa, Italy }

\date{\today}

\begin{abstract}
In this work we study how the non-Markovian character of the dynamics can affect the thermodynamic performance of a quantum thermal engine, by analysing the maximum power output of Carnot and Otto cycles  departing from the quasi-static and infinite-time-thermalization regime respectively, introducing techniques for their control optimization in general dynamical models.
In our model, non-Markovianity is introduced by allowing some degrees of freedom of the reservoirs to be taken into account explicitly and share correlations with the engine by Hamiltonian coupling. 
It is found that the non-Markovian effects   can fasten the control and improve the power output.
\end{abstract}

\maketitle

\section{Introduction}
Quantum Thermodynamics \cite{quantumthermodynamics,gooldreview,andersreview} was born and rapidly grew in the last decades.
Fuelled by high experimental control of quantum systems and engineering at microscopic scales, one of the central goals of physicists is to push the limits of conventional thermodynamics, and the extension of standard models and cycles to include quantum effects and small ensemble sizes. 
Beyond the drive to clarify fundamental physical issues, these models may also turn out to be relevant from a more practical point of view: it is expected that industrial need for miniaturisation of technologies will benefit from the understanding of quantum thermodynamic processes. In both biology, for example, and nanotechnology, where the benefits from a cooling at the atomic scales are clear, refrigerators models \cite{refrigerator1,refrigerator2} based on quantum thermal machines could find actual application. Moreover, proposals for experimental realisations of quantum engines were made considering various physical platforms, and many were actually realised \cite{pekola-SET,pekolareview,liquid-NMR,ion-trap_Jarzynski,single-atom_heat-engine,optomechanical_h.e.,single-atom_prop,ronzani2018tunable,chen1994maximum,rezek2006irreversible,watanabe2017quantum,scully2011quantum,correa2013performance,
dorfman2013photosynthetic,brunner2014entanglement,optomechanical_h.e.,campisi2016power,brandner2017universal}.

Thermodynamics is, \textit{par excellence}, a theory involving non-isolated systems, and it must take into account the interaction and evolution induced by external degrees of freedom on a working medium. The description of open quantum systems \cite{breuer-petruccione} needs however, especially in cases where the number of degrees of freedom of the surroundings is big, an effective description on the local degrees of freedom by means of some approximation or assumption. The most important class of simplified dynamics of open systems goes under the name of {Markovian dynamics}. From the physical point of view, Markovianity is associated to systems interacting with large, unperturbed environments that ``spread away the information" contained in the system, while on the formal side different definitions of quantum Markovianity \cite{RHP,BLP} were introduced in the literature. We stand by the  approach (although the model we will consider is non-Markovian even for stronger definitions of quantum Markovianity \cite{RHP,BLP}) which identifies the Markovian character of a quantum process with its CP-divisibility \cite{RHP-article} hence 
admitting  a first order Master Equation (ME) that can be casted in the Gorini-Kossakowski-Sudarshan-Lindblad form (GKSL)~\cite{G-K-S,Lindblad}. 

Recent works have started to investigate how the breaking of the Markovianity in quantum dynamics can affect control and performance of quantum thermodynamic systems, motivated both by the necessity to overcome the approximation on very small systems, and by the speculation of non-Markovianity possibly being an actual resource in practical tasks, see e.g. Refs.~\cite{n-M_work-resourceth,n-M_work-finland2,n-M_work-inthequestofoptimalcontrol,n-M_work-giova,n-M_work-finland,n-M_work-exploiting,n-M_work-india,n-M_work-china,basilewitsch2017beating,pezzutto2018out}.
We contribute here considering two archetypical classes of thermal engines, i.e. the quantum Carnot cycle and the quantum Otto cycle \cite{quan2007quantum,karimi2016otto,kosloff2017quantum,watanabe2017quantum,rezek2006irreversible,single-atom_prop} which use as working medium a two-level (qubit) system  coupled to two thermal reservoirs while being externally driven. 
For these models we simulate non-Markovian effects by splitting the degrees of freedom of the system environmental baths into a local contribution, which we treat dynamically, and a remote component which instead
is described in terms of an effective GKSL Master Equation that tends to drive the rest of the model into thermal equilibrium. 
In this configuration it can be shown that the coupling with the local bath components 
 ignites the non-Markovian behaviour of the model whose effects can then be tested in terms of
the engine performance. In particular, performing an optimization on the external driving, we show that, both 
in the Carnot and Otto scheme, 
the maximum power extractable improves with respect to the Markovian limit.
To do this we first discuss both cycles in the
finite-time regime; to solve the dynamics and optimize the control for the
Carnot cycle, we use the powerful technique introduced in \cite{slowdriving} (Slow-Driving approximation, or S-D), which efficiently solves the approximate dynamics
of a system slowly perturbed from thermalization. For the Otto case we use exact solutions.

The article is structured as follows: \\
Sec.~\ref{sec:qthermal_machine_control}  and Sec.~\ref{sec:cycles} are devoted to 
introduce the technical tools we use to derive the  results of Sec.~\ref{sec:n-M_model}. 
Specifically in Sec.~\ref{sec:qthermal_machine_control} 
we discuss the physics of an externally controlled, quantum thermal machine introducing the notation 
in Sec.~\ref{sec:ii1}, drawing general thermodynamic considerations in Sec.~\ref{sec:ii2new} and
reviewing some basic facts about the S-D approximation method~\cite{slowdriving} in Sec.~\ref{sec:ii2}. 
In Sec.~\ref{sec:cycles} instead we analyze 
 the performances of some thermodynamic cycles. 
 In particular  Sec.~\ref{sec:carnot} is devoted to study 
 the quantum Carnot cycle in the quasi-static
approximation and its first order S-D corrections, recovering some known results in a slightly broader context. 
Sec.~\ref{sec:otto} instead focuses on the Otto cycle. 
In Sec.~\ref{sec:n-M_model} we finally introduce the specific non-Markovian model. 
Using the preceding section results, we then show how the power output of the cycles
gets affected both for the Carnot machine (Sec.~\ref{non-MarkovCARNOT})  and for the Otto machine
(Sec.~\ref{sec:n-M_otto}). In Sec.~\ref{sec:interpret_n-M} an argument is presented
to interpret the results obtained, focusing on why the information flow induced by non-Markovianity can fasten the speed of thermalization.
Comments and conclusions are presented in 
 Sec.~\ref{sec:comment} 
while the Appendix contains some technical derivations.
\begin{figure}
\includegraphics[width=0.48\textwidth]{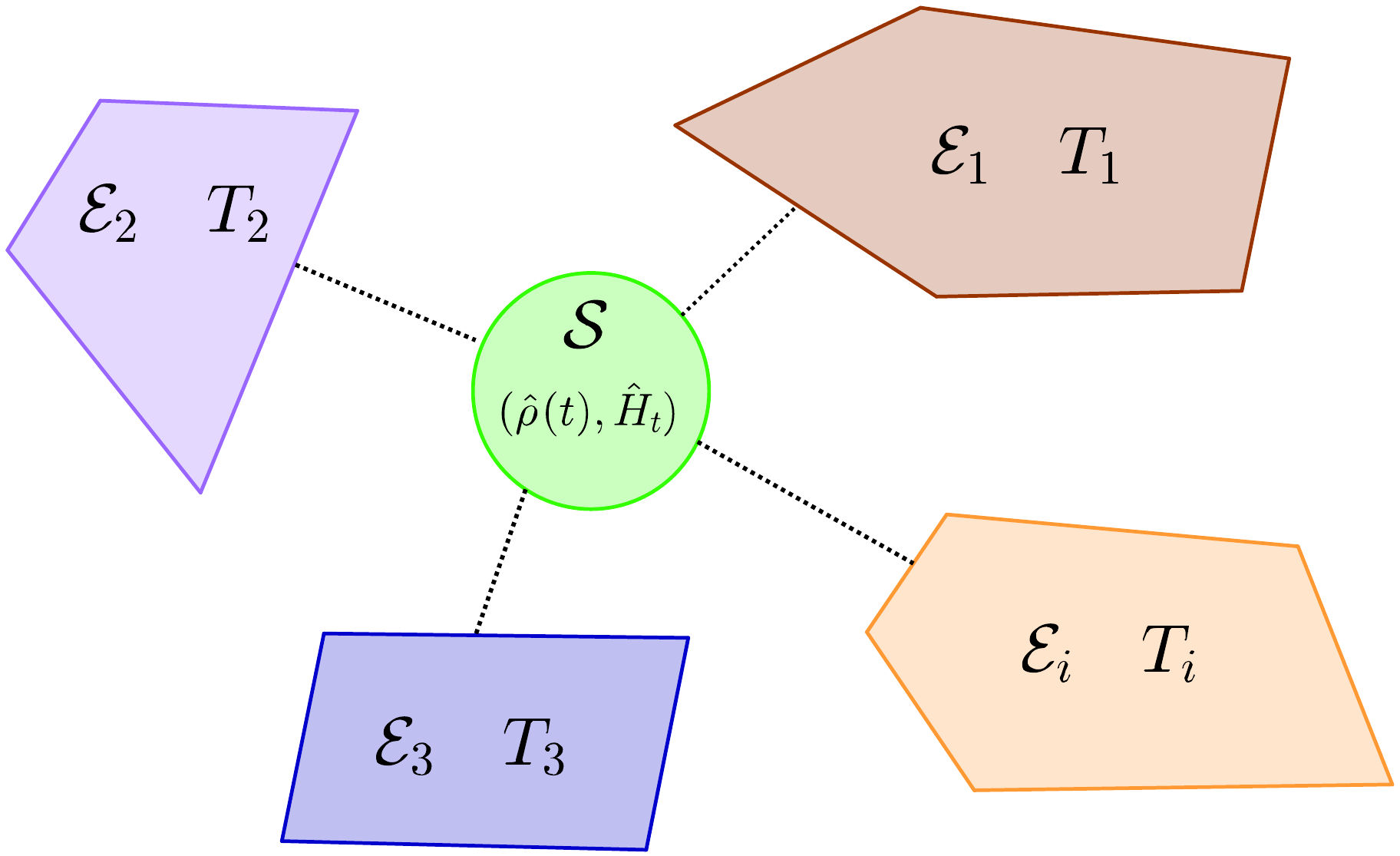}
\caption{General schematics of a quantum thermal machine with working medium described by the quantum system $\mathcal{S}$, characterized by local Hamiltonian ${\hat{H}_t}$, connected with different thermal baths $\mathcal{E}_i$, described by temperatures $T_i$ and coupled with ${\cal S}$.}
\label{fig:generalthermo}
\end{figure}

\section{Quantum thermal machines in the Markovian regime}
\label{sec:qthermal_machine_control}
In this section we review some basic facts about quantum thermal machines in the Markovian
regime, setting the notation and developing the tools that we shall later employ for analysing the
non-Markovian case. 

\subsection{The setup} \label{sec:ii1} 
Consider a quantum working medium ${\cal S}$
characterized by a time-dependent internal Hamiltonian $\hat{H}_t$ which can be externally controlled via some  classical pulses.  
As schematically shown in Fig.~\ref{fig:generalthermo} ${\cal S}$ is coupled to a collection of external thermal baths $\{{\cal E}_j\}$ characterized by temperatures $T_j$,
which are also externally controlled to allow selective activation and deactivation. 
In particular we shall assume at each time $t$ only one of the baths is actively coupled with the working medium. Accordingly,
enforcing the Markovian character in the system-bath interactions, 
we  describe the evolution of 
 ${\cal S}$ in terms of  a Master Equation~\cite{breuer-petruccione} associated with a step-continuous generator
${\cal L}_t$ which, on the time interval ${\cal I}_j$ where only the  $j$-th bath interaction is
active, writes
\begin{equation}
\label{eq:general_QTmachineEq}
\dot{\hat{\rho}}(t)={\cal L}_t[ \hat{\rho}(t)] := -i[\hat{H}_t,\hat{\rho}(t)]_- + \mathcal{D}_t^{(j)}[\hat{\rho}(t)] \ ,
\end{equation}
where $\hat{\rho}(t)$ is the density matrix of ${\cal S}$ at time $t$, 
 $[\cdots, \cdots]_-$ is the commutator symbol, and where finally $\mathcal{D}_t^{(j)}$ is the GKSL dissipator~\cite{G-K-S,Lindblad} mimicking the interaction with ${\cal E}_j$ 
(hereafter  for easy of notation we set both the Plank and the Boltzmann constant equal to one, i.e. $\hbar=k_{\rm B}=1$).  

As indicated by the notation the  $\mathcal{D}_t^{(j)}$s exhibit an explicit time dependence
which, in a weak-coupling regime, we assume to be a direct consequence of the
modulations affecting the system 
Hamiltonian, i.e.
\begin{eqnarray} \mathcal{D}_t^{(j)} = \mathcal{D}^{(j)}(\hat{H}_t)\;. \label{FFD}\end{eqnarray} 
Furthermore, in order to impose proper thermalization conditions on the scheme 
we require $\mathcal{D}_t^{(j)}$ to  admit 
the instantaneous Gibbs state
\begin{equation}
\label{def:gibbs_state}
\hat{\Omega}^{(j)}_{\hat{H}_t}:=\frac{e^{-\beta_j \hat{H}_t}}{\Tr[e^{-\beta_j \hat{H}_t}]}\;, 
\end{equation} 
with $\beta_j:=1/T_j$ being the associated
inverse temperature, as unique fixed point, i.e. 
\begin{equation}\label{STAB} 
\mathcal{D}_t^{(j)}[\hat{\rho}]=0 \; \;  \Leftrightarrow \; \; \hat{\rho}= \hat{\Omega}_{t}^{(j)}\;.
\end{equation}
Notice that the functional dependence of $\hat{\Omega}^{(j)}_{\hat{H}_t}$ with respect to
$\hat{H}_t$, ensures that the requirement Eq.~(\ref{STAB}) is fully compatible with (\ref{FFD}) and it implies that for $t\in{\cal I}_j$, $\hat{\Omega}^{(j)}_{\hat{H}_t}$ is also the unique
fixed point of the full generator ${\cal L}_t$, i.e. 
\begin{equation}\label{STAB1} 
\mathcal{L}_t[\hat{\rho}]=0 \; \;  \Leftrightarrow \; \; \hat{\rho}= \hat{\Omega}^{(j)}_{\hat{H}_t}\;.
\end{equation}
Explicit examples of dissipators $\mathcal{D}_t^{(j)}$ obeying the above constraints
are presented in Appendix~\ref{APP-DISS}, here we only remark that they have been extensively used in the characterization of  equilibration processes induced by fermionic or bosonic baths, see e.g. Refs.~\cite{optimalcontrol,gardiner,breuer-petruccione,esposito-EMP-bounds}. In the absence of Hamiltonian modulations (i.e. for  $\hat{H}_t=\hat{H}$ constant),  Eqs.~(\ref{STAB}) and 
(\ref{STAB1}) ensure that 
 if ${\cal S}$ is left in contact with the $j$-th bath,
 it will be forced by 
(\ref{eq:general_QTmachineEq}) to asymptotically reach thermal equilibrium at temperature $T_j$,~i.e. 
\begin{eqnarray} \label{TERMA} 
\lim_{t\rightarrow \infty} \hat{\rho}(t)= \hat{\Omega}^{(j)}_{\hat{H}}\;, 
\end{eqnarray} 
irrespectively from the initial condition of the problem. 

\subsection{Energy exchanges and thermodynamic consistency} \label{sec:ii2new} 
Within the above theoretical framework the
internal energy $E(t)$ of ${\cal S}$  can be identified with the expectation value of $\hat{H}_t$ on $\hat{\rho}(t)$, i.e. 
\begin{eqnarray} \label{DEFINTE} 
E(t):=\Tr[\hat{\rho}(t) \hat{H}_t]\ .
\end{eqnarray}
Its infinitesimal variation 
comprises two terms which, following the canonical approach of Refs.~\cite{alickiwork,anders-giova,kieuwork,andersreview}, are associated
 respectively with a work  (performed on ${\cal S}$) contribution
 \begin{eqnarray} \label{WORK1} 
d W(t) :=\Tr[\hat{\rho}(t) d \hat{H}_t]\ ,
\end{eqnarray}
and with a heat  (absorbed by ${\cal S}$) contribution
\begin{eqnarray}
  dQ_{j}(t)&:=&\Tr\Big[ \hat{H}_t d\hat{\rho}(t) \Big] \label{DEFQj0}   \\
  &=&   \Tr \Big[ \hat{H}_t \mathcal{D}_t^{(j)}[\hat{\rho}(t)]  \Big] \; dt \;, \label{DEFQJ} 
\end{eqnarray}
where in the second identity we make explicit use of Eq.~(\ref{eq:general_QTmachineEq}), 
 ${\cal E}_j$ being the only bath that
is coupled  with ${\cal S}$ at time $t$.
It is worth stressing that the consistency of the above identifications is explicitly  justified by the Markovian character of
the thermalizing process we are considering.
To see this let us introduce the functional~\cite{parrondo,esposito-2ndlaw} 
\begin{equation} \label{FREEENERGY}
F(\hat{\rho}(t),\hat{H}_t):=E(t)- S(\hat{\rho}(t))/\beta_j \;,
\end{equation}
where $S(\hat{\rho}(t)):= -\mbox{Tr}[ \hat{\rho}(t) \ln \hat{\rho}(t)]$
is the von Neumann entropy of $\hat{\rho}(t)$.
 Exploiting the formal connection between informational and
 thermodynamical entropy, the quantity~(\ref{FREEENERGY}) can
 be identified with  the counterpart of the
 free energy functional of classical equilibrium thermodynamics.
One can easily verify that it obeys  the identity 
\begin{eqnarray} 
 \label{eq:relativeentropy-freeenergy}
F(\hat{\rho}(t),\hat{H}_t)-F(\hat{\Omega}^{(j)}_{\hat{H}_t},\hat{H}_t)= 
S(\hat{\rho}(t) \parallel\hat{\Omega}^{(j)}_{\hat{H}_t})/\beta_j\;,
\end{eqnarray}
where $S(\hat{\rho}_1 \parallel\hat{\rho}_2):=
-S(\hat{\rho}_1)- \mbox{Tr}[ \hat{\rho}_1 \ln \hat{\rho}_2]$ is
the relative entropy  functional~\cite{holevo}. The latter is
know to be
decreasing when the same  completely positive mapping acts on both its argument: 
accordingly, given that  the dynamical generator ${\cal L}_t$ of
Eq.~(\ref{eq:general_QTmachineEq}) 
 is guaranteed to grant complete positive evolution
 and using the invariance (\ref{STAB1}) of 
$\hat{\Omega}^{(j)}_{\hat{H}_t}$ 
 we can claim that $S({\cal L}_t[\hat{\rho}(t)] \parallel
\hat{\Omega}^{(j)}_{\hat{H}_t})\leq 0$.
Inserting this into (\ref{eq:relativeentropy-freeenergy}) 
we can establish that the time derivative of the l.h.s. must be upper bounded by the quantity 
$S(\hat{\rho}(t) \parallel\frac{d}{dt} 
\hat{\Omega}^{(j)}_{\hat{H}_t})/\beta_j$, 
which after proper reordering of the various terms leads  
to  the inequality 
\begin{eqnarray}
d F(\hat{\rho}(t),\hat{H}_t) \leq d W(t)  \Longleftrightarrow \beta_j dQ_{j}(t) \leq dS(\hat{\rho}(t))\;,
\end{eqnarray} 
that is an instance of the 2nd Law of thermodynamics
providing an operational justification for the definitions
 (\ref{WORK1}) and~(\ref{DEFQj0}).

\subsection{Thermodynamic cycles} \label{sec:ii2} 

Integrating Eq.~(\ref{eq:general_QTmachineEq}) we can now analyze the work production rates, their associated efficiencies, and the corresponding
heat fluxes,  of thermodynamic 
cycles where the system ${\cal S}$ is externally driven by an assigned modulation
of the Hamiltonian $\hat{H}_t$  while being put in selective
contact with the baths ${\cal E}_j$s -- see below.
Unfortunately the presence of 
 Hamiltonian modulations makes typically Eq.~(\ref{eq:general_QTmachineEq})  hard to solve. Yet
assuming the time scale at which (\ref{TERMA}) takes places to be short enough, 
one expects  ${\cal S}$ to have enough time to adiabatically follow the instantaneous fixed points of Eq.~(\ref{def:gibbs_state}), obtaining 
\begin{eqnarray} \label{ZEROTH} 
\hat{\rho}(t) \simeq \hat{\Omega}^{(j)}_{\hat{H}_t}\;. 
\end{eqnarray} 
This is the standard quasi-static regime where the working medium is always at thermal  equilibrium with one of the baths. 
Departing from this scenario one enters 
 the regime of {Finite Time Thermodynamics} (FTT)~\cite{andresenFTT}, where
 the time-scales on which the external controls responsible for the modulations of $\hat{H}_t$
 occur, begin to compete with the thermalization times. 
 In what follows we shall study this complex regime by adopting the 
{Slow-Driving (S-D) approximation}  technique introduced in Ref.~\cite{slowdriving}.
The latter is a perturbative approach 
which can be applied to study deviations from Eq.~(\ref{ZEROTH}) in the limit 
of slow variation of~$\mathcal{L}_t$.
As we detail in Appendix~\ref{appendixNEW}, the S-D approximation  can be used as a way for putting on firm ground some of the assumptions
typically adopted in FFT analysis.
It accounts in expressing 
the solution of Eq.~(\ref{eq:general_QTmachineEq}) as an 
expansion series with a  
 perturbation parameter  given by the ratio $\tau_R/\tau$ between 
the typical timescale $\tau \sim \|\dot{\mathcal{L}}_t/\mathcal{L}_t\|$ associated with the variation of the dynamics generator,  and  the typical relaxation time $\tau_R$ 
governing the convergence of the limit~(\ref{TERMA}).  
At the lowest orders one has 
\begin{eqnarray} \label{expansion} 
\hat{\rho}(t) = \hat{\rho}^{(0)}(t)+\hat{\rho}^{(1)}(t) + ... \;,\end{eqnarray}
 with $\hat{\rho}^{(0)}(t):= \hat{\Omega}^{(j)}_{\hat{H}_t}$ being the zero-th order term, while the first order
 correction $\hat{\rho}^{(1)}(t)$ is obtained as~\cite{slowdriving}
\begin{eqnarray}
\label{eq:rho^1}
\hat{\rho}^{(1)}(t)=(\mathcal{L}_t\mathcal{P})^{-1}[ \dot{\hat{\rho}}^{(0)}(t)]\,,
\end{eqnarray}
where $\mathcal{P}$ is  the projector on the null-trace subspace of linear operators (its presence being
 required to make $\mathcal{L}_t$ invertible, under the assumption of unique null eigenstate).
Therefore, by direct substitution in Eq.~(\ref{DEFQJ}) we get 
\begin{eqnarray} \label{ddf} 
d Q_{j}(t)\simeq 
 d Q_j^{(0)}(t) +  dQ_j^{(1)}(t) \;,
 \end{eqnarray} 
 where
 \begin{eqnarray}  \label{ORDzero} 
 d Q_j^{(0)}(t) :=
 \Tr\Big[ \hat{H}_t d\hat{\rho}^{(0)}(t) \Big] = d S^{(0)}(t)/\beta_j\;,
\end{eqnarray}
is the quasi-static contribution which, by using the fact that 
 $\hat{\rho}^{(0)}(t)$ is the Gibbs state $\hat{\Omega}^{(j)}_{\hat{H}_t}$,
we  expressed in terms of the infinitesimal increment the von Neumann entropy 
$S^{(0)}(t):=-\mbox{Tr}[ \hat{\rho}^{(0)}(t) \ln \hat{\rho}^{(0)}(t)]$ of the latter, 
 and where 
 \begin{eqnarray}  \label{DDF1} 
 d Q_j^{(1)}(t) =  
    \Tr\Big[ \hat{H}_t d\hat{\rho}^{(1)}(t) \Big]\;, 
 \end{eqnarray}
 is the  first order correction term.

 \section{Thermodynamic cycles optimisation} \label{sec:cycles} 
In this section we will show how it is possible to optimize the control on a quantum engine in order to maximize is performance, i.e. its power output, addressing the paradigmatic case of  Quantum Carnot 
and Otto cycles performed on a two-level (qubit) system ${\cal S}$ which evolves under the influence of a hot bath H and a cold bath C, the modulation of its Hamiltonian being associated with control pulses that act on its energy gap
$\epsilon(t)\geq 0$, i.e.
\begin{equation}\label{energy} 
\hat{H}_t={\epsilon}(t) \hat{\sig}^z /2 \ ,
\end{equation}
with $\hat{\sig}^z$ being the third Pauli matrix, with eigenstates $|0\rangle$ and $|1\rangle$. It's not difficult to generalise these cycles (in the quasi-static regime) to more general Hamiltonians.
While in deriving the above considerations we shall make explicit reference to the expressions we developed in Sec.~\ref{sec:qthermal_machine_control} for the Markovian regime, we stress that the results we obtain also hold for non-Markovian dynamics, as we shall use them later in Sec.~\ref{sec:n-M_model} to analyse the non-Markovian model we present.

\subsection{Quantum Carnot Cycle} \label{sec:carnot} 
\begin{figure}
\centering
\includegraphics[width=0.48\textwidth]{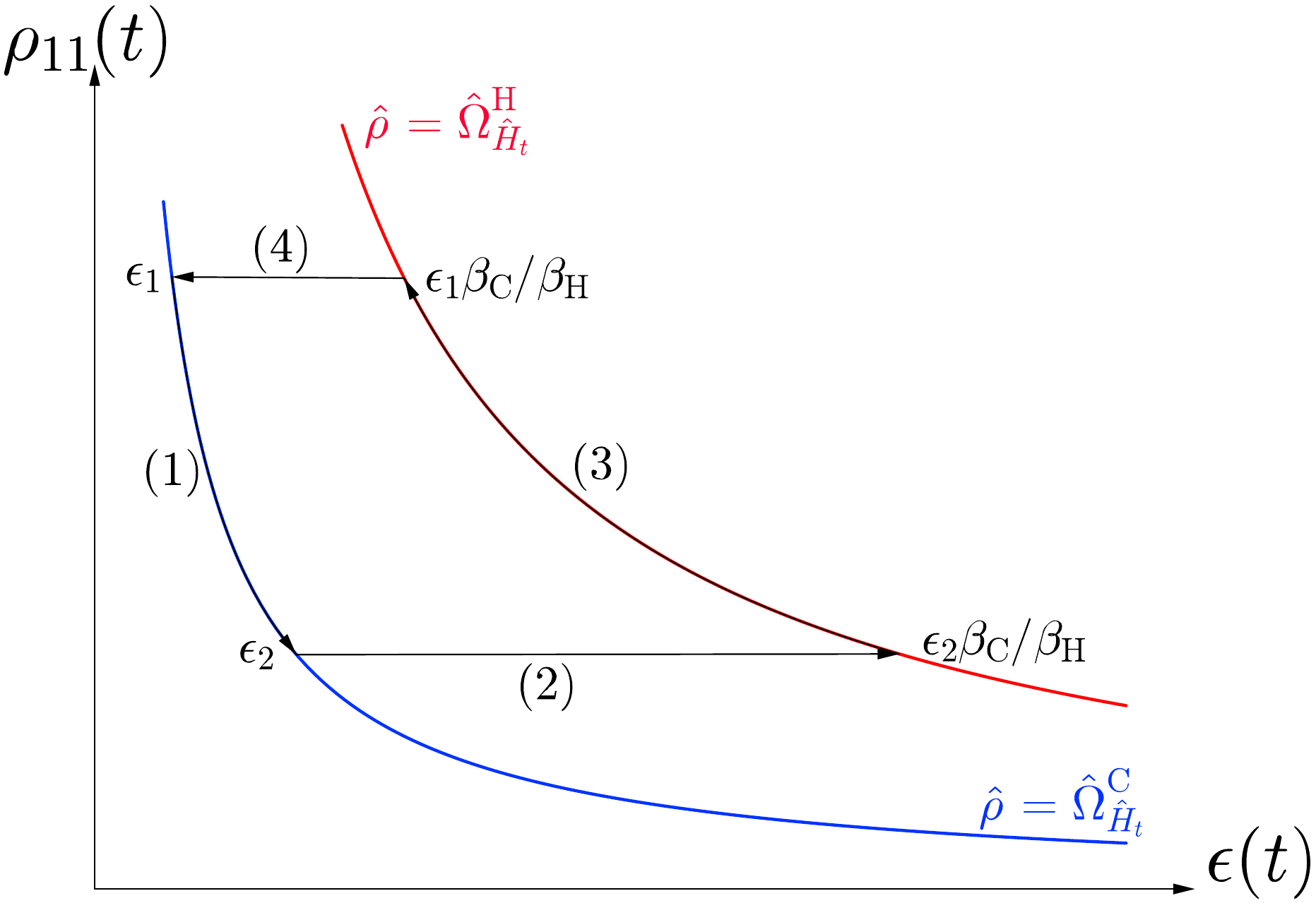}
\caption{(Color online) Pictorial representation of the (quasi-static) Quantum Carnot cycle  in the configuration space $\rho_{11}(t):= \langle 1| \hat{\rho}(t) |1\rangle$ vs. the energy gap $\epsilon(t)$ in the ideal quasi-static limit. Steps 1) and 3) correspond to the isothermal transformations where ${\cal S}$ is kept in contact with the cold   bath $\rm C$   and hot bath $\rm H$, respectively; steps 2) and 4) 
instead represent sudden quenches of the Hamiltonian gap.  }
\label{fig:Carnot}
\end{figure}
A Quantum Carnot cycle is identified with a 4 steps process inspired directly by its classical counterpart, that is two isothermal strokes where the Hamiltonian of ${\cal S}$ is modulated while keeping the system in thermal contact with one of the two baths, alternated with two iso-entropic (adiabatic) strokes, where instead the Hamiltonian undergoes to instantaneous sudden switches (quenches).
In the ideal quasi-static limit~(\ref{ZEROTH})
 the operations are performed slowly enough to allow the system to be in thermal equilibrium at every instant, i.e. 
states which for the Hamiltonian~(\ref{energy}) 
 can be expressed as 
\begin{equation} \label{GIBBO1}  
\hat{\Omega}^{({j})}_{\hat{H}_t}  = \hat{\Omega}^{({j})}_{\epsilon(t)} := 
p_j(\epsilon(t))  |0\rangle\langle 0| + (1-p_j(\epsilon(t))) |1\rangle\langle 1|\;,
\end{equation} 
with 
\begin{eqnarray} 
 p_j(\epsilon)  : = 
\label{def:p_gibbs}
\dfrac{1}{1+e^{-\beta_j \epsilon}}
\;,
\end{eqnarray} 
being the associated  ground state population.
 
  In this case the 4 steps  of the cycle are  as in Figure \ref{fig:Carnot}:
\begin{itemize}
\item [1)] while being coupled to the cold reservoir C, the energy gap is modified continuously and monotonically, from the initial value $\epsilon_{1}$
 to $\epsilon_{2}\geq \epsilon_1$ (more precisely we require $\epsilon(t)$ to be continuous and differentiable with first order derivative which is not negative);
\item [2)] with the system isolated from the reservoirs, a quench is now performed by suddenly taking
the gap from  $\epsilon_{2}$  to 
\begin{eqnarray}
\label{EPS3}  \epsilon_3 := \epsilon_{2}{\beta_{\rm C}}/{\beta_{\rm H}} \;,\end{eqnarray} 
which by construction is larger than $\epsilon_2$, i.e. $
\epsilon_3\geq \epsilon_2$; 
\item [3)] while being coupled to the hot reservoir H, the energy gap is then modified continuously, and monotonically, from 
$\epsilon_{3}$ to 
\begin{eqnarray}
\epsilon_{4} \label{EPS4} 
:=\epsilon_{1}{\beta_{\rm C}}/{\beta_{\rm H}} \;, 
\end{eqnarray}  that automatically fulfils the constraint 
$\epsilon_1 \leq \epsilon_4 \leq \epsilon_{2}{\beta_{\rm C}}/{\beta_{\rm H}}=
\epsilon_{3}$ (again, more precisely we require $\epsilon(t)$ to be continuous and differentiable with 
first order derivative that is non-positive); 
\item [4)] finally isolating the system a quench is performed to restore 
the gap at the initial value $\epsilon_{1}$.
\end{itemize}
It is worth pointing out that the continuity requirement of $\epsilon(t)$ during the steps 1) and 3)
is inserted in order to make sure that one could later on apply the S-D expansion
which needs to have a zero-th order contribution of term differentiable -- see Eq.~(\ref{expansion}). More specifically in what follows we shall
require $\epsilon(t)$ to have null first order derivative at the extrema of the isotherms. This is a technical assumption which we introduce in order to ensure the solution of the dynamics \eqref{expansion} to be continuous and differentiable also in proximity of the quenches (where it  coincides with the Gibbs state \eqref{GIBBO1}), which  in turn implies that  no first order correction \eqref{eq:rho^1} at the extremal points of the isothermal strokes has to be expected. 
The monotonicity behaviour of $\epsilon(t)$ during the steps 1) and 3) is instead motivated by 
energetic considerations. As a matter of fact  
having set the gap to evolve monotonically from $ \epsilon_1\geq 0$ to 
$\epsilon_2\geq \epsilon_1$, we can ensure that at each instant of step 1) the system
always releases heat to the cold bath without absorbing it: 
this can be easily verified by observing that the von Neumann entropy of a Gibbs
state~(\ref{GIBBO1})
writes  
\begin{equation} 
S(\hat{\Omega}^{({j})}_{\epsilon}) = -p_j(\epsilon) \ln p_j(\epsilon)  - (1-p_j(\epsilon))  \ln (1-p_j(\epsilon))\;,
\end{equation} 
which is monotonically decreasing with $\epsilon$, and from the fact that 
 at the lowest order in the expansion (\ref{ORDzero}) 
  the associated incremental heat can be expressed as 
\begin{eqnarray} \label{ffr1} 
d Q_{\rm C}(t)\simeq d Q^{(0)}_{\rm C}(t) &=& 
 \frac{ d S(\hat{\Omega}^{({\rm C})}_{\epsilon(t)}) }{\beta_{\rm C}}
 \leq 0\;,  \quad \forall t\in{\cal I}_{\rm C} \;.
\end{eqnarray}
Similarly having ensured that in step 3)  the value of the gap decreases monotonically 
from $\epsilon_3$ to $\epsilon_4\geq \epsilon_3$, we can guarantee that
the heat in the process is always absorbed from the bath H, i.e. 
\begin{eqnarray}\label{ffr2} 
d Q_{\rm H}(t)\simeq d Q^{(0)}_{\rm H}(t) &=& 
\frac{ d S(\hat{\Omega}^{({\rm H})}_{\epsilon(t)})}
{\beta_{\rm H}}  \geq 0\;,  \quad \forall t\in{\cal I}_{\rm H} \;.
\end{eqnarray}
Thanks to these properties,  and by the observation that
of course no heat is exchanged between ${\cal S}$ and the baths during the steps 2) and 4), 
 the total heat absorbed   by the working medium in a cycle can
be obtained by integrating (\ref{ffr2}) over the full duration of step 3), i.e. 
\begin{eqnarray}\label{DD1} 
Q_{\rm ABS} &=& \int_{{\cal I}_{\rm H}} d Q_{\rm H}(t)  \\
&\simeq& \int_{{\cal I}_{\rm H}} d Q^{(0)}_{\rm H}(t)   =  \nonumber
\frac{ S(\hat{\Omega}^{({\rm H})}_{\epsilon_4}) - S(\hat{\Omega}^{({\rm H})}_{\epsilon_3})}
{\beta_{\rm H}} \geq 0\;,  
\end{eqnarray}
while the total released heat is given by
\begin{eqnarray}\label{DD2} 
Q_{\rm REL}&=&  \int_{{\cal I}_{\rm C}} d Q_{\rm C}(t) \\
&\simeq& \int_{{\cal I}_{\rm C}} d Q^{(0)}_{\rm C}(t)  \nonumber   = 
\frac{ S(\hat{\Omega}^{({\rm H})}_{\epsilon_2}) - S(\hat{\Omega}^{({\rm H})}_{\epsilon_1})}
{\beta_{\rm C}} \leq 0\;.
\end{eqnarray}
Notice also that  the constraints~(\ref{EPS3}) and (\ref{EPS4}) impose 
\begin{eqnarray} \label{CONTINUITY} 
p_{\rm H}(\epsilon_3)=p_{\rm C}(\epsilon_2) \; \qquad 
p_{\rm H}(\epsilon_4)=p_{\rm C}(\epsilon_1) \;,
\end{eqnarray} 
which implies 
$\hat{\Omega}^{({\rm H})}_{\epsilon_3}= \hat{\Omega}^{({\rm C})}_{\epsilon_2}$ and $\hat{\Omega}^{({\rm H})}_{\epsilon_4}= \hat{\Omega}^{({\rm C})}_{\epsilon_1}$.
Accordingly by direct inspection of (\ref{DD1}) and (\ref{DD2}) 
we obtain  the  fundamental identity 
\begin{eqnarray}\label{DELTAG} 
\Delta Q^{(0)}_{\rm H}  = - \frac{\beta_C}{\beta_H}
\Delta Q^{(0)}_{\rm C}  \;,
\end{eqnarray} 
which we
expressed in terms of the simplified notation $\Delta Q^{(0)}_{j} := \int_{{\cal I}_{j}} d 
Q^{(0)}_{j}(t)$.
Now, since the work produced by ${\cal S}$ on a cycle can be identified with $Q_{\rm ABS} +Q_{\rm REL}$ by invoking the
internal energy conservation,  
 the efficiency 
 (work done over heat absorbed) of the process can be shown to correspond to the
 Carnot efficiency $\eta_{\text{c}}:=1-\frac{\beta_{\rm H}}{\beta_{\rm C}}$. Indeed 
\begin{equation}
{\eta}:=\frac{Q_{\rm ABS} +Q_{\rm REL}}{Q_{\rm ABS}}
\simeq 1+\frac{\Delta Q^{(0)}_{\rm C}
}{\Delta Q^{(0)}_{\rm H} }
= \eta_{\text{c}} \;, \label{ddfCARNOT} 
\end{equation}
the last identity following directly from (\ref{DELTAG}). It is worth stressing that
Eqs.~(\ref{ffr1}) and (\ref{ddfCARNOT}) 
are universal results that do not depend on the specific 
 structure of the generators
$\mathcal{D}_t^{(j)}$ entering the system ME. This is a consequence of the 
{quasi-static} approximation~(\ref{ZEROTH})
 in which,  as in classical thermodynamics, complete thermalization is allowed at any time in contact with a thermal source: in this regime no explicit dynamics as in Eq.~(\ref{eq:general_QTmachineEq}) is needed to describe the thermodynamics of the engine, neither the
 exact temporal dependence of the control $\epsilon(t)$, except the  properties of the equilibrium state (\ref{def:gibbs_state}) and the knowledge of the Hamiltonian at the turning points of the protocol. 
 All this of course 
 holds true 
as long as we can neglect the first-order contributions in the S-D expansion~(\ref{expansion}). 
To account for them we now 
use (\ref{ddf}) to refine Eqs~(\ref{DD1}) and (\ref{DD2}), writing 
 $Q_{\rm REL} \simeq \Delta Q^{(0)}_{\rm C} + \Delta Q^{(1)}_{\rm C}$
 and $Q_{\rm ABS} \simeq \Delta Q^{(0)}_{\rm H} + \Delta Q^{(1)}_{\rm H}$
 with 
 \begin{eqnarray} \Delta Q^{(1)}_{j} := \int_{{\cal I}_{j}} d  \label{DEFDQ1} 
Q^{(1)}_{j}(t)\;, \end{eqnarray} obtaining 
\begin{eqnarray}
{\eta}&=&1+\frac{Q_{\rm REL}}{Q_{\rm ABS}} \simeq 1 +
 \frac{\Delta Q^{(0)}_{\rm C}}{\Delta Q^{(0)}_{\rm H}}
\frac{1+ \Delta Q^{(1)}_{\rm C}/\Delta Q^{(0)}_{\rm C}}{1+ \Delta Q^{(1)}_{\rm H}/\Delta Q^{(0)}_{\rm H}}\nonumber \\ \label{EFFIC} 
&=& 1 - (1- \eta_{\text{c}}) \frac{1+ \alpha_{\rm C}}{1+ \alpha_{\rm H}}\;,
\end{eqnarray}
where in the last identity we employed  (\ref{ddf}) to express the ratio 
$\Delta Q^{(0)}_{\rm C}/\Delta Q^{(0)}_{\rm H}$ in terms of 
 the Carnot efficiency and for $j\in \{ {\rm H,C}\}$ introduced
the parameter 
\begin{eqnarray} \label{DEFALPHA} 
\alpha_j := \Delta Q^{(1)}_{j}/\Delta Q^{(0)}_{j}\;,
\end{eqnarray}  
to gauge the ratio between the first and the zero-th order heat contributions
associated with the $j$-th bath. In a similar fashion we can also express the power $P$  associated with
the work production per cycle. Indicating hence with 
$\tau_{\rm H}$ and $\tau_{\rm C}$ the durations of the transformations 1) and 3) (the only 
being time-consuming given that step 2) and 4) are assumed to be instantaneous), we write 
\begin{eqnarray} 
\label{def:power}
P&:=&\frac{{Q_{\rm ABS}}+{Q_{\rm REL}}}{\tau_{\rm C}+\tau_{\rm H}} \simeq  
\frac{{\Delta Q^{(0)}_{\rm C}}+{\Delta Q^{(0)}_{\rm H}}+{\Delta Q^{(1)}_{\rm C}}+{\Delta Q^{(1)}_{\rm H}}}{\tau_{\rm C}+\tau_{\rm H}} \nonumber \\
&=&{\Delta Q^{(0)}_{\rm H}} \frac{\eta_{\text{c}}+ \alpha_{\rm H} -
(\beta_{\rm H}/\beta_{\rm C})  \alpha_{\rm C} }{\tau_{\rm C}+\tau_{\rm H}} \;,
\end{eqnarray}
where  we used Eqs.~(\ref{DELTAG}) and (\ref{DEFALPHA}).

\subsubsection{Performance optimization in the S-D regime} \label{sec:perf} 
To proceed with our analysis we need to provide some details on the system ME
and in particular on the GKSL dissipators which define it. 
As a preliminary step, however we observe that thanks to our choice~(\ref{energy})
we can express Eq.~(\ref{DEFQJ}) as 
\begin{eqnarray}
  dQ_{j}(t)=\frac{1}{2} \epsilon(t)[d {\rho}_{11}(t) - d {\rho}_{00}(t) ]=  
- \epsilon(t)d {\rho}_{00}(t)\ ,\ \
    \label{DEFQJ1} 
\end{eqnarray}
where for $k,k' =0,1$,  ${\rho}_{kk'}(t):= \langle k|\hat{\rho}(t)|k'\rangle$ are the matrix
 elements of $\hat{\rho}(t)$ with respect to the eigenbasis of $\hat{H}_t$ and where
 in the second identity we use the normalization condition $\mbox{Tr}[ \hat{\rho}(t)]=1$
 to write $d {\rho}_{11}(t)= - d {\rho}_{00}(t)$.
 Due to linearity Eq.~(\ref{DEFQJ1}) applies to all orders of the S-D expansion~(\ref{ddf}), implying in particular that Eqs.~(\ref{ORDzero}), (\ref{DDF1}) take the form
 \begin{eqnarray}
  dQ^{(0)}_{j}(t)&=& - 
\epsilon(t)d {\rho}^{(0)}_{00}(t)\;, \nonumber \\   dQ^{(1)}_{j}(t)&=& -
\epsilon(t)d {\rho}^{(1)}_{00}(t)\;,
    \label{DEFQJ2} 
\end{eqnarray}
where ${\rho}^{(0)}_{00}(t)$ and ${\rho}^{(1)}_{00}(t)$ are respectively the zero-th and first order contribution to the population of the ground state of ${\cal S}$. 
The first of these two terms is nothing but the function~(\ref{def:p_gibbs}), i.e.
${\rho}^{(0)}_{00}(t) = p_j(\epsilon(t))$. 
The second instead can be determined exploiting~Eq.~(\ref{eq:rho^1}).
In particular due to the linearity of operators in
Eq.~(\ref{eq:rho^1}) and the one-parameter dependence of $\hat{\rho}^{(0)}$ it is possible to draw, in full generality, the following formal connection between ${\rho}^{(1)}_{00}(t)$ and the function 
$p_j(\epsilon(t))$ which, effectively,  becomes the real control parameter of the setting.
Specifically we get
\begin{equation}
\label{eq:qubit_ground_general}
{\rho}^{(1)}_{00}(t)=-A_j[p_j(\epsilon(t))] \;  \frac{d}{dt} {p}_j(\epsilon(t))\;,
\end{equation}
where $A_j$, which we 
dub the \emph{S-D amplitude} of the problem,
quantifies how large is the first order correction determining the relaxation timescale of the setup. In general, besides
 depending on the the parameters of the model, the S-D amplitude is an explicit 
 functional  of ${p}_j(\epsilon(t))$, e.g. as in the case of dissipators $\mathcal{D}_t^{(j)}$ associated with
Bosonic baths defined by Eq.~(\ref{DIS1}) with rates as in (\ref{BOSONICRATE})
for which we get $A_j=(2p_j(\epsilon(t))-1)/\Gamma_j$. 
 When considering instead as dissipators $\mathcal{D}_t^{(j)}$ the super-operators defined in Eq.~(\ref{PRIMO}) or those associated with fermionic baths defined by Eq.~(\ref{DIS1}) with rates as in (\ref{FERMIRATE}), one gets  an S-D amplitude which is constant, i.e. 
 \begin{eqnarray} \label{ddsf} 
 A_j={1}/{\Gamma_j}\;,\end{eqnarray}  with $\Gamma_j$ being a fundamental constant of the model.  In what follows, for the sake of simplicity we shall focus on this special case: 
 our finding however can be approximatively applied to all those configurations where, for all
 $t \in {\cal I}_{j}$,  
 $A_j$ is a slowly varying functional of $p_j(\epsilon(t))$.

With the help of the above identities
 we can hence  cast (\ref{DEFDQ1}) as
\begin{eqnarray} 
\Delta Q_j^{(1)}
&=&\frac{A_j}{\beta_j}\int_{{\cal I}_j} d t \;  
\ln\big(\tfrac{p_j(t)}{1-p_j(t)}\big)\;  \ddot{p}_j(t)
\nonumber \\ \label{eq:heats1}
&=& - \frac{A_j}{\beta_j}
 \int_{{\cal I}_j} d t \;   \frac{ [\dot{p}_j(t)]^2}{p_j(t) (1-p_j(t))} \;, 
\end{eqnarray} 
where in the first identity we used Eq.~(\ref{def:p_gibbs}) to write $\epsilon(t)$ in terms of $p_j(t):=p_j(\epsilon(t))$, i.e.
$\epsilon(t) = \frac{1}{\beta_j} \ln\big(\tfrac{p_j(t)}{1-p_j(t)}\big)$, 
and in the second we adopted integration by parts exploiting the fact that 
at the extrema of the isotherms steps the control functions have been set to  have null first order derivative. 
Equation~(\ref{eq:heats1}) 
should be compared with the zero-th order term $\Delta Q_j^{(0)}$ which we have already computed in the previous section and which, expressed in terms $p_j(t)$,  results to be the integral of an exact differential that
depends only on the initial and final values $p_{j}^{(in)}$ 
 and $p_{j}^{(fin)}$ assumed on  
the interval ${\cal I}_j$, i.e. 
\begin{eqnarray} 
\Delta Q_j^{(0)}&=&-\frac{1}{\beta_j}\int_{{\cal I}_j}   dt \;\ln\big(\tfrac{p_j(t)}{1-p_j(t)}\big)\; 
 \dot{p}_j(t) \nonumber \\
 &=&
 -\frac{1}{\beta_j}\int_{{\cal I}_j}   dp \;\ln\big(\tfrac{p}{1-p}\big) \nonumber  \\ &=&
 \frac{1}{\beta_j} \left( \ln (1-p) + p \ln \frac{p}{1-p}\right)\label{eq:heats0}
 \Big\rvert_{p_{j}^{(in)}}^{p_{j}^{(fin)}}\;, 
 \end{eqnarray} 
 the last identity being an alternative way of expressing the entropy increment of the Gibbs state
 (\ref{GIBBO1}). 

Our next problem is  to determine which choices of $\epsilon(t)$, or equivalently of $p_j(t)$, can be used in order to guarantee better performances with respect to the quasi-static regime. 
To begin with it is worth stressing that from Eq.~(\ref{eq:heats1}) 
it follows that  
 for all choices of the control functions  the first order correction term to the heat is always
 negative semi-definite, i.e.
\begin{eqnarray} \label{Q1neg}
\Delta Q_j^{(1)} \leq 0\;,
\end{eqnarray} 
which in turn implies 
\begin{eqnarray} 
\alpha_{\rm H} \leq 0\;, \qquad \qquad \alpha_{\rm C} \geq 0\label{IMPIMP} \;,\end{eqnarray} 
due to the positivity of $\Delta Q_{\rm H}^{(0)}$ and the negativity of $\Delta Q_{\rm C}^{(0)}$ (incidentally we observe that ~\eqref{Q1neg} continues to hold by the same argument even if $A_j(p_j)$ is not constant but explicitly dependent on the control $p_j(t)$).
  The first consequence of Eq.~(\ref{IMPIMP}) is the fact that the efficiency $\eta$
 of Eq.~(\ref{EFFIC})  cannot be larger than $\eta_{\text c}$,
  as one expects from the second principle of thermodynamics (formally speaking
  to show that $\eta \leq \eta_{\text c}$ we also need $|\alpha_j|\ll 1$ which however
  is always implicit assumed by the perturbative character of the S-D approach). 
  At the level of the power~(\ref{def:power}) we notice instead that 
  first order corrections explicitly depend on features which one may try to optimize with proper choices of the controls. 
  For this purpose
looking at the expression (\ref{eq:heats1}) we can isolate different contributions:
\begin{itemize}
\item \textbf{Control speed:}
 keeping the same shape (and extrema) for the driving protocol, we can modify its duration via the mapping $\tau_j \rightarrow\lambda\tau_j$ with $\lambda>0$.
 By a simple change of variable $t\rightarrow t/\lambda$ in Eq.~(\ref{eq:heats1}), it is immediate to find that this induces the following rescaling 
\begin{equation}
\label{eq:speedscaling-b}
\Delta Q_j^{(1)} \xrightarrow[{t\rightarrow t/\lambda}]{} \Delta Q_j^{(1)}/\lambda\;, 
\end{equation}
while, of course, the zero-order terms $\Delta Q_j^{(0)}$  are unaffected; 
\item \textbf{Control shape:} over a fixed time length, we can clearly optimize 
with respect to the shape of the function $\epsilon(t)$, i.e. with respect to the 
function $p_j(t)$ under the constraint i), ii) and iii). Once more this will induce
a modification of $\Delta Q_j^{(1)}$ while leaving unaffected the zero-order contribution terms;
\item \textbf{S-D amplitude selection:} 
this is the main figure of merit after control optimisation. It merely consists in selecting
different kind of bath-system interactions in order to influence the value of 
$\Delta Q_j^{(1)}$ via its dependence upon the S-D amplitude $A_j$ (this optimization will be 
specifically analyzed in the study of non-Markovian models).
\end{itemize}

Let us first analyze how the power $P$ is affected by   Speed Control optimization.
Using the scaling relations (\ref{eq:speedscaling-b}) 
we find that
Eq.~(\ref{def:power}) changes as 
\begin{equation}
P \xrightarrow[{t\rightarrow t/\lambda_{\rm H,C}}]{}
{\Delta Q^{(0)}_{\rm H}} \frac{\eta_{\text{c}}- |\alpha_{\rm H}|/\lambda_{\rm H} - 
(\beta_{\rm H}/\beta_{\rm C})  \alpha_{\rm C}/\lambda_{\rm C} }{\tau_{\rm C}\lambda_{\rm C}+\tau_{\rm H}\lambda_{\rm H}} \;,\label{powerresc} \end{equation}
while the associated efficiency 
\begin{eqnarray}  \label{ETAD} 
\eta \xrightarrow[{t\rightarrow t/\lambda_{\rm H,C}}]{} 
 1 - (1- \eta_{\text{c}}) \frac{1+ \alpha_{\rm C}/\lambda_{\rm C}}{1- |\alpha_{\rm H}|/\lambda_{\rm H}}\;,
\end{eqnarray} 
where we used (\ref{IMPIMP})  to rewrite $-\alpha_{\rm H}=|\alpha_{\rm H}|$, and where 
$\lambda_{\rm C},\lambda_{\rm H}>0$ represent the stretching of 
the intervals ${\cal I}_{\rm C}$ and ${\cal I}_{\rm H}$, respectively. 
A simple analytical study reveals that the function~(\ref{powerresc})  admits a maximum for 
\begin{align}
\lambda_{\rm C}&=\frac{2\alpha_{\rm C} \beta_{\rm H}}{{\eta}_{\text{c}}\beta_{\rm C}}\bigg(1+\sqrt{\frac{|\alpha_{\rm H}| \beta_{\rm C} \tau_{\rm C}}{\alpha_{\rm C} \beta_{\rm H}\tau_{\rm H}}}\bigg) \ ,\\
\lambda_{\rm H}&=\lambda_{\rm C} \; \frac{\tau_{\rm C}}{\tau_{\rm H}} \; 
\sqrt{\frac{|\alpha_{\rm H}| \beta_{\rm C}}{\alpha_{\rm C} \beta_{\rm H}}}  \ . \label{LAMBDA} 
\end{align}
Replacing these values into (\ref{powerresc}) and (\ref{ETAD}) 
 it is possible then to express the maximum power $P_{\max}$ and the correspondent efficiency at maximum power (EMP) that we indicate with the symbol $\eta^*$. For the sake of simplicity let us now suppose $\alpha_{\rm C}=|\alpha_{\rm H}|=\alpha$ and
 $\tau_{\rm C}=\tau_{\rm H}=\tau$, a regime attained for instance
 under symmetric bath couplings and driving  assumptions, i.e. 
 posing $A_{\rm C}=A_{\rm H}$  and requiring $\epsilon(t)$ during the hot isotherm to be the time-reversal of the cold isotherm -- see however Ref.~\cite{slowdriving} for an explicit treatment of the cases where this hypothesis is relaxed.  In this case, 
 using 
 Eq.~(\ref{DELTAG}) and 
the quasi-static relation $\Delta S_j^{(0)}=\Delta Q_j^{(0)}/T_j$ by direct integration of \eqref{ORDzero},
 we find 
\begin{eqnarray} 
P_{\max} 
\label{res:pmaxS-W}
&=& \Delta Q_{\rm H}^{(0)}\Delta Q_{\rm C}^{(0)} \dfrac{(\sqrt{\beta_{\rm C}}-\sqrt{\beta_{\rm H}})^2 
}{A F[q_x]} \nonumber\\&=& 
(\Delta S^{(0)})^2 \dfrac{(\sqrt{T_{\rm C}}-\sqrt{T_{\rm H}})^2 
}{4A {\cal F}[q_x]}\ ,
\end{eqnarray}
where we introduced the adimensional  functional
\begin{equation}
\label{def:F}
{\cal F}[q_x] := 
\Big| \int_0^1 dx \;  \ln\left(\tfrac{q_x}{1-q_x}\right)
\tfrac{d^2{q}_x}{d^2x} \Big|\;, 
\end{equation}
the variable $x:=t/\tau$ being a rescaled temporal coordinate and  $q_x:= p(x\tau)$.
Regarding the EMP instead we get  
\begin{equation}
\eta^*=1-\sqrt{\dfrac{\beta_{\rm H}}{\beta_{\rm C}}} = 1-\sqrt{\dfrac{T_{\rm C}}{T_{\rm H}}}\ ,
\end{equation}
which is the Curzon-Ahlborn efficiency~\cite{curzon-ahlborn}, see also Appendix~\ref{appendixNEW}.

Equation (\ref{res:pmaxS-W}) implies that the maximum power
$P_{\max}$ is inversely proportional to S-D amplitude, hence 
the larger values of $A_j$ is, the worse the effects on thermodynamic performance are (note that also the efficiency \eqref{EFFIC} worsen for larger $A_j$s).
Regarding the shape-pulse optimization instead, remembering that
the zero-th order terms $\Delta Q_{\rm H}^{(0)}$, and $\Delta Q_{\rm C}^{(0)}$ are not affected
such choice, we observe that larger values of $P_{\max}$ are attained by minimizing 
the term ${\cal F}[q_x]$ 
appearing at the denominator for all possible choices of a monotonic, continuous, differentiable function $q_x \in [0,1]$, i.e. 
\begin{eqnarray} \nonumber 
P_{\max}= (\Delta S^{(0)})^2 \dfrac{(\sqrt{T_{\rm C}}-\sqrt{T_{\rm H}})^2
}{4A\; {\cal F}_{\min}}\ .
\end{eqnarray}
For each assigned  initial and final values $q_0$ and $q_1$ of $q_x$, the problem can
be solved by a variational study of the integrand, leading to solutions of the form  
$q_x=\frac{1+\cos({\Omega}(x+\varphi))}{2}$, cf. Appendix \ref{app:optimalshape}.
The resulting value of $P_{\max}$ obtained with such a driving reaches  the maximal performances for $q_1=q_0+\varepsilon$, for which it is possible to obtain an analytic expression valid in the $\lim \varepsilon\rightarrow0$, that is
\begin{eqnarray}
P_{\max}
&=&  \xi \dfrac{(\sqrt{T_{\rm H}}-\sqrt{T_{\rm C}})^2}{A}\;,
 \label{MAXP1}
\end{eqnarray}
which we expressed in terms of the bath temperatures $T_{\rm H}$ and 
$T_{\rm C}$, with $\xi$ being the numerical constant
\begin{eqnarray}
\xi := \max_{q_0}\left\{ \left[ \ln\big(\tfrac{q_0}{1-q_0}\big)\right]^2\tfrac{q_0(1-q_0)}{4}\right\}
\simeq 0.11\;,
\end{eqnarray} 
the maximum being reached for 
 $q_0 \simeq 0.92$, which thus corresponds to the optimal thermal ground state population around which the cycle shall be performed, or in terms of the energy gap $\epsilon/T\simeq 2.4$. 
Note that this result accounts in 
taking  $q_0\sim q_1$ which formally corresponds to performing a quasi-Otto cycles \cite{t.b.p.}, as it has been found for the exact optimal control of Carnot cycle in Ref.~\cite{optimalcontrol}(with a specific dissipator) and \cite{erdman2018maximum}. 

\subsection{Otto cycle}\label{sec:otto} 
\begin{figure}
\centering
\includegraphics[width=0.48\textwidth]{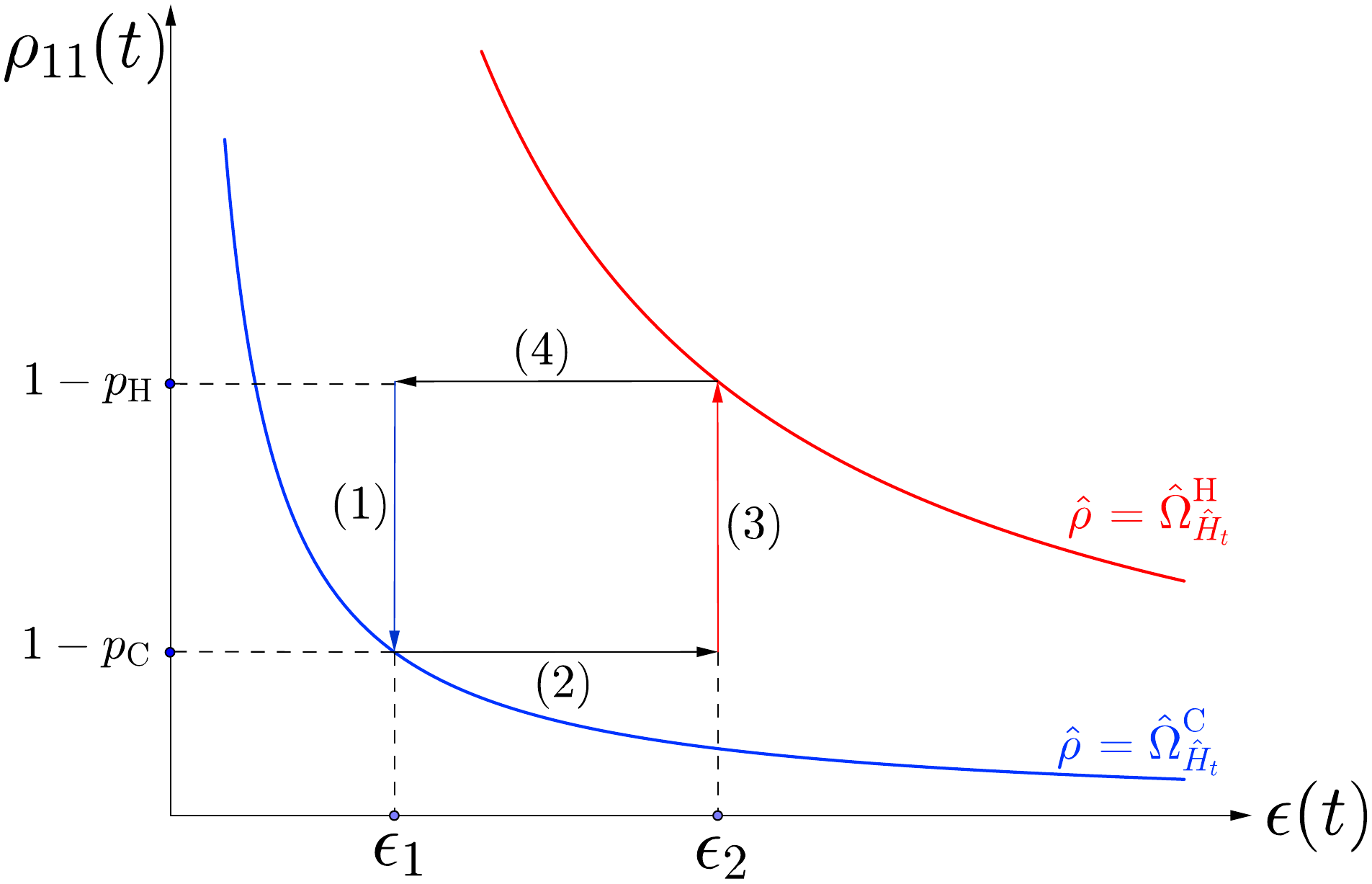}
\caption{(Color online) Pictorial representation of the Quantum Otto cycle in the configuration space $\rho_{11}(t):= \langle 1| \hat{\rho}(t) |1\rangle$ vs. the energy gap $\epsilon(t)$  in the ITT limit. Steps 1) and 3) correspond to thermalizations  where ${\cal S}$ is kept in contact with the cold   bath $\rm C$   and hot bath $\rm H$, respectively; steps 2) and 4) 
instead represent sudden quenches of the Hamiltonian gap.}
\label{fig:Otto}
\end{figure}

Again taking inspiration by the classical version translated in our setup, the Otto cycle is composed by two isoentropic (adiabatic) strokes alternated with two thermalizations (classically isochores).  Considering the same qubit engine used for the description of the Carnot Cycle, the 4 steps can be summarized as in Fig.~\ref{fig:Otto}:
\begin{itemize}
\item [1)] starting from an initial state $\hat{\rho}_1$,
 keeping fixed the gap $\epsilon_1$ the system is let thermalize in contact with the cold reservoir C; 
\item [2)] after isolating the system from the bath, a quench is performed taking $
\epsilon_1\rightarrow \epsilon_2 (>\epsilon_1)$; 
\item [3)] while the gap is fixed, the system is let thermalize in contact with the hot reservoir H; 
\item [4)] a final quench restores $\epsilon_2\rightarrow \epsilon_1$.
\end{itemize}
Unless considering infinitesimal transformations where 
$\epsilon_2\simeq \epsilon_1$, it is clear that 
at variance with the Carnot cycle, in the Otto cycle  the working medium ${\cal S}$ is always in a 
out-of-equilibrium state. Accordingly  in this case the S-D approximation technique~\cite{slowdriving} 
cannot be applied.

As for the Carnot cycle, the system exchange heat with the baths only  during the steps 1) and 3).
In particular 
 exploiting the fact that now during the thermalization the Hamiltonian is kept constant we have 
\begin{eqnarray}
\Delta Q_{\rm C}&=&  \int_{{\cal I}_{\rm C}} d Q_{\rm C}(t) 
=  \Tr[\Delta\hat{\rho}_{\rm C}\; {\hat{H}_1}]\;, 
\nonumber \\
\label{DD2OTTO} 
\Delta Q_{\rm H} &=& \int_{{\cal I}_{\rm H}} d Q_{\rm H}(t)  
=\Tr[\Delta\hat{\rho}_{\rm H}
\; {\hat{H}_2}]\;, 
\end{eqnarray}
where for $j={\rm H,C}$, $\Delta\hat{\rho}_{j}$ represent the increment experienced by  the system density during the associated step. Equations \eqref{DD2OTTO} are valid for any Otto cycle, but we can specify them for our case $\hat{H}_{1,2}=\frac{\epsilon_{1,2}}{2}\hat{\sigma}^z$. 
If we also allow  infinite time for the thermalization  stages (ITT limit),
the states at the end  of the steps 1) and 4) are 
 the thermal states $\hat{\Omega}^{({\rm C})}_{\epsilon_1}$ and $\hat{\Omega}^{({\rm H})}_{\epsilon_2}$,
 respectively, but in general they do not need to.
In this case  we have
\begin{eqnarray}  \label{IMPOII} 
\Delta\hat{\rho}_{\rm C} = - \Delta\hat{\rho}_{\rm H} = \hat{\Omega}^{({\rm C})}_{\epsilon_1}-\hat{\Omega}^{({\rm H})}_{\epsilon_2}  \;, 
\end{eqnarray} 
which replaced into~(\ref{DD2OTTO}) 
yields the identities
\begin{eqnarray} 
 \Delta Q^{(0)}_{\rm H} 
&:=& \epsilon_2 (p_{\rm C}-p_{\rm H})
\;, \\
 \Delta Q^{(0)}_{\rm C} &:=&\epsilon_1 (p_{\rm H}-p_{\rm C})
= -\frac{\epsilon_1}{\epsilon_2} \Delta Q^{(0)}_{\rm H}\;,
\end{eqnarray} 
where we use the upper index ``$(0)$" to  indicate that 
these are the heat exchanged in the IIT regime, and we used asymptotic ground state probabilities for the two isochores, defined as in Eq.\eqref{def:p_gibbs}
\begin{equation} 
\label{PITT} 
p_{\rm C}=\frac{1}{e^{-\beta_{\rm C} \epsilon_1}+1}\ , \quad \quad p_{\rm H}=\frac{1}{e^{-\beta_{\rm H} \epsilon_2}+1}\ .
\end{equation} 
If we further assume 
 the constraint 
\begin{equation}\label{CONSTR1} 
\beta_{\rm C} \epsilon_1 \geq \beta_{\rm H} {\epsilon_2} \;,
\end{equation}  
 $\Delta Q^{(0)}_{\rm H}$ turns out to be positive while  $\Delta Q^{(0)}_{\rm C}$ is negative.
 The
 absorbed and released heat contributions can hence be identified as
   \begin{eqnarray} 
 Q_{\rm ABS}&=& \Delta Q^{(0)}_{\rm H} 
 \geq 0 \;, \\
 Q_{\rm REL}&=& \Delta Q^{(0)}_{\rm C} = - 
 \frac{\epsilon_1}{\epsilon_2}
 Q_{\rm ABS}\;,
\end{eqnarray} 
leading to an efficiency  
\begin{equation}
\eta_{\rm o}=\frac{Q_{\rm ABS} + Q_{\rm REL}  }{Q_{\rm ABS} }= 1-\frac{\epsilon_1}{\epsilon_2} \;,
\end{equation}
which thanks to (\ref{CONSTR1}) is 
 smaller than the corresponding Carnot efficiency (\ref{ddfCARNOT}).
 Departing from the ITT regime, corrections can be computed analogously to what
 done for the Carnot cycle when considering non quasi-static cycles. Specifically 
 we can write 
 \begin{eqnarray}
{\eta}&=&1+\frac{Q_{\rm REL}}{Q_{\rm ABS}} \simeq 1 +
 \frac{\Delta Q^{(0)}_{\rm C}}{\Delta Q^{(0)}_{\rm H}}
\frac{1+ \Delta Q^{(1)}_{\rm C}/\Delta Q^{(0)}_{\rm C}}{1+ \Delta Q^{(1)}_{\rm H}/\Delta Q^{(0)}_{\rm H}}\nonumber \\ \label{EFFICotto} 
&=& 1 - \frac{\epsilon_1}{\epsilon_2}  \frac{1+ \alpha_{\rm C}}{1+ \alpha_{\rm H}} 
= 1 -( 1-\eta_{\text{o}})   \frac{1+ \alpha_{\rm C}}{1+ \alpha_{\rm H}} \;,
\end{eqnarray}
 where now the $\Delta Q^{(1)}_{j}$s refer to first order corrections associated with 
 the finite  thermalization times, while the 
 $\alpha_{j}$s are the associated ratios ~$\Delta Q^{(1)}_{j}/\Delta Q^{(0)}_{j}$ 
 analogous to those introduced in Eq.~(\ref{DEFALPHA}) for the S-D corrections of the Carnot cycle. 
 In a similar way the power of the cycle can be expressed as in Eq.~(\ref{def:power}) yielding
 \begin{eqnarray} 
\label{def:powerotto}
P&=&\frac{{Q_{\rm ABS}}+{Q_{\rm REL}}}{\tau_{\rm C}+\tau_{\rm H}} \simeq  
\frac{{\Delta Q^{(0)}_{\rm C}}+{\Delta Q^{(0)}_{\rm H}}+{\Delta Q^{(1)}_{\rm C}}+{\Delta Q^{(1)}_{\rm H}}}{\tau_{\rm C}+\tau_{\rm H}} \nonumber \\
&=&{\Delta Q^{(0)}_{\rm H}} \frac{\eta_{\text{o}}+ \alpha_{\rm H} -
(\epsilon_1/\epsilon_2)  \alpha_{\rm C} }{\tau_{\rm C}+\tau_{\rm H}} \;,
\end{eqnarray}
where $\tau_{\rm C}$ and $\tau_{\rm H}$ are the finite temporal durations of the steps
 1) and 3) respectively, which for $\alpha_j=0$ gives the quasi-static result
 \begin{equation}
\label{eq:quasi-static_ottopow}
P^{(0)}:= {\Delta Q^{(0)}_{\rm H}} \frac{\eta_{\text{o}}}{\tau_{\rm C}+\tau_{\rm H}}=\dfrac{(\epsilon_2-\epsilon_1)(p_{\rm C}-p_{\rm H})}{\tau_{\rm C}+\tau_{\rm H}}\;.
\end{equation}

\subsubsection{Exact FTT Otto Cycle}
An application of the perturbative analysis~(\ref{def:powerotto}) 
 in the case of a general engine evolving under the action of the dissipation model~(\ref{PRIMO}),
 is presented in Appendix~\ref{sec:perfOTTO} . Due to the simplicity of the scheme however, this approach can be replaced by the exact 
 finite-time solution of the problem, which we are going to present in the following.

Departing from the ITT regime the two
thermalizations (isochores) of the Otto cycle become inevitably partial. 
To account for this effect, 
we represent the  ground state populations of the working medium ${\cal S}$ at time
$t$ after the beginning of the isochore with the $j$-th bath, 
\begin{equation}
\label{eq:general_partial_therm}
\rho_{00}(t): = \langle 0| \hat{\rho}(t)|0\rangle=p_j+\Delta_j f_j(t)\;,
\end{equation}
where $p_j$ is the equilibrium probability~(\ref{PITT}) one would get in the strict ITT regime, 
and where $\Delta_j$ quantifies how out of equilibrium is the system at the beginning of the isochore. In this expression  $f_j(t)$  is a function of $t$ that depends on the explicit details of the dynamics and which, by construction must fulfil the conditions  $f_j(0)=1$ and $\lim_{t\rightarrow\infty} f_j(t)=0$ to ensure proper thermalization in the ITT regime.  
The parameters $\Delta_{\rm H}$ and $\Delta_{\rm C}$ are not completely independent and can be connected via  the temporal durations, $\tau_{\rm C}$ and $\tau_{\rm H}$, of the two isochore.
Indeed by invoking continuity conditions for the density matrix of \s between the two isothermal strokes, 
we obtain
\begin{eqnarray}
p_{\rm C}+\Delta_{\rm C}f_{\rm C}(\tau_{\rm C})=p_{\rm H}+\Delta_{\rm H} \ ,  \nonumber \\
p_{\rm H}+\Delta_{\rm H}f_{\rm H}(\tau_{\rm H})=p_{\rm C}+\Delta_{\rm C}\ , \label{eq:fo1}
\end{eqnarray}
which in particular imply
\begin{equation}
\label{eq:deltah}
p_{\rm C}-p_{\rm H}=\Delta_{\rm H}\bigg(\frac{1-f_{\rm C}(\tau_{\rm C})f_{\rm H}(\tau_{\rm H})}{1-f_{\rm C}(\tau_{\rm C})}\bigg)\ .
\end{equation}

From~Eq.~(\ref{DD2OTTO}) it follows now that  the relative heat exchanged during the isochore can now be expressed as
\begin{eqnarray} 
\Delta Q_{\rm H} &=-\epsilon_2 \Delta_{\rm H} \int_0^{\tau_{\rm H}}\dot{f}_{\rm H} &=-\epsilon_2\Delta_{\rm H}(f_{\rm H}(\tau_{\rm H})-1)\ , \\
\Delta Q_{\rm C} &=-\epsilon_1 \Delta_{\rm C} \int_0^{\tau_{\rm C}}\dot{f}_{\rm C} &=-\epsilon_1\Delta_{\rm C}(f_{\rm C}(\tau_{\rm C})-1)\ ,
\end{eqnarray} 
which yields a power equal to
\begin{equation} \label{eq:ottopower} 
P
= \dfrac{(\epsilon_2-\epsilon_1)(p_{\rm C}-p_{\rm H})}{\tau_{\rm C}+\tau_{\rm H}} 
 \;
\dfrac{(1-f_{\rm H}(\tau_{\rm H}))(1-f_{\rm C}(\tau_{\rm C}))}{1-f_{\rm C}(\tau_{\rm C})f_{\rm H}(\tau_{\rm H})}\;,
\end{equation}
where in the second line we used (\ref{eq:deltah}) and the expression~(\ref{eq:quasi-static_ottopow})
for the power of the cycle under ITT conditions (notice that as expected when 
 $f_j(\tau_j)=0$ then $P$ reduces to the value $P^{(0)}$ of Eq.~\eqref{eq:quasi-static_ottopow}).  
This is the exact expression for $P$ which depends on the explicit form of $f_j$. It is worth
observing that  the first numerator $(\epsilon_2-\epsilon_1)(p_{\rm C}-p_{\rm H})$ of Eq.~\eqref{eq:ottopower}
depends only on the model temperatures and gaps, hence fixing the efficiency it is possible to maximize the remaining independently, choosing the optimal length of the strokes.

\section{Quantum Thermal Machines in the Non-Markovian regime}
\label{sec:n-M_model}
As we have explicitly discussed in
 Sec.~\ref{sec:ii2new} 
the Markovian character of the system dynamics guarantees that the 2nd Law of thermodynamics is satisfied in the open quantum systems setting, i.e. the unavoidable loss of free energy of systems interacting with large baths. With this in mind, it is easy to realise that modelling the coupling of an engine with a non-Markovian bath may result as a pumping of free energy from the environment. For this reason, any acclaimed boost of performance in such a setting can be considered trivial, or even meaningless if not justified physically (e.g. using non-equilibrium baths). Therefore we choose to model the dynamics of the reservoir coupling in an overall Markovian framework, picturing an environment which contains some degrees of freedom who share correlations and interact with the working medium, such as to make its local dynamics non-Markovian; in this way we avoid any unjustified external resource draining. In such a set up indeed, Refs. \cite{eisert1,eisert2,eisert3} show how this mechanism can only have detrimental effects from the point of view of quasi-static Thermodynamics: that is, without external free energy injections, the 2nd Law assures the Carnot efficiency is the maximal one. Nothing instead, has been stated from the point of view of FTT: even if lowering the maximal efficiency, there is still question on the effects on power and EMP, which are, from the practical point of view, much more interesting than pure maximal efficiency. Here we try to fill this gap, finding indeed that non-Markovian dynamics may have positive effects.

\subsection{The model} \label{sec:model3} 

\begin{figure*}
\includegraphics[width=0.48\textwidth]{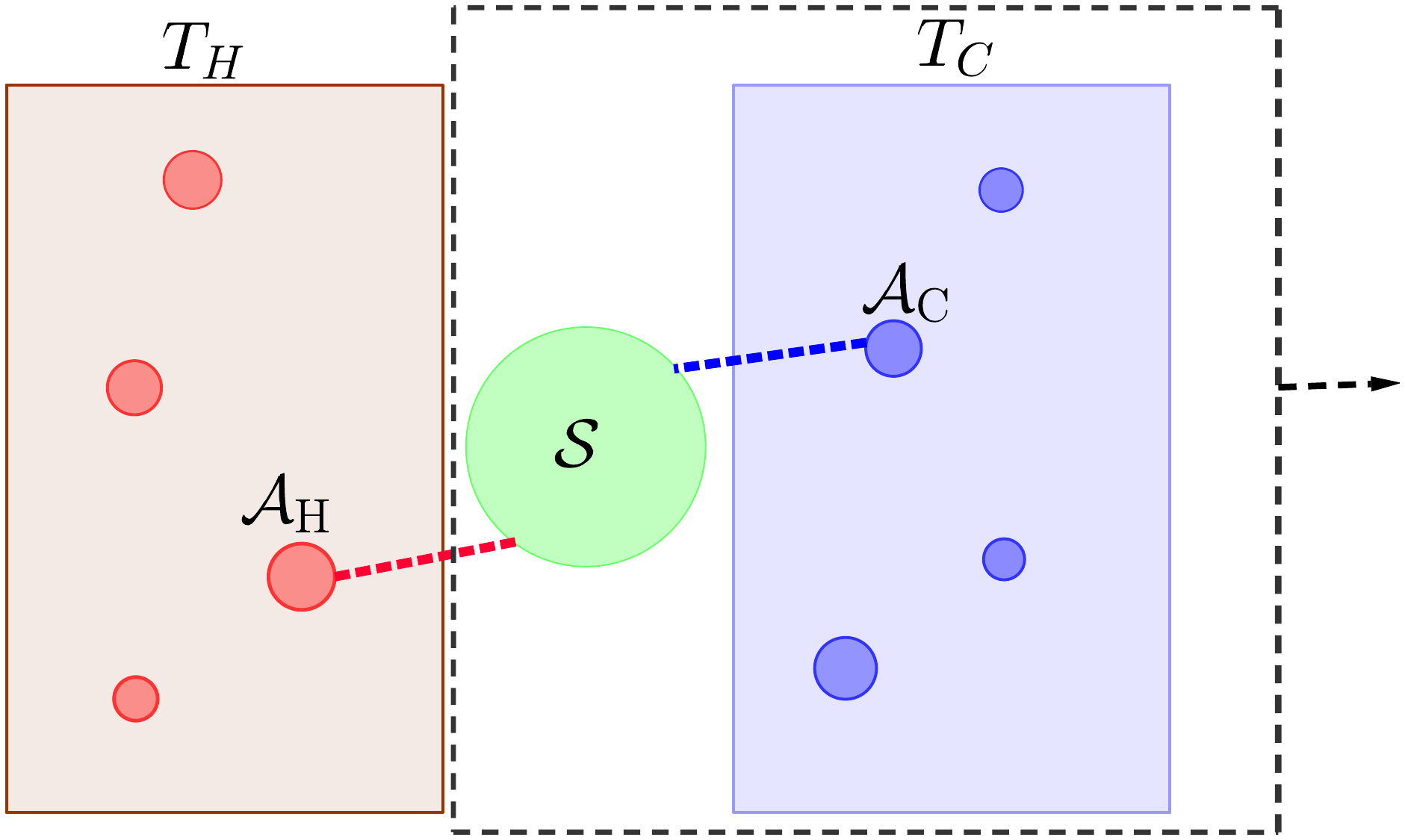}
\includegraphics[width=0.48\textwidth]{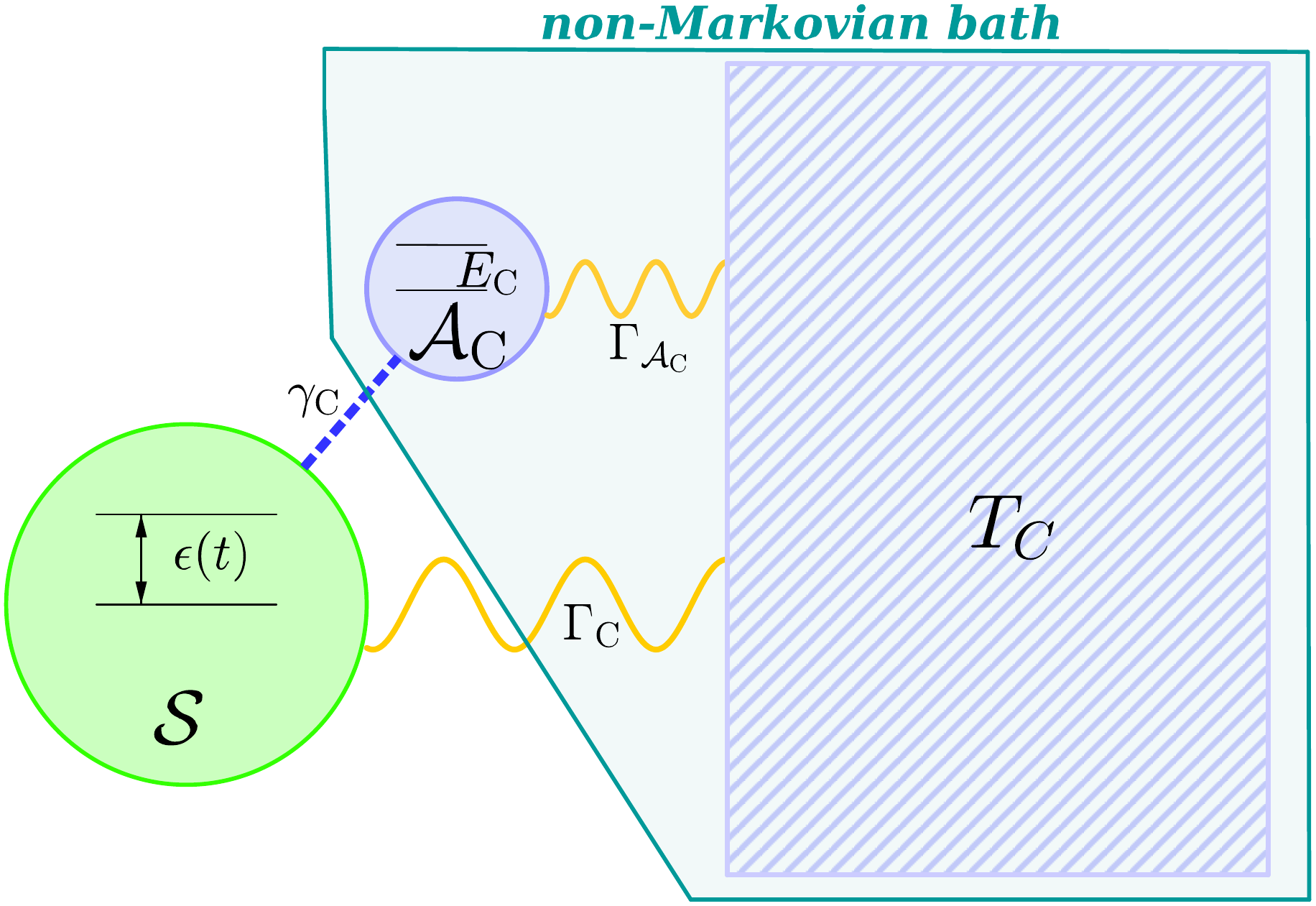}
\caption{(Color online) Left: schematic of the non-Markovian model: ${\cal A}_{\rm H}$ and ${\cal A}_{\rm C}$ are local degrees of freedom
of the hot and cold bath respectively.  Right: specifics of the model, representation of the interaction between 
${\cal S}$ and the non-Markovian cold bath.}
\label{fig:singlebath}
\end{figure*}

In order to account for non-Markovian effects we consider a modification of the set-up of Sec.~\ref{sec:ii1},
with the one schematically sketched in Fig.~\ref{fig:singlebath} where both the  hot  reservoir H and the cold reservoir C  include a  local and a remote component.
The first, represented by the qubit ancillary subsystems  ${\cal A}_{\rm H}$  and ${\cal A}_{\rm C}$ of the figure, corresponds to degrees of freedom characterized by local Hamiltonian terms
\begin{eqnarray} 
\hat{H}_{{\cal A}_j} : = {E}_j \hat{\sig}^z_{{\cal A}_j}  /2\;, 
\end{eqnarray} 
which are directly connected with \s through dedicated coupling (energy exchanging) Hamiltonians which we assume to have the 
form
\begin{equation}
\label{eq:interaction}
\hat{V}_j:=\gamma_j(\hat{\sig}^+\ten\hat{\sig}_{{\cal A}_j}^-+\hat{\sig}^-\ten\hat{\sig}_{{\cal A}_j}^+)\;, 
\end{equation}
$\hat{\sig}_{{\cal A}_j}^{\pm}$  and $\hat{\sig}^{\pm}$ indicating the lowering/raising operators of  ${\cal A}_j$ and ${\cal S}$ respectively. 
 The remote components of the baths, instead,
are associated with standard local GKSL dissipators ($\mathcal{D}_t^{(j)}$ for \s, 
and $\mathcal{D}_{{\cal A}_{\rm C}}$ and 
$\mathcal{D}_{{\cal A}_{\rm H}}$ for ${\cal A}_{\rm H}$ and ${\cal A}_{\rm H}$, respectively)
inducing local thermalization toward their associated Gibbs states,
i.e. the usual 
canonical state  $\hat{\Omega}^{(j)}_{\hat{H}_t}$ of (\ref{def:gibbs_state}) for ${\cal S}$, and 
for $j={\rm H,C}$, 
\begin{equation}
\label{def:gibbs_stateA}
\hat{\Omega}^{(j)}_{\hat{H}_{{\cal A}_j}}:=
\frac{e^{-\beta_j {\hat{H}_{{\cal A}_j}} }}{\Tr[e^{-\beta_j {\hat{H}_{{\cal A}_j}} } ]}\;, 
\end{equation} 
 for ${\cal A}_j$. 
Once more, we shall consider cyclic operations where the gap of the local Hamiltonian of ${\cal S}$ is externally modulated as in Eq.~(\ref{energy}), and the system, at each given time, is selectively coupled to one and only one of the two baths.
 Accordingly we describe the dynamics of joint density matrix $\hat{\mathbf{R}}(t)$
the compound  formed by ${\cal S}$, ${\cal A}_{\rm C}$, and ${\cal A}_{\rm H}$,  
in terms of a standard  Markovian evolution which has the same form 
of Eq.~(\ref{eq:general_QTmachineEq}), the 
 non-Markovian character  of the local dynamics of \s being obtained instead by tracing away
the ancillas, i.e. 
 $\hat{\rho}(t) = \mbox{Tr}_{\cal A}[\hat{\mathbf{R}}(t)]$ follows trajectories that no longer exhibit 
 the divisibility condition that instead is granted to  $\hat{\mathbf{R}}(t)$ -- see Appendix~\ref{app:n-M_measure}.
  Furthermore, in order to simplify the analysis we shall also enforce the approximation,
    that 
on the time intervals ${\cal I}_{\rm C}$ (resp. ${\cal I}_{\rm H}$) during which  the working medium is coupled with 
the cold bath ${\rm C}$ (resp. ${\rm H}$), the other ancilla ${\cal A}_{\rm H}$ (resp. ${\cal A}_{\rm C}$), that is temporarily  decoupled from 
${\cal S}$, relaxes to thermal equilibrium with the remote counterpart of ${\rm H}$ (resp. ${\rm C}$), i.e. 
\begin{eqnarray} 
\hat{\mathbf{R}}(t) &\simeq& \hat{\mathbf{R}}_{\rm C} (t) \otimes \hat{\Omega}^{(\rm H)}_{\hat{H}_{{\cal A}_{\rm H}}}\;, 
 \qquad \forall t\in {\cal I}_{\rm C}\;, \nonumber \\
 \hat{\mathbf{R}}(t) &\simeq& \hat{\mathbf{R}}_{\rm H} (t) \otimes \hat{\Omega}^{(\rm C)}_{\hat{H}_{{\cal A}_{\rm C}}}\;, 
 \qquad \forall t\in {\cal I}_{\rm H}\;, \label{FACTOR} 
\end{eqnarray} 
with $\hat{\mathbf{R}}_{\rm C}(t)$ and $\hat{\mathbf{R}}_{\rm H}(t)$ that describe the reduced density matrix of of ${\cal S}{\cal A}_{\rm C}$ 
and 
${\cal S}{\cal A}_{\rm H}$, respectively
 -- the assumption being consistent with the first order S-D approximation, where ultimately one only needs to determine the quasi-static trajectories of the system~\cite{relax_hyp}. 
The evolution
of  $\hat{\mathbf{R}}_{j}(t)$ is finally expressed as
\begin{equation}
\dot{\hat{\mathbf{R}}}_j(t)= 
\mathcal{\bf L}_t^{(j)}[\hat{\mathbf{R}}_j(t)]:
=-i[\hat{\bf H}_t^{(j)} ,\hat{\mathbf{R}}_j(t)]_- +\mathcal{\bf D}^{(j)}_t[\hat{\mathbf{R}}_j(t)]\ , \label{MENM} 
\end{equation}
with Hamiltonian 
\begin{eqnarray} \label{HDEF}
\hat{\bf H}_t^{(j)} : = \hat{H}_t + \hat{H}_{{\cal A}_{j}} + \hat{V}_j\;, 
\end{eqnarray} 
and dissipator
\begin{equation}\mathcal{\bf D}^{(j)}_t:=\mathcal{D}_t^{(j)}\otimes {I}_{{\cal A}_j}+
 {I} \otimes\mathcal{D}_{{\cal A}_j} \;,  \label{DISDEF} 
 \end{equation}
whose local contributions on ${\cal S}$ and on ${\cal A}_j$
will be assumed  to have the simple form~(\ref{PRIMO}), i.e.
\begin{eqnarray}
\mathcal{D}_t^{(j)} [\cdots]&=&\Gamma_j(\hat{\Omega}^{(j)}_{\hat{H}_t}-\cdots)\ ,
\label{CHOICE}  \\
\mathcal{D}_{{\cal A}_j}[\cdots]&=&\Gamma_{{\cal A}_j} (\hat{\Omega}_{\hat{H}_{{\cal A}_j}}-\cdots)\ ,
\end{eqnarray}
where $\hat{\Omega}^{(j)}_{\hat{H}_t}$ and $\hat{\Omega}^{(j)}_{\hat{H}_{{\cal A}_j}}$ are the Gibbs
states of  (\ref{def:gibbs_state}) and (\ref{def:gibbs_stateA}) while 
$\Gamma_j$ and $\Gamma_{{\cal A}_j}$ are rates. 
As evident from the above expressions, we are assuming control
on the energy gap of the working medium but not on the one of ${\cal A}_j$
which formally is just an element of the bath.
We finally stress that $\gamma_j$ in Eq.~\eqref{eq:interaction} is the parameter defining the non-Markovianity of the model: this follows from the fact that the Markovian regime is recovered in the limit~$\gamma_j\rightarrow0$ (separable dynamics) and from the fact that, as shown in Appendix \ref{app:n-M_measure}, the non-Markovian measure by Breuer, Laine, Piilo \cite{BLP} is monotonously increasing in $\gamma_j$. As we shall see in the next section a similar dependence can be observed on the optimized power output of a Carnot and Otto engine providing hence a clear indication of the fact that the non-Markovian character of the dynamics can be beneficial to these figures of merit.

\subsection{Non-Markovian Carnot cycle performance} \label{non-MarkovCARNOT} 
In what follows we focus on the quasi-resonant case where the gap modulations of ${\cal S}$ on the interval
${\cal I}_j$ are 
such that the system is almost at resonance with ${\cal A}_j$, i.e.
\begin{eqnarray} \epsilon(t) \simeq E_j \;, \qquad\qquad  \forall t\in {\cal I}_j\; \label{RESONANT} \end{eqnarray} 
(note that the optimal Quasi-Otto trajectories found in Section~\ref{sec:carnot} are obtained in the limit of $\epsilon(t)$ being infinitesimally modulated).
In this regime the stationary state of $\mathcal{\bf L}^{(j)}_t$ is approximatively equal 
to the tensor product of the individual thermal states associated with  the two 
 local dissipators, i.e.
 \begin{equation}\label{STABnm} 
\mathcal{\bf L}^{(j)}_t[\hat{\mathbf{R}}_j]=0 \; \;  \Leftrightarrow \; \; \hat{\mathbf{R}}_j
\simeq \hat{\Omega}^{(j)}_{\hat{H}_t}\ten\hat{\Omega}^{(j)}_{\hat{H}_{{\cal A}_j}}\;,
\end{equation}
ensuring thermodynamic consistency of the model and being in agreement with Eq.~(\ref{FACTOR}).  
We hence identify the heat absorbed by  ${\cal S}$ from the $j$-th bath
as in Eq.~(\ref{DEFQj0}) we obtain
\begin{eqnarray}
  dQ_{j}(t)&=&\Tr\Big[ \hat{H}_t d\hat{\rho}(t) \Big] =      \label{DEFQJnNM} 
  \Tr\Big[ \hat{H}_t d\hat{\mathbf{R}}_j (t) \Big]  \;, 
\end{eqnarray}
where in the first identity we used the fact that $\hat{\rho}(t)$ is the partial trace with respect to ${\cal A}_j$ of $\hat{\mathbf{R}}_j(t)$.
We then expand this quantity as in Eq.~(\ref{ddf}) by 
invoking  the S-D approximation 
$\hat{\mathbf{R}}_j(t) \simeq \hat{\mathbf{R}}^{(0)}_j(t) + \hat{\mathbf{R}}^{(1)}_j(t) + \cdots$, where
$\hat{\mathbf{R}}^{(0)}_j(t)$ 
is the quasi-static solution which according to (\ref{STABnm}) 
is the state 
$\hat{\Omega}^{(j)}_{\hat{H}_t}\ten\hat{\Omega}^{(j)}_{\hat{H}_{{\cal A}_j}}$, while
$\hat{\mathbf{R}}^{(1)}_j(t)$ is the first order correction term which 
according to Eq.~(\ref{eq:rho^1}) we identify with the operator 
 \begin{eqnarray}
\label{eq:rho^1222}
\hat{\mathbf{R}}^{(1)}_j(t)=(\mathcal{\bf L}^{(j)}_t \mathcal{ P})^{-1}[ \dot{\hat{\mathbf{R}}}^{(0)}_j(t)]\,.
\end{eqnarray}
Due to the factorization of the fixed point (\ref{STABnm}) 
 the  zero-th term contribution $dQ^{(0)}_{j}(t)$ of $ dQ_{j}(t)$ is still provided by the increment of the von Neumann entropy of the
 Gibbs states $\hat{\Omega}^{(j)}_{\hat{H}_t}$ 
  as in Eq.~(\ref{ORDzero}). On the contrary  $dQ^{(1)}_{j}(t)$
can still be cast in the form~(\ref{DEFQJ2})  where now 
${\rho}^{(1)}_{00}(t)= \langle 0| \mbox{Tr}_{{\cal A}_j}[\hat{\mathbf{R}}^{(1)}_j(t)]|0\rangle$
that  in the limit~(\ref{RESONANT}) can be expressed as in
Eq.~(\ref{eq:qubit_ground_general}) with an S-D amplitude
$A_j$ that can be found (see Appendix~\ref{sec:non-resonant_sol}) equal to 
\begin{equation}
\label{res:zerodeltaA1}
A_j=
 \dfrac{1/\Gamma_j}{(c_j+1)^2}\bigg(2+\dfrac{c_j(c_j-2)}{(c_j+2y_j^2)}+\dfrac{c_j^2(c_j+1)^2-c_j^3}{(c_j+2y_j^2)^2}\bigg)
\ ,
\end{equation}
where we introduce the quantities
\begin{eqnarray} \label{DEFcy} 
c_j:= \Gamma_{{\cal A}_j}/\Gamma_{j} \;, \qquad y_j:= \gamma_j/\Gamma_j \;.
\end{eqnarray}
To evaluate the effect of non-Markovianity on the
maximum power associated with a Carnot cycle we can then 
follow the same analysis we performed in Sec.~\ref{sec:perf}.
In particular under symmetrization of
the bath couplings and driving conditions (i.e.  choosing $A_{\rm C} =A_{\rm H}=A$ and imposing $\epsilon(t)$ along the cold
isotherm to be the time reversal of the one along the hot isotherm),
we can directly use Eq.~(\ref{MAXP1}), which makes it clear that
to get 
higher power performance  we should target the low values of the $A_j$s. 

\begin{figure}
\includegraphics[width=0.48\textwidth]{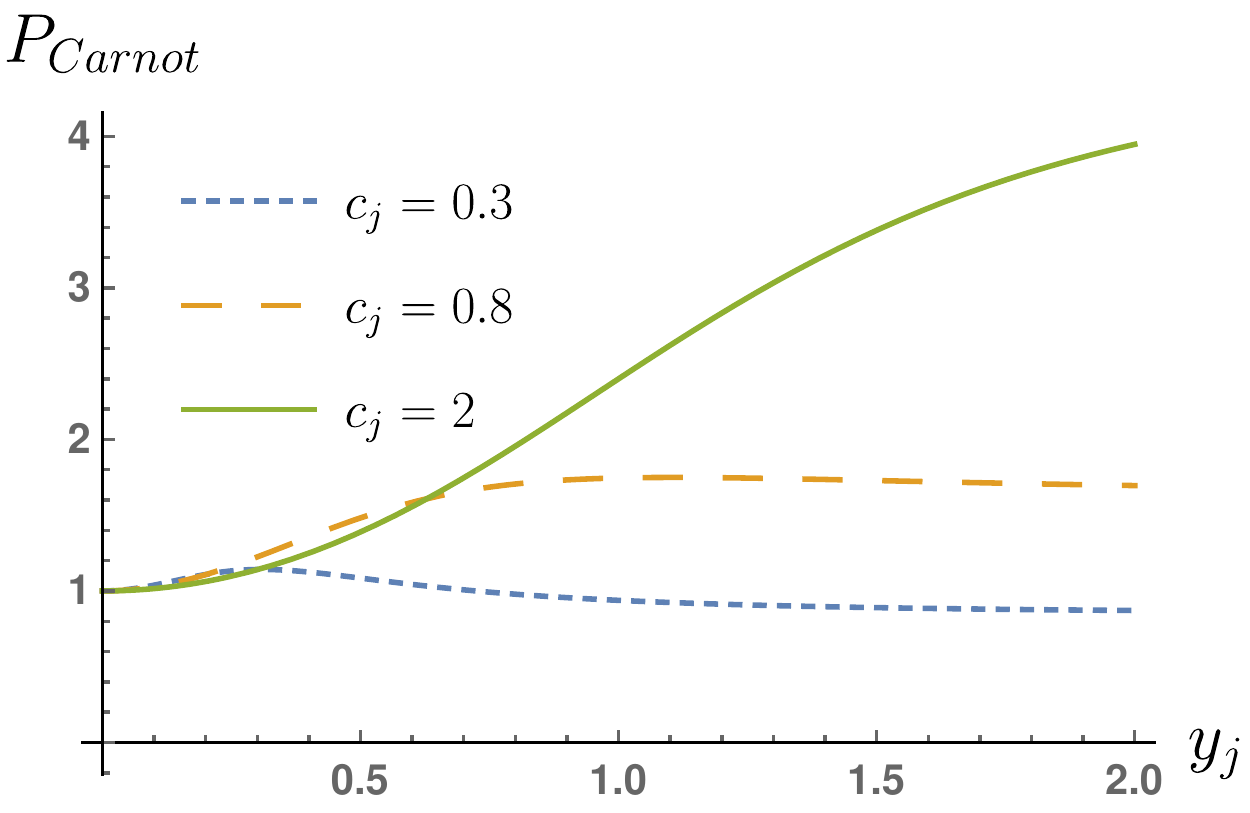}
\includegraphics[width=0.48\textwidth]{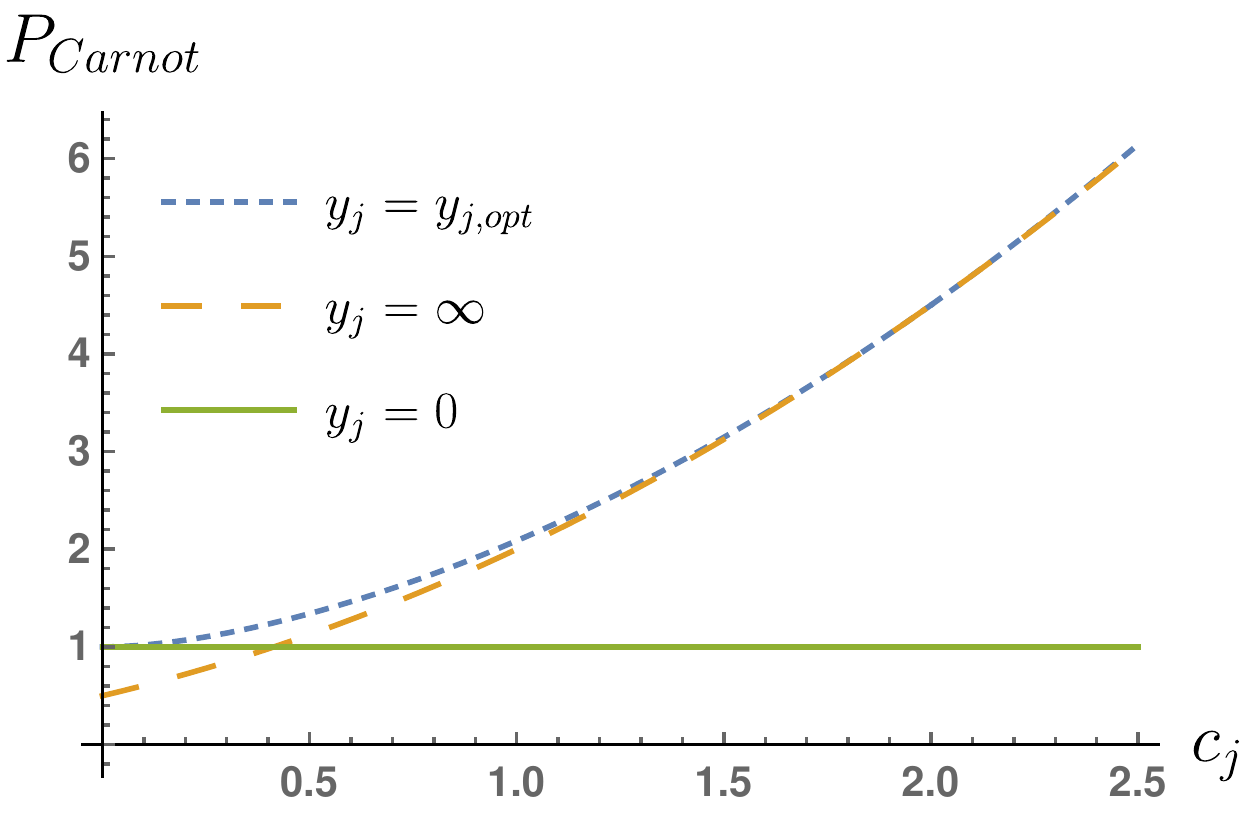}
\caption{(Color online) Upper panel: 
Maximum power of the Carnot cycle normalized to its Markovian value, in terms of the parameter $y_j$ that accounts for the non-Markovian effects in the model.
The plot corresponds to $A_j^{-1}$ of Eq.~(\ref{res:zerodeltaA1}) in units of $\Gamma_j$.
For  $0\leq c_j \leq \sqrt{2}-1$: there exists an optimal $y_j$ but for $y_j\rightarrow\infty$, $A_j^{-1}$ is inferior to the initial value  $\Gamma_j$.
For $\sqrt{2}-1\leq c_j\leq 2$: there exists an optimal $y$ and for $y_j\rightarrow\infty$, $A_j^{-1}$ remains bigger than the Markov case.
For $2\leq c_j$: there is no finite optimal value for $y_j$, but $A_j^{-1}$ continues growing to its asymptotic value.
Lower panel: comparison of $A_j^{-1}$  (in units of $\Gamma_j$) as a function of $c_j$ having fixed $y_j$ equal
to the optimal choice~\eqref{eq:optimalcoup} (blue) with the Markovian case  $\gamma_j=0$ (green) and strong-coupling case $\gamma_j=\infty$ (yellow).
}
\label{fig:Anonresonant}
\end{figure}

First of all we notice that for $\gamma_j=0$ (i.e. $y=0$) correctly reduces to $A_j={1}/{\Gamma_j}$,
which is the value~(\ref{ddsf})
one would obtained in the Markovian limit in the presence of the dissipator~(\ref{CHOICE}).
In the strong coupling limit
$\gamma_j\gg \Gamma_j$ (i.e. $y_j\gg 1$), instead we get 
\begin{eqnarray}
A_j^{\rm (strong)} := \lim_{y_j\rightarrow \infty} A_j=
(2/\Gamma_j)/(c_j+1)^2\;, 
\label{Ajasymp} 
\end{eqnarray} 
 which gets smaller than 
 the non-Markovian limit 
for values of $c_j$ above  the critical threshold $\sqrt{2}-1\simeq 0.414$. 
Another important  value is $c_j=2$ that determines the sign of the second addend in the parenthesis in the r.h.s. of Eq.~(\ref{res:zerodeltaA1}) (the third addend being always positive). In fact  we find that for 
$c_j<2$,  $A_j$ attains its  minimum value, smaller than the Markovian $1/\Gamma_j$,  at 
\begin{equation}
\label{eq:optimalcoup}
y_{j,opt}^2=\frac{c_j^2(2c_j+3)}{2(2-c_j)}\ ,
\end{equation}
otherwise the optimal value is infinite, in the sense of $A_j$ monotonously decreasing with $y_j$.
These results on the dependence of $A_j$ on the model parameters are summed up in the Fig.~\ref{fig:Anonresonant}, 
where we plot $A_j^{-1}$ in adimensional units, which thanks to Eq.s~\eqref{res:pmaxS-W} and \eqref{MAXP1} corresponds to the maximum power attainable by the Carnot cycle, normalized to its Markovian value.
In each case we see how the presence of the coupling $\gamma_j$, i.e. of non-Markovian effects leads to the an improvement of the maximum power $P_{\max}$ of the Carnot cycle.

\subsection{Non-Markovian Otto cycle performance}
\label{sec:n-M_otto}
\begin{figure}
\centering
\includegraphics[width=0.48\textwidth]{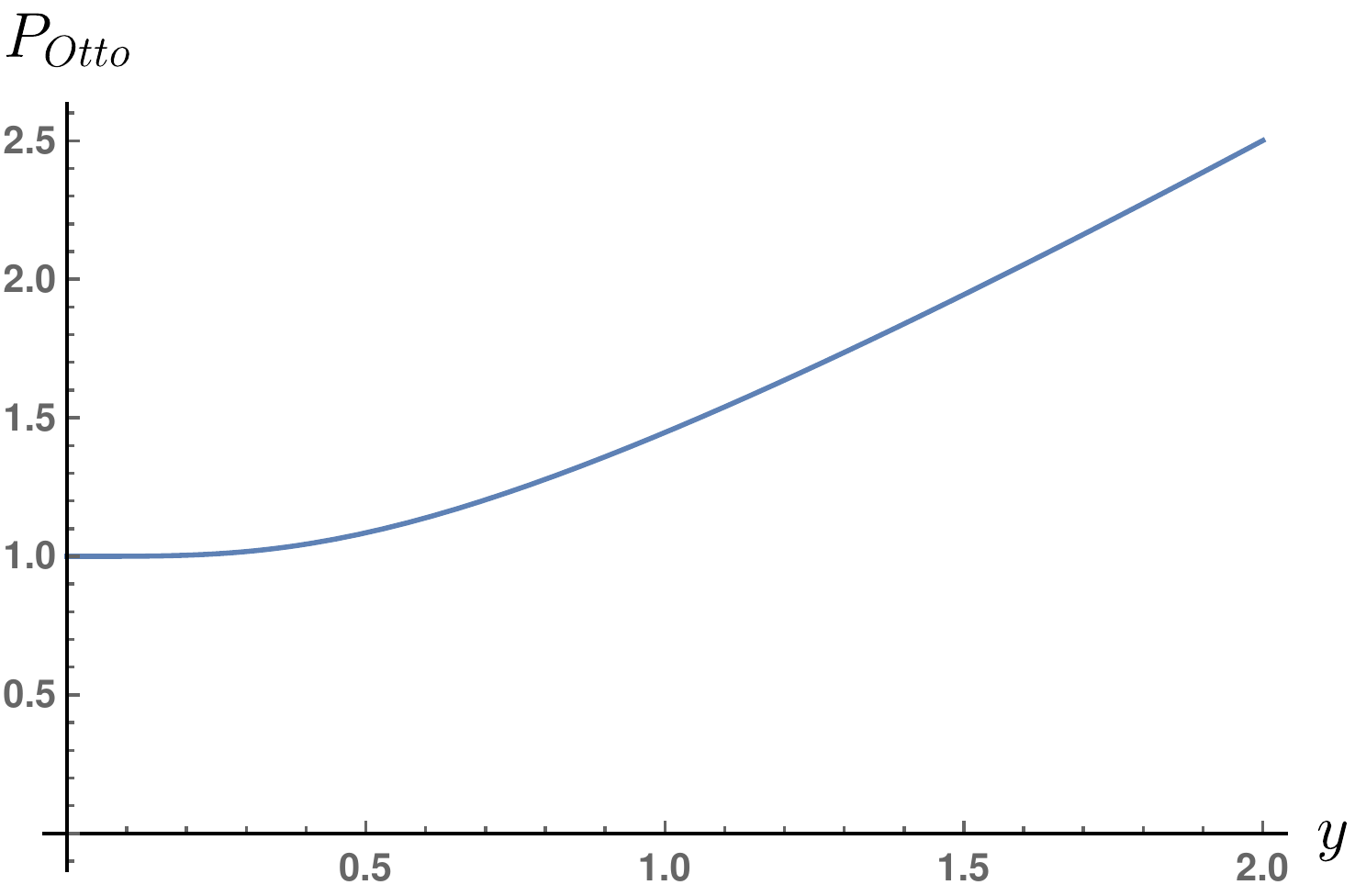}
\caption{Maximum power for the non-Markovian Otto cycle~(\ref{POWEROTTONM}), as a function of $\gamma/\Gamma$, 
in units of $\frac{\Gamma}{4}(E_{\rm H}-E_{\rm C})(p_{\rm C}-p_{\rm H})$ (Markovian value).
}
\label{fig:powerkmax}
\end{figure}

To study the performance of an Otto cycle for the non-Markovian model introduced, we can apply the exact power result \eqref{eq:ottopower}, which in turn is characterized by the function $f_j$ describing the relaxation of the ground state during the two isochores ($j=\rm H,C$) (cf. Eq.~\eqref{eq:general_partial_therm}). 
In this case we restrict ourself to case  $\Gamma_{{\cal{A}}_j}=\Gamma_j$ in resonance conditions between the system \s and the ancillas, i.e. $\epsilon_1=E_{\rm C}$ and $\epsilon_2=E_{\rm H}$. As shown in Appendix \ref{sec:analytical_resonant_sol}, under these conditions
(\ref{RESONANT}), the model
 allows for simple 
analytical solution of the form
\begin{equation}
\label{eq:f_k-adim}
f_j(t)= \frac{e^{- t}}{2}+e^{(-\frac{3}{2}+\frac{\kappa_j}{2})t}\bigg(\frac{1+\kappa_j}{4\kappa_j}\bigg)- e^{(-\frac{3}{2}-\frac{\kappa_j}{2}) t}\bigg(\frac{1-\kappa_j}{4\kappa_j}\bigg)\ ,
\end{equation}
where to simplify the notation the time has been expressed in units of $\Gamma_j^{-1}$ and where $\kappa_j:=\sqrt{1-16(\gamma_j/\Gamma_j)^2}$. 
As for the previous example we enforce symmetric conditions in which the couplings to the thermal baths have the same Lindbladian form and strength, i.e. $\Gamma_{\rm H}=\Gamma_{\rm C}$ and $\gamma_{\rm H}=\gamma_{\rm C}$, implying
$f_{\rm C}(t)=f_{\rm H}(t)=f(t,\kappa)$.
Under this assumption it is not difficult to prove (see Appendix~\ref{app:sym_otto}) that the maximum value of the power obtainable from~\eqref{eq:ottopower} is found on the bisector $\tau_{\rm C}=\tau_{\rm H}=\tau$. 
Specifically with this choice we get 
\begin{equation}
P= \frac{(E_{\rm H}-E_{\rm C})(p_{\rm C}-p_{\rm H})}{2\tau} 
\frac{1-f(\tau,\kappa)}{1+f(\tau,\kappa)} \;, \label{POWEROTTONM} 
\end{equation} 
which for each value of $\kappa(y)$ has a maximum for a finite value of the duration $\bar{\tau}(y)$.
In Fig.~\ref{fig:powerkmax} we plot the obtained maximum as a function of $y=\gamma/\Gamma$, normalized to the $\gamma=0$ case, for which Eq.~\eqref{POWEROTTONM} is easily seen to have a maximum equal to $\frac{\Gamma}{4}(E_{\rm H}-E_{\rm C})(p_{\rm C}-p_{\rm H})$. 
As in the case of the Carnot cycle we see once more that  increasing the strength of the non-Markovian coupling parameter $\gamma$ the power of the Otto engine dramatically increases. 

\subsection{Free-energy analysis}
\label{sec:interpret_n-M}

Ruling out the possibility of using non-Markovian effects
to improve the
 efficiency  in  the  model
~\cite{eisert1,eisert2,eisert3}, the power boost we reported above  for the Carnot and Otto cycle can only be seen as a consequence of the latter in the reduction of 
 thermalization timescales. We show here an argument to explain why it happens and how it is related to the non-Markovian building of correlations between the engine \s and the $j$-th bath. 
  
\begin{figure}
\centering
\includegraphics[width=0.48\textwidth]{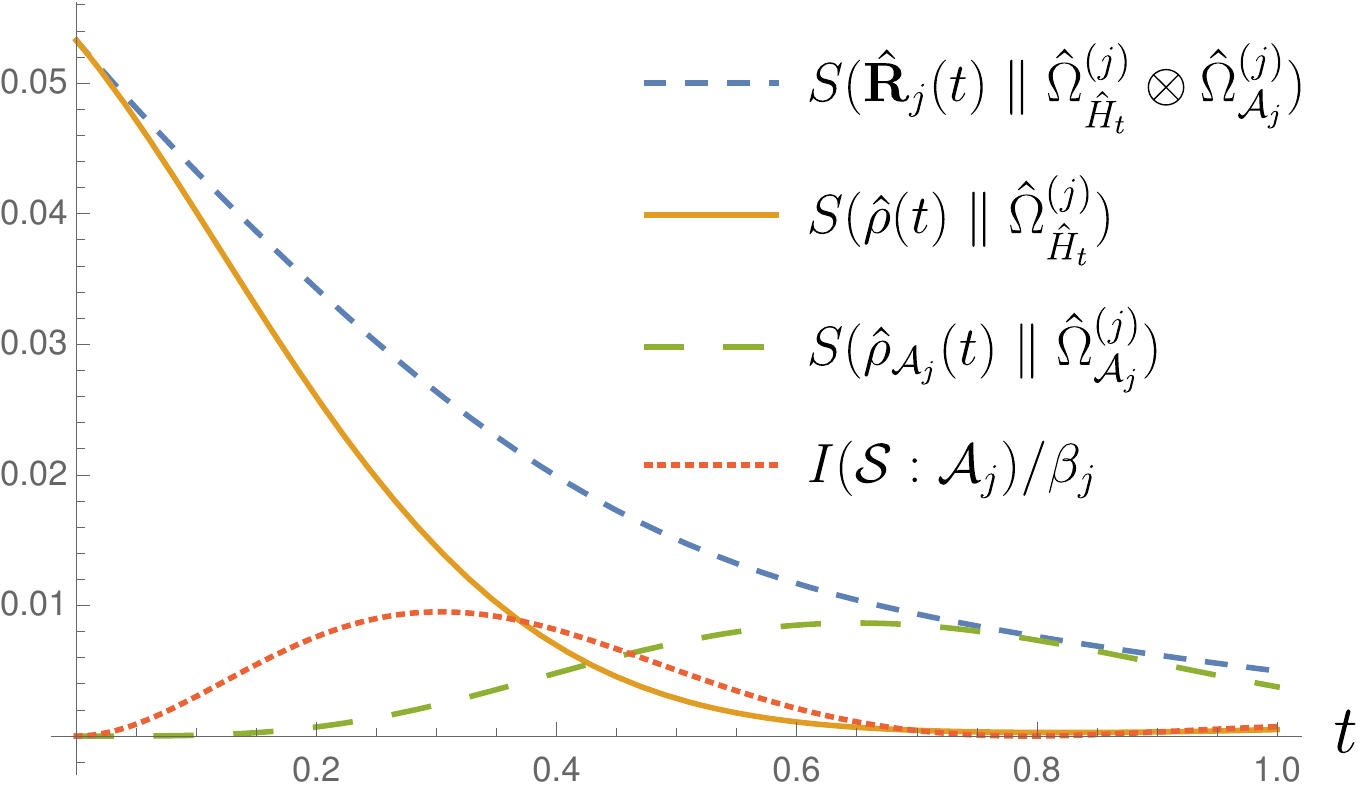}
\caption{(Color online) The total free energy 
(normalized to 0)
 $F_{{\cal S}{\cal A}_j}=S(\hat{\mathbf{R}}_j(t)\parallel \hat{\Omega}^{(j)}_{\hat{H}_t}\ten\hat{\Omega}^{(j)}_{{\cal A}_j})$ (upper line, dashed blue), with its decomposition in the local free energies and the correlation term
 $S(\hat{\rho}(t)\parallel\hat{\Omega}^{(j)}_{\hat{H}_t})+S(\hat{\rho}_{{\cal A}_j}(t)\parallel\hat{\Omega}^{(j)}_{{\cal A}_j})+I(\mathcal{S}:\mathcal{A}_j)/\beta_j$,
 decreasing in time (in units of $\Gamma^{-1}$). The free energy of \s (yellow) decays faster due to the free energy of ${\cal A}_j$ (green, dashed) and the system-bath correlations (red, dotted) growing. Energies are in units of $\epsilon=E_j$. $T_j=2.5$. The initial state is given by $\hat{\rho}(0)\ten\hat{\Omega}^{(j)}_{\hat{H}_{\mathcal{A}_j}}$ with ground state populations ${\rho}_{00}(0)=0.7$, and $[\hat{\Omega}^{(j)}_{\hat{H}_{\mathcal{A}_j}}]_{00}=0.6$ respectively, and the non-Markovian parameter $y_j=2$.}
\label{fig:explain}
\end{figure}

When attaching \s  to the $j$-th bath, the former is out of equilibrium, while according to Eq.~(\ref{FACTOR}) 
 $\mathcal{A}_j$  is already thermalized: we can hence describe
 their initial state as 
 $\hat{\rho}(0)\ten\hat{\Omega}^{(j)}_{\hat{H}_{\mathcal{A}_j}}$. A possible way to quantify the rapidity of \s thermalizing is to compute the relative entropy $S(\hat{\rho}(t)\parallel\hat{\Omega}^{(j)}_{\hat{H}_t})$ and see how fast it diminishes. We recall that according to \eqref{eq:relativeentropy-freeenergy} this quantity also measures the excess of free energy present in the system from the corresponding Gibbs state at temperature $T_j$. 
 Indicating hence with $\hat{\mathbf{R}}_j(t)$ the joint state of ${\cal S}$
 and ${\cal A}_j$ at time $t$ we notice that 
\begin{eqnarray}
S(\hat{\mathbf{R}}_j(t)\parallel \hat{\Omega}^{(j)}_{\hat{H}_t}\ten\hat{\Omega}^{(j)}_{{\cal A}_j})&=&
S(\hat{\rho}(t)\parallel\hat{\Omega}^{(j)}_{\hat{H}_t})+S(\hat{\rho}_{{\cal A}_j}(t)\parallel\hat{\Omega}^{(j)}_{{\cal A}_j})\nonumber \\
&+&I(\mathcal{S}:\mathcal{A}_j)/\beta_j\ ,
\end{eqnarray}
where $I(\mathcal{S}:\mathcal{A}_j)\geq 0$ is the mutual information~\cite{holevo} between the two systems at time $t$ (in the above
derivation we explicitly use the fact that
the mean value of the interaction term stays null due to
the quasi-resonant conditions assumptions).
Hence the variation of the free energy of \s can be expressed as 
\begin{eqnarray}
\Delta S(\hat{\rho}(t)\parallel\hat{\Omega}^{(j)}_{\hat{H}_t})&=&
\Delta S(\hat{\mathbf{R}}_j(t)\parallel \hat{\Omega}^{(j)}_{\hat{H}_t}\ten\hat{\Omega}^{(j)}_{{\cal A}_j})\\ \nonumber 
&-&\Delta S(\hat{\rho}_{{\cal A}_j}(t)\parallel\hat{\Omega}^{(j)}_{{\cal A}_j})-\Delta I(\mathcal{S}:\mathcal{A}_j)/\beta_j\ . \label{DDE1} 
\end{eqnarray}
Now we observe that the last two terms provide a negative contributions to $\Delta S(\hat{\rho}(t)\parallel\hat{\Omega}^{(j)}_{\hat{H}_t})$. Indeed being the relative entropy and mutual information positive definite, and being the initial state 
of ${\cal S}$ and ${\cal A}_j$ factorized by hypothesis,
 they are initially null, meaning that they can only increase with time, thus bringing negative contribution to the r.h.s. of (\ref{DDE1}), that is 
 faster thermalization of ${\cal S}$. For a comparison, in case \s and $\mathcal{A}_j$ do not interact (that is $\gamma=0$ in our model) mutual information would remain zero and $\hat{\rho}_{{\cal A}_j}$ would stay thermal making these two terms exactly null.
In support of the above analysis we report in Fig.~\ref{fig:explain} an example of an evolution of the above quantities. It is possible to see how the free energy of \s decreases faster than the total free energy, due to "suction of free energy" by $\mathcal{A}_j$ and the correlations $I(\mathcal{S}:\mathcal{A}_j)$ building.

\section{Conclusions}
\label{sec:comment}
The  purpose of the work is to assess the effects of non-Markovian dynamics on quantum thermal machines.  We considered a simple class of models which allows shared correlation between the system and some degrees of belonging to the baths, preserving the global evolution as Markovian. This avoids resource pumping from the baths, and cannot induce advantages from the quasi-static point of view. 
Exploiting the S-D technique we studied the thermodynamic performance for a (finite-time) Carnot cycle:
results indicate that the maximum power can indeed be boosted by the presence of this non-Markovian mechanism.  Exact results obtained by studying  Otto cycles confirm this trend. Noting that in general the S-D amplitude $A$ is related to the relaxation time of the system, we are naturally led to interpret this positive effect as an acceleration of the thermalization timescale of ${\cal S}$, in presence of its possible interaction with the local components of the baths (cf. Fig.\ref{fig:singlebath}). Again this is intuitively reasonable, having \s a new channel of thermalization which passes through $\mathcal{A}_j$, and we showed explicitly in Sec.~\ref{sec:interpret_n-M} how this effect is related to the non-Markovian feature of the baths building correlations with the working medium.

Preliminary to the results, we also showed how to optimize the control for a 2-level engine performing a Carnot cycle (in the low-dissipation regime) or an Otto cycle (exactly).

A possible extension of the research would be to generalize these results for systems beyond qubits, with variable number of levels
or in the geometrical picture introduced in \cite{scandi-perarnau,t.b.p.}. 

\section*{Acknowledgements}
The authors wish to thank V.Cavina, P.Erdman, A.Mari for useful discussions.
P.A. is supported by the Spanish MINECO (QIBEQI FIS2016-80773-P, and  Severo Ochoa  SEV-2015-0522), Generalitat de Catalunya (SGR1381 and CERCA Programme), Fundacio Privada Cellex.

\medskip
\bibliographystyle{apsrev4-1}
\bibliography{BIB.bib}

\begin{thebibliography}{62}%
\makeatletter
\providecommand \@ifxundefined [1]{%
 \@ifx{#1\undefined}
}%
\providecommand \@ifnum [1]{%
 \ifnum #1\expandafter \@firstoftwo
 \else \expandafter \@secondoftwo
 \fi
}%
\providecommand \@ifx [1]{%
 \ifx #1\expandafter \@firstoftwo
 \else \expandafter \@secondoftwo
 \fi
}%
\providecommand \natexlab [1]{#1}%
\providecommand \enquote  [1]{``#1''}%
\providecommand \bibnamefont  [1]{#1}%
\providecommand \bibfnamefont [1]{#1}%
\providecommand \citenamefont [1]{#1}%
\providecommand \href@noop [0]{\@secondoftwo}%
\providecommand \href [0]{\begingroup \@sanitize@url \@href}%
\providecommand \@href[1]{\@@startlink{#1}\@@href}%
\providecommand \@@href[1]{\endgroup#1\@@endlink}%
\providecommand \@sanitize@url [0]{\catcode `\\12\catcode `\$12\catcode
  `\&12\catcode `\#12\catcode `\^12\catcode `\_12\catcode `\%12\relax}%
\providecommand \@@startlink[1]{}%
\providecommand \@@endlink[0]{}%
\providecommand \url  [0]{\begingroup\@sanitize@url \@url }%
\providecommand \@url [1]{\endgroup\@href {#1}{\urlprefix }}%
\providecommand \urlprefix  [0]{URL }%
\providecommand \Eprint [0]{\href }%
\providecommand \doibase [0]{http://dx.doi.org/}%
\providecommand \selectlanguage [0]{\@gobble}%
\providecommand \bibinfo  [0]{\@secondoftwo}%
\providecommand \bibfield  [0]{\@secondoftwo}%
\providecommand \translation [1]{[#1]}%
\providecommand \BibitemOpen [0]{}%
\providecommand \bibitemStop [0]{}%
\providecommand \bibitemNoStop [0]{.\EOS\space}%
\providecommand \EOS [0]{\spacefactor3000\relax}%
\providecommand \BibitemShut  [1]{\csname bibitem#1\endcsname}%
\let\auto@bib@innerbib\@empty
\bibitem [{\citenamefont {Gemmer}\ \emph {et~al.}(2009)\citenamefont {Gemmer},
  \citenamefont {Michel},\ and\ \citenamefont
  {Mahler}}]{quantumthermodynamics}%
  \BibitemOpen
  \bibfield  {author} {\bibinfo {author} {\bibfnamefont {J.}~\bibnamefont
  {Gemmer}}, \bibinfo {author} {\bibfnamefont {M.}~\bibnamefont {Michel}}, \
  and\ \bibinfo {author} {\bibfnamefont {G.}~\bibnamefont {Mahler}},\
  }\href@noop {} {\emph {\bibinfo {title} {Quantum Thermodynamics: Emergence of
  Thermodynamic Behavior Within Composite Quantum Systems}}},\ Lecture Notes in
  Physics\ (\bibinfo  {publisher} {Springer Berlin Heidelberg},\ \bibinfo
  {year} {2009})\BibitemShut {NoStop}%
\bibitem [{\citenamefont {Goold}\ \emph {et~al.}(2016)\citenamefont {Goold},
  \citenamefont {Huber}, \citenamefont {Riera}, \citenamefont {del Rio},\ and\
  \citenamefont {Skrzypczyk}}]{gooldreview}%
  \BibitemOpen
  \bibfield  {author} {\bibinfo {author} {\bibfnamefont {J.}~\bibnamefont
  {Goold}}, \bibinfo {author} {\bibfnamefont {M.}~\bibnamefont {Huber}},
  \bibinfo {author} {\bibfnamefont {A.}~\bibnamefont {Riera}}, \bibinfo
  {author} {\bibfnamefont {L.}~\bibnamefont {del Rio}}, \ and\ \bibinfo
  {author} {\bibfnamefont {P.}~\bibnamefont {Skrzypczyk}},\ }\href@noop {}
  {\bibfield  {journal} {\bibinfo  {journal} {Journal of Physics A:
  Mathematical and Theoretical}\ }\textbf {\bibinfo {volume} {49}},\ \bibinfo
  {pages} {143001} (\bibinfo {year} {2016})}\BibitemShut {NoStop}%
\bibitem [{\citenamefont {Vinjanampathy}\ and\ \citenamefont
  {Anders}(2016)}]{andersreview}%
  \BibitemOpen
  \bibfield  {author} {\bibinfo {author} {\bibfnamefont {S.}~\bibnamefont
  {Vinjanampathy}}\ and\ \bibinfo {author} {\bibfnamefont {J.}~\bibnamefont
  {Anders}},\ }\href {\doibase 10.1080/00107514.2016.1201896} {\bibfield
  {journal} {\bibinfo  {journal} {Contemporary Physics}\ }\textbf {\bibinfo
  {volume} {57}},\ \bibinfo {pages} {545} (\bibinfo {year} {2016})}\BibitemShut
  {NoStop}%
\bibitem [{\citenamefont {Linden}\ \emph {et~al.}(2010)\citenamefont {Linden},
  \citenamefont {Popescu},\ and\ \citenamefont {Skrzypczyk}}]{refrigerator1}%
  \BibitemOpen
  \bibfield  {author} {\bibinfo {author} {\bibfnamefont {N.}~\bibnamefont
  {Linden}}, \bibinfo {author} {\bibfnamefont {S.}~\bibnamefont {Popescu}}, \
  and\ \bibinfo {author} {\bibfnamefont {P.}~\bibnamefont {Skrzypczyk}},\
  }\href@noop {} {\bibfield  {journal} {\bibinfo  {journal} {Physical Review
  Letters}\ }\textbf {\bibinfo {volume} {105}},\ \bibinfo {pages} {130401}
  (\bibinfo {year} {2010})}\BibitemShut {NoStop}%
\bibitem [{\citenamefont {Skrzypczyk}\ \emph {et~al.}(2011)\citenamefont
  {Skrzypczyk}, \citenamefont {Brunner}, \citenamefont {Linden},\ and\
  \citenamefont {Popescu}}]{refrigerator2}%
  \BibitemOpen
  \bibfield  {author} {\bibinfo {author} {\bibfnamefont {P.}~\bibnamefont
  {Skrzypczyk}}, \bibinfo {author} {\bibfnamefont {N.}~\bibnamefont {Brunner}},
  \bibinfo {author} {\bibfnamefont {N.}~\bibnamefont {Linden}}, \ and\ \bibinfo
  {author} {\bibfnamefont {S.}~\bibnamefont {Popescu}},\ }\href@noop {}
  {\bibfield  {journal} {\bibinfo  {journal} {Journal of Physics A:
  Mathematical and Theoretical}\ }\textbf {\bibinfo {volume} {44}},\ \bibinfo
  {pages} {492002} (\bibinfo {year} {2011})}\BibitemShut {NoStop}%
\bibitem [{\citenamefont {Koski}\ \emph {et~al.}(2014)\citenamefont {Koski},
  \citenamefont {Maisi}, \citenamefont {Pekola},\ and\ \citenamefont
  {Averin}}]{pekola-SET}%
  \BibitemOpen
  \bibfield  {author} {\bibinfo {author} {\bibfnamefont {J.~V.}\ \bibnamefont
  {Koski}}, \bibinfo {author} {\bibfnamefont {V.~F.}\ \bibnamefont {Maisi}},
  \bibinfo {author} {\bibfnamefont {J.~P.}\ \bibnamefont {Pekola}}, \ and\
  \bibinfo {author} {\bibfnamefont {D.~V.}\ \bibnamefont {Averin}},\
  }\href@noop {} {\bibfield  {journal} {\bibinfo  {journal} {Proceedings of the
  National Academy of Sciences}\ }\textbf {\bibinfo {volume} {111}},\ \bibinfo
  {pages} {13786} (\bibinfo {year} {2014})}\BibitemShut {NoStop}%
\bibitem [{\citenamefont {Pekola}(2015)}]{pekolareview}%
  \BibitemOpen
  \bibfield  {author} {\bibinfo {author} {\bibfnamefont {J.~P.}\ \bibnamefont
  {Pekola}},\ }\href@noop {} {\bibfield  {journal} {\bibinfo  {journal} {Nature
  Physics}\ }\textbf {\bibinfo {volume} {11}},\ \bibinfo {pages} {118}
  (\bibinfo {year} {2015})}\BibitemShut {NoStop}%
\bibitem [{\citenamefont {Batalh{\~a}o}\ \emph {et~al.}(2014)\citenamefont
  {Batalh{\~a}o}, \citenamefont {Souza}, \citenamefont {Mazzola}, \citenamefont
  {Auccaise}, \citenamefont {Sarthour}, \citenamefont {Oliveira}, \citenamefont
  {Goold}, \citenamefont {De~Chiara}, \citenamefont {Paternostro},\ and\
  \citenamefont {Serra}}]{liquid-NMR}%
  \BibitemOpen
  \bibfield  {author} {\bibinfo {author} {\bibfnamefont {T.~B.}\ \bibnamefont
  {Batalh{\~a}o}}, \bibinfo {author} {\bibfnamefont {A.~M.}\ \bibnamefont
  {Souza}}, \bibinfo {author} {\bibfnamefont {L.}~\bibnamefont {Mazzola}},
  \bibinfo {author} {\bibfnamefont {R.}~\bibnamefont {Auccaise}}, \bibinfo
  {author} {\bibfnamefont {R.~S.}\ \bibnamefont {Sarthour}}, \bibinfo {author}
  {\bibfnamefont {I.~S.}\ \bibnamefont {Oliveira}}, \bibinfo {author}
  {\bibfnamefont {J.}~\bibnamefont {Goold}}, \bibinfo {author} {\bibfnamefont
  {G.}~\bibnamefont {De~Chiara}}, \bibinfo {author} {\bibfnamefont
  {M.}~\bibnamefont {Paternostro}}, \ and\ \bibinfo {author} {\bibfnamefont
  {R.~M.}\ \bibnamefont {Serra}},\ }\href@noop {} {\bibfield  {journal}
  {\bibinfo  {journal} {Physical Review Letters}\ }\textbf {\bibinfo {volume}
  {113}},\ \bibinfo {pages} {140601} (\bibinfo {year} {2014})}\BibitemShut
  {NoStop}%
\bibitem [{\citenamefont {An}\ \emph {et~al.}(2015)\citenamefont {An},
  \citenamefont {Zhang}, \citenamefont {Um}, \citenamefont {Lv}, \citenamefont
  {Lu}, \citenamefont {Zhang}, \citenamefont {Yin}, \citenamefont {Quan},\ and\
  \citenamefont {Kim}}]{ion-trap_Jarzynski}%
  \BibitemOpen
  \bibfield  {author} {\bibinfo {author} {\bibfnamefont {S.}~\bibnamefont
  {An}}, \bibinfo {author} {\bibfnamefont {J.-N.}\ \bibnamefont {Zhang}},
  \bibinfo {author} {\bibfnamefont {M.}~\bibnamefont {Um}}, \bibinfo {author}
  {\bibfnamefont {D.}~\bibnamefont {Lv}}, \bibinfo {author} {\bibfnamefont
  {Y.}~\bibnamefont {Lu}}, \bibinfo {author} {\bibfnamefont {J.}~\bibnamefont
  {Zhang}}, \bibinfo {author} {\bibfnamefont {Z.-Q.}\ \bibnamefont {Yin}},
  \bibinfo {author} {\bibfnamefont {H.}~\bibnamefont {Quan}}, \ and\ \bibinfo
  {author} {\bibfnamefont {K.}~\bibnamefont {Kim}},\ }\href@noop {} {\bibfield
  {journal} {\bibinfo  {journal} {Nature Physics}\ }\textbf {\bibinfo {volume}
  {11}},\ \bibinfo {pages} {193} (\bibinfo {year} {2015})}\BibitemShut
  {NoStop}%
\bibitem [{\citenamefont {Ro{\ss}nagel}\ \emph {et~al.}(2016)\citenamefont
  {Ro{\ss}nagel}, \citenamefont {Dawkins}, \citenamefont {Tolazzi},
  \citenamefont {Abah}, \citenamefont {Lutz}, \citenamefont {Schmidt-Kaler},\
  and\ \citenamefont {Singer}}]{single-atom_heat-engine}%
  \BibitemOpen
  \bibfield  {author} {\bibinfo {author} {\bibfnamefont {J.}~\bibnamefont
  {Ro{\ss}nagel}}, \bibinfo {author} {\bibfnamefont {S.~T.}\ \bibnamefont
  {Dawkins}}, \bibinfo {author} {\bibfnamefont {K.~N.}\ \bibnamefont
  {Tolazzi}}, \bibinfo {author} {\bibfnamefont {O.}~\bibnamefont {Abah}},
  \bibinfo {author} {\bibfnamefont {E.}~\bibnamefont {Lutz}}, \bibinfo {author}
  {\bibfnamefont {F.}~\bibnamefont {Schmidt-Kaler}}, \ and\ \bibinfo {author}
  {\bibfnamefont {K.}~\bibnamefont {Singer}},\ }\href@noop {} {\bibfield
  {journal} {\bibinfo  {journal} {Science}\ }\textbf {\bibinfo {volume}
  {352}},\ \bibinfo {pages} {325} (\bibinfo {year} {2016})}\BibitemShut
  {NoStop}%
\bibitem [{\citenamefont {Zhang}\ \emph
  {et~al.}(2014{\natexlab{a}})\citenamefont {Zhang}, \citenamefont {Bariani},\
  and\ \citenamefont {Meystre}}]{optomechanical_h.e.}%
  \BibitemOpen
  \bibfield  {author} {\bibinfo {author} {\bibfnamefont {K.}~\bibnamefont
  {Zhang}}, \bibinfo {author} {\bibfnamefont {F.}~\bibnamefont {Bariani}}, \
  and\ \bibinfo {author} {\bibfnamefont {P.}~\bibnamefont {Meystre}},\
  }\href@noop {} {\bibfield  {journal} {\bibinfo  {journal} {Physical Review
  Letters}\ }\textbf {\bibinfo {volume} {112}},\ \bibinfo {pages} {150602}
  (\bibinfo {year} {2014}{\natexlab{a}})}\BibitemShut {NoStop}%
\bibitem [{\citenamefont {Abah}\ \emph {et~al.}(2012)\citenamefont {Abah},
  \citenamefont {Rossnagel}, \citenamefont {Jacob}, \citenamefont {Deffner},
  \citenamefont {Schmidt-Kaler}, \citenamefont {Singer},\ and\ \citenamefont
  {Lutz}}]{single-atom_prop}%
  \BibitemOpen
  \bibfield  {author} {\bibinfo {author} {\bibfnamefont {O.}~\bibnamefont
  {Abah}}, \bibinfo {author} {\bibfnamefont {J.}~\bibnamefont {Rossnagel}},
  \bibinfo {author} {\bibfnamefont {G.}~\bibnamefont {Jacob}}, \bibinfo
  {author} {\bibfnamefont {S.}~\bibnamefont {Deffner}}, \bibinfo {author}
  {\bibfnamefont {F.}~\bibnamefont {Schmidt-Kaler}}, \bibinfo {author}
  {\bibfnamefont {K.}~\bibnamefont {Singer}}, \ and\ \bibinfo {author}
  {\bibfnamefont {E.}~\bibnamefont {Lutz}},\ }\href@noop {} {\bibfield
  {journal} {\bibinfo  {journal} {Physical review letters}\ }\textbf {\bibinfo
  {volume} {109}},\ \bibinfo {pages} {203006} (\bibinfo {year}
  {2012})}\BibitemShut {NoStop}%
\bibitem [{\citenamefont {Ronzani}\ \emph {et~al.}(2018)\citenamefont
  {Ronzani}, \citenamefont {Karimi}, \citenamefont {Senior}, \citenamefont
  {Chang}, \citenamefont {Peltonen}, \citenamefont {Chen},\ and\ \citenamefont
  {Pekola}}]{ronzani2018tunable}%
  \BibitemOpen
  \bibfield  {author} {\bibinfo {author} {\bibfnamefont {A.}~\bibnamefont
  {Ronzani}}, \bibinfo {author} {\bibfnamefont {B.}~\bibnamefont {Karimi}},
  \bibinfo {author} {\bibfnamefont {J.}~\bibnamefont {Senior}}, \bibinfo
  {author} {\bibfnamefont {Y.-C.}\ \bibnamefont {Chang}}, \bibinfo {author}
  {\bibfnamefont {J.~T.}\ \bibnamefont {Peltonen}}, \bibinfo {author}
  {\bibfnamefont {C.}~\bibnamefont {Chen}}, \ and\ \bibinfo {author}
  {\bibfnamefont {J.~P.}\ \bibnamefont {Pekola}},\ }\href@noop {} {\bibfield
  {journal} {\bibinfo  {journal} {Nature Physics}\ }\textbf {\bibinfo {volume}
  {14}},\ \bibinfo {pages} {991} (\bibinfo {year} {2018})}\BibitemShut
  {NoStop}%
\bibitem [{\citenamefont {Chen}(1994)}]{chen1994maximum}%
  \BibitemOpen
  \bibfield  {author} {\bibinfo {author} {\bibfnamefont {J.}~\bibnamefont
  {Chen}},\ }\href@noop {} {\bibfield  {journal} {\bibinfo  {journal} {Journal
  of Physics D: Applied Physics}\ }\textbf {\bibinfo {volume} {27}},\ \bibinfo
  {pages} {1144} (\bibinfo {year} {1994})}\BibitemShut {NoStop}%
\bibitem [{\citenamefont {Rezek}\ and\ \citenamefont
  {Kosloff}(2006)}]{rezek2006irreversible}%
  \BibitemOpen
  \bibfield  {author} {\bibinfo {author} {\bibfnamefont {Y.}~\bibnamefont
  {Rezek}}\ and\ \bibinfo {author} {\bibfnamefont {R.}~\bibnamefont
  {Kosloff}},\ }\href@noop {} {\bibfield  {journal} {\bibinfo  {journal} {New
  Journal of Physics}\ }\textbf {\bibinfo {volume} {8}},\ \bibinfo {pages} {83}
  (\bibinfo {year} {2006})}\BibitemShut {NoStop}%
\bibitem [{\citenamefont {Watanabe}\ \emph {et~al.}(2017)\citenamefont
  {Watanabe}, \citenamefont {Venkatesh}, \citenamefont {Talkner},\ and\
  \citenamefont {del Campo}}]{watanabe2017quantum}%
  \BibitemOpen
  \bibfield  {author} {\bibinfo {author} {\bibfnamefont {G.}~\bibnamefont
  {Watanabe}}, \bibinfo {author} {\bibfnamefont {B.~P.}\ \bibnamefont
  {Venkatesh}}, \bibinfo {author} {\bibfnamefont {P.}~\bibnamefont {Talkner}},
  \ and\ \bibinfo {author} {\bibfnamefont {A.}~\bibnamefont {del Campo}},\
  }\href@noop {} {\bibfield  {journal} {\bibinfo  {journal} {Physical Review
  Letters}\ }\textbf {\bibinfo {volume} {118}},\ \bibinfo {pages} {050601}
  (\bibinfo {year} {2017})}\BibitemShut {NoStop}%
\bibitem [{\citenamefont {Scully}\ \emph {et~al.}(2011)\citenamefont {Scully},
  \citenamefont {Chapin}, \citenamefont {Dorfman}, \citenamefont {Kim},\ and\
  \citenamefont {Svidzinsky}}]{scully2011quantum}%
  \BibitemOpen
  \bibfield  {author} {\bibinfo {author} {\bibfnamefont {M.~O.}\ \bibnamefont
  {Scully}}, \bibinfo {author} {\bibfnamefont {K.~R.}\ \bibnamefont {Chapin}},
  \bibinfo {author} {\bibfnamefont {K.~E.}\ \bibnamefont {Dorfman}}, \bibinfo
  {author} {\bibfnamefont {M.~B.}\ \bibnamefont {Kim}}, \ and\ \bibinfo
  {author} {\bibfnamefont {A.}~\bibnamefont {Svidzinsky}},\ }\href@noop {}
  {\bibfield  {journal} {\bibinfo  {journal} {Proceedings of the National
  Academy of Sciences}\ }\textbf {\bibinfo {volume} {108}},\ \bibinfo {pages}
  {15097} (\bibinfo {year} {2011})}\BibitemShut {NoStop}%
\bibitem [{\citenamefont {Correa}\ \emph {et~al.}(2013)\citenamefont {Correa},
  \citenamefont {Palao}, \citenamefont {Adesso},\ and\ \citenamefont
  {Alonso}}]{correa2013performance}%
  \BibitemOpen
  \bibfield  {author} {\bibinfo {author} {\bibfnamefont {L.~A.}\ \bibnamefont
  {Correa}}, \bibinfo {author} {\bibfnamefont {J.~P.}\ \bibnamefont {Palao}},
  \bibinfo {author} {\bibfnamefont {G.}~\bibnamefont {Adesso}}, \ and\ \bibinfo
  {author} {\bibfnamefont {D.}~\bibnamefont {Alonso}},\ }\href@noop {}
  {\bibfield  {journal} {\bibinfo  {journal} {Physical Review E}\ }\textbf
  {\bibinfo {volume} {87}},\ \bibinfo {pages} {042131} (\bibinfo {year}
  {2013})}\BibitemShut {NoStop}%
\bibitem [{\citenamefont {Dorfman}\ \emph {et~al.}(2013)\citenamefont
  {Dorfman}, \citenamefont {Voronine}, \citenamefont {Mukamel},\ and\
  \citenamefont {Scully}}]{dorfman2013photosynthetic}%
  \BibitemOpen
  \bibfield  {author} {\bibinfo {author} {\bibfnamefont {K.~E.}\ \bibnamefont
  {Dorfman}}, \bibinfo {author} {\bibfnamefont {D.~V.}\ \bibnamefont
  {Voronine}}, \bibinfo {author} {\bibfnamefont {S.}~\bibnamefont {Mukamel}}, \
  and\ \bibinfo {author} {\bibfnamefont {M.~O.}\ \bibnamefont {Scully}},\
  }\href@noop {} {\bibfield  {journal} {\bibinfo  {journal} {Proceedings of the
  National Academy of Sciences}\ }\textbf {\bibinfo {volume} {110}},\ \bibinfo
  {pages} {2746} (\bibinfo {year} {2013})}\BibitemShut {NoStop}%
\bibitem [{\citenamefont {Brunner}\ \emph {et~al.}(2014)\citenamefont
  {Brunner}, \citenamefont {Huber}, \citenamefont {Linden}, \citenamefont
  {Popescu}, \citenamefont {Silva},\ and\ \citenamefont
  {Skrzypczyk}}]{brunner2014entanglement}%
  \BibitemOpen
  \bibfield  {author} {\bibinfo {author} {\bibfnamefont {N.}~\bibnamefont
  {Brunner}}, \bibinfo {author} {\bibfnamefont {M.}~\bibnamefont {Huber}},
  \bibinfo {author} {\bibfnamefont {N.}~\bibnamefont {Linden}}, \bibinfo
  {author} {\bibfnamefont {S.}~\bibnamefont {Popescu}}, \bibinfo {author}
  {\bibfnamefont {R.}~\bibnamefont {Silva}}, \ and\ \bibinfo {author}
  {\bibfnamefont {P.}~\bibnamefont {Skrzypczyk}},\ }\href@noop {} {\bibfield
  {journal} {\bibinfo  {journal} {Physical Review E}\ }\textbf {\bibinfo
  {volume} {89}},\ \bibinfo {pages} {032115} (\bibinfo {year}
  {2014})}\BibitemShut {NoStop}%
\bibitem [{\citenamefont {Campisi}\ and\ \citenamefont
  {Fazio}(2016)}]{campisi2016power}%
  \BibitemOpen
  \bibfield  {author} {\bibinfo {author} {\bibfnamefont {M.}~\bibnamefont
  {Campisi}}\ and\ \bibinfo {author} {\bibfnamefont {R.}~\bibnamefont
  {Fazio}},\ }\href@noop {} {\bibfield  {journal} {\bibinfo  {journal} {Nature
  Communications}\ }\textbf {\bibinfo {volume} {7}},\ \bibinfo {pages} {11895}
  (\bibinfo {year} {2016})}\BibitemShut {NoStop}%
\bibitem [{\citenamefont {Brandner}\ \emph {et~al.}(2017)\citenamefont
  {Brandner}, \citenamefont {Bauer},\ and\ \citenamefont
  {Seifert}}]{brandner2017universal}%
  \BibitemOpen
  \bibfield  {author} {\bibinfo {author} {\bibfnamefont {K.}~\bibnamefont
  {Brandner}}, \bibinfo {author} {\bibfnamefont {M.}~\bibnamefont {Bauer}}, \
  and\ \bibinfo {author} {\bibfnamefont {U.}~\bibnamefont {Seifert}},\
  }\href@noop {} {\bibfield  {journal} {\bibinfo  {journal} {Physical Review
  Letters}\ }\textbf {\bibinfo {volume} {119}},\ \bibinfo {pages} {170602}
  (\bibinfo {year} {2017})}\BibitemShut {NoStop}%
\bibitem [{\citenamefont {Breuer}\ and\ \citenamefont
  {Petruccione}(2002)}]{breuer-petruccione}%
  \BibitemOpen
  \bibfield  {author} {\bibinfo {author} {\bibfnamefont {H.-P.}\ \bibnamefont
  {Breuer}}\ and\ \bibinfo {author} {\bibfnamefont {F.}~\bibnamefont
  {Petruccione}},\ }\href@noop {} {\emph {\bibinfo {title} {The theory of open
  quantum systems}}}\ (\bibinfo  {publisher} {Oxford University Press on
  Demand},\ \bibinfo {year} {2002})\BibitemShut {NoStop}%
\bibitem [{\citenamefont {Rivas}\ \emph {et~al.}(2014)\citenamefont {Rivas},
  \citenamefont {Huelga},\ and\ \citenamefont {Plenio}}]{RHP}%
  \BibitemOpen
  \bibfield  {author} {\bibinfo {author} {\bibfnamefont {A.}~\bibnamefont
  {Rivas}}, \bibinfo {author} {\bibfnamefont {S.~F.}\ \bibnamefont {Huelga}}, \
  and\ \bibinfo {author} {\bibfnamefont {M.~B.}\ \bibnamefont {Plenio}},\
  }\href@noop {} {\bibfield  {journal} {\bibinfo  {journal} {Reports on
  Progress in Physics}\ }\textbf {\bibinfo {volume} {77}},\ \bibinfo {pages}
  {094001} (\bibinfo {year} {2014})}\BibitemShut {NoStop}%
\bibitem [{\citenamefont {Breuer}\ \emph {et~al.}(2016)\citenamefont {Breuer},
  \citenamefont {Laine}, \citenamefont {Piilo},\ and\ \citenamefont
  {Vacchini}}]{BLP}%
  \BibitemOpen
  \bibfield  {author} {\bibinfo {author} {\bibfnamefont {H.-P.}\ \bibnamefont
  {Breuer}}, \bibinfo {author} {\bibfnamefont {E.-M.}\ \bibnamefont {Laine}},
  \bibinfo {author} {\bibfnamefont {J.}~\bibnamefont {Piilo}}, \ and\ \bibinfo
  {author} {\bibfnamefont {B.}~\bibnamefont {Vacchini}},\ }\href@noop {}
  {\bibfield  {journal} {\bibinfo  {journal} {Reviews of Modern Physics}\
  }\textbf {\bibinfo {volume} {88}},\ \bibinfo {pages} {021002} (\bibinfo
  {year} {2016})}\BibitemShut {NoStop}%
\bibitem [{\citenamefont {Rivas}\ \emph {et~al.}(2010)\citenamefont {Rivas},
  \citenamefont {Huelga},\ and\ \citenamefont {Plenio}}]{RHP-article}%
  \BibitemOpen
  \bibfield  {author} {\bibinfo {author} {\bibfnamefont {A.}~\bibnamefont
  {Rivas}}, \bibinfo {author} {\bibfnamefont {S.~F.}\ \bibnamefont {Huelga}}, \
  and\ \bibinfo {author} {\bibfnamefont {M.~B.}\ \bibnamefont {Plenio}},\
  }\href@noop {} {\bibfield  {journal} {\bibinfo  {journal} {Physical Review
  Letters}\ }\textbf {\bibinfo {volume} {105}},\ \bibinfo {pages} {050403}
  (\bibinfo {year} {2010})}\BibitemShut {NoStop}%
\bibitem [{\citenamefont {Gorini}\ \emph {et~al.}(1976)\citenamefont {Gorini},
  \citenamefont {Kossakowski},\ and\ \citenamefont {Sudarshan}}]{G-K-S}%
  \BibitemOpen
  \bibfield  {author} {\bibinfo {author} {\bibfnamefont {V.}~\bibnamefont
  {Gorini}}, \bibinfo {author} {\bibfnamefont {A.}~\bibnamefont {Kossakowski}},
  \ and\ \bibinfo {author} {\bibfnamefont {E.~C.~G.}\ \bibnamefont
  {Sudarshan}},\ }\href@noop {} {\bibfield  {journal} {\bibinfo  {journal}
  {Journal of Mathematical Physics}\ }\textbf {\bibinfo {volume} {17}},\
  \bibinfo {pages} {821} (\bibinfo {year} {1976})}\BibitemShut {NoStop}%
\bibitem [{\citenamefont {Lindblad}(1976)}]{Lindblad}%
  \BibitemOpen
  \bibfield  {author} {\bibinfo {author} {\bibfnamefont {G.}~\bibnamefont
  {Lindblad}},\ }\href@noop {} {\bibfield  {journal} {\bibinfo  {journal}
  {Communications in Mathematical Physics}\ }\textbf {\bibinfo {volume} {48}},\
  \bibinfo {pages} {119} (\bibinfo {year} {1976})}\BibitemShut {NoStop}%
\bibitem [{\citenamefont {Bhattacharya}\ \emph {et~al.}(2018)\citenamefont
  {Bhattacharya}, \citenamefont {Bhattacharya},\ and\ \citenamefont
  {Majumdar}}]{n-M_work-resourceth}%
  \BibitemOpen
  \bibfield  {author} {\bibinfo {author} {\bibfnamefont {S.}~\bibnamefont
  {Bhattacharya}}, \bibinfo {author} {\bibfnamefont {B.}~\bibnamefont
  {Bhattacharya}}, \ and\ \bibinfo {author} {\bibfnamefont {A.}~\bibnamefont
  {Majumdar}},\ }\href@noop {} {\bibfield  {journal} {\bibinfo  {journal}
  {arXiv preprint arXiv:1803.06881}\ } (\bibinfo {year} {2018})}\BibitemShut
  {NoStop}%
\bibitem [{\citenamefont {Bylicka}\ \emph {et~al.}(2016)\citenamefont
  {Bylicka}, \citenamefont {Tukiainen}, \citenamefont {Chru{\'s}ci{\'n}ski},
  \citenamefont {Piilo},\ and\ \citenamefont {Maniscalco}}]{n-M_work-finland2}%
  \BibitemOpen
  \bibfield  {author} {\bibinfo {author} {\bibfnamefont {B.}~\bibnamefont
  {Bylicka}}, \bibinfo {author} {\bibfnamefont {M.}~\bibnamefont {Tukiainen}},
  \bibinfo {author} {\bibfnamefont {D.}~\bibnamefont {Chru{\'s}ci{\'n}ski}},
  \bibinfo {author} {\bibfnamefont {J.}~\bibnamefont {Piilo}}, \ and\ \bibinfo
  {author} {\bibfnamefont {S.}~\bibnamefont {Maniscalco}},\ }\href@noop {}
  {\bibfield  {journal} {\bibinfo  {journal} {Scientific reports}\ }\textbf
  {\bibinfo {volume} {6}},\ \bibinfo {pages} {27989} (\bibinfo {year}
  {2016})}\BibitemShut {NoStop}%
\bibitem [{\citenamefont {Mirkin}\ \emph {et~al.}(2017)\citenamefont {Mirkin},
  \citenamefont {Poggi},\ and\ \citenamefont
  {Wisniacki}}]{n-M_work-inthequestofoptimalcontrol}%
  \BibitemOpen
  \bibfield  {author} {\bibinfo {author} {\bibfnamefont {N.}~\bibnamefont
  {Mirkin}}, \bibinfo {author} {\bibfnamefont {P.}~\bibnamefont {Poggi}}, \
  and\ \bibinfo {author} {\bibfnamefont {D.}~\bibnamefont {Wisniacki}},\
  }\href@noop {} {\bibfield  {journal} {\bibinfo  {journal} {arXiv preprint
  arXiv:1711.10551}\ } (\bibinfo {year} {2017})}\BibitemShut {NoStop}%
\bibitem [{\citenamefont {Mukherjee}\ \emph {et~al.}(2015)\citenamefont
  {Mukherjee}, \citenamefont {Giovannetti}, \citenamefont {Fazio},
  \citenamefont {Huelga}, \citenamefont {Calarco},\ and\ \citenamefont
  {Montangero}}]{n-M_work-giova}%
  \BibitemOpen
  \bibfield  {author} {\bibinfo {author} {\bibfnamefont {V.}~\bibnamefont
  {Mukherjee}}, \bibinfo {author} {\bibfnamefont {V.}~\bibnamefont
  {Giovannetti}}, \bibinfo {author} {\bibfnamefont {R.}~\bibnamefont {Fazio}},
  \bibinfo {author} {\bibfnamefont {S.~F.}\ \bibnamefont {Huelga}}, \bibinfo
  {author} {\bibfnamefont {T.}~\bibnamefont {Calarco}}, \ and\ \bibinfo
  {author} {\bibfnamefont {S.}~\bibnamefont {Montangero}},\ }\href@noop {}
  {\bibfield  {journal} {\bibinfo  {journal} {New Journal of Physics}\ }\textbf
  {\bibinfo {volume} {17}},\ \bibinfo {pages} {063031} (\bibinfo {year}
  {2015})}\BibitemShut {NoStop}%
\bibitem [{\citenamefont {Raja}\ \emph {et~al.}(2018)\citenamefont {Raja},
  \citenamefont {Borrelli}, \citenamefont {Schmidt}, \citenamefont {Pekola},\
  and\ \citenamefont {Maniscalco}}]{n-M_work-finland}%
  \BibitemOpen
  \bibfield  {author} {\bibinfo {author} {\bibfnamefont {S.~H.}\ \bibnamefont
  {Raja}}, \bibinfo {author} {\bibfnamefont {M.}~\bibnamefont {Borrelli}},
  \bibinfo {author} {\bibfnamefont {R.}~\bibnamefont {Schmidt}}, \bibinfo
  {author} {\bibfnamefont {J.~P.}\ \bibnamefont {Pekola}}, \ and\ \bibinfo
  {author} {\bibfnamefont {S.}~\bibnamefont {Maniscalco}},\ }\href@noop {}
  {\bibfield  {journal} {\bibinfo  {journal} {Physical Review A}\ }\textbf
  {\bibinfo {volume} {97}},\ \bibinfo {pages} {032133} (\bibinfo {year}
  {2018})}\BibitemShut {NoStop}%
\bibitem [{\citenamefont {Reich}\ \emph {et~al.}(2015)\citenamefont {Reich},
  \citenamefont {Katz},\ and\ \citenamefont {Koch}}]{n-M_work-exploiting}%
  \BibitemOpen
  \bibfield  {author} {\bibinfo {author} {\bibfnamefont {D.~M.}\ \bibnamefont
  {Reich}}, \bibinfo {author} {\bibfnamefont {N.}~\bibnamefont {Katz}}, \ and\
  \bibinfo {author} {\bibfnamefont {C.~P.}\ \bibnamefont {Koch}},\ }\href@noop
  {} {\bibfield  {journal} {\bibinfo  {journal} {Scientific Reports}\ }\textbf
  {\bibinfo {volume} {5}},\ \bibinfo {pages} {12430} (\bibinfo {year}
  {2015})}\BibitemShut {NoStop}%
\bibitem [{\citenamefont {Thomas}\ \emph {et~al.}(2018)\citenamefont {Thomas},
  \citenamefont {Siddharth}, \citenamefont {Banerjee},\ and\ \citenamefont
  {Ghosh}}]{n-M_work-india}%
  \BibitemOpen
  \bibfield  {author} {\bibinfo {author} {\bibfnamefont {G.}~\bibnamefont
  {Thomas}}, \bibinfo {author} {\bibfnamefont {N.}~\bibnamefont {Siddharth}},
  \bibinfo {author} {\bibfnamefont {S.}~\bibnamefont {Banerjee}}, \ and\
  \bibinfo {author} {\bibfnamefont {S.}~\bibnamefont {Ghosh}},\ }\href@noop {}
  {\bibfield  {journal} {\bibinfo  {journal} {arXiv preprint arXiv:1801.00744}\
  } (\bibinfo {year} {2018})}\BibitemShut {NoStop}%
\bibitem [{\citenamefont {Zhang}\ \emph
  {et~al.}(2014{\natexlab{b}})\citenamefont {Zhang}, \citenamefont {Huang},\
  and\ \citenamefont {Yi}}]{n-M_work-china}%
  \BibitemOpen
  \bibfield  {author} {\bibinfo {author} {\bibfnamefont {X.}~\bibnamefont
  {Zhang}}, \bibinfo {author} {\bibfnamefont {X.}~\bibnamefont {Huang}}, \ and\
  \bibinfo {author} {\bibfnamefont {X.}~\bibnamefont {Yi}},\ }\href@noop {}
  {\bibfield  {journal} {\bibinfo  {journal} {Journal of Physics A:
  Mathematical and Theoretical}\ }\textbf {\bibinfo {volume} {47}},\ \bibinfo
  {pages} {455002} (\bibinfo {year} {2014}{\natexlab{b}})}\BibitemShut
  {NoStop}%
\bibitem [{\citenamefont {Basilewitsch}\ \emph {et~al.}(2017)\citenamefont
  {Basilewitsch}, \citenamefont {Schmidt}, \citenamefont {Sugny}, \citenamefont
  {Maniscalco},\ and\ \citenamefont {Koch}}]{basilewitsch2017beating}%
  \BibitemOpen
  \bibfield  {author} {\bibinfo {author} {\bibfnamefont {D.}~\bibnamefont
  {Basilewitsch}}, \bibinfo {author} {\bibfnamefont {R.}~\bibnamefont
  {Schmidt}}, \bibinfo {author} {\bibfnamefont {D.}~\bibnamefont {Sugny}},
  \bibinfo {author} {\bibfnamefont {S.}~\bibnamefont {Maniscalco}}, \ and\
  \bibinfo {author} {\bibfnamefont {C.~P.}\ \bibnamefont {Koch}},\ }\href@noop
  {} {\bibfield  {journal} {\bibinfo  {journal} {New Journal of Physics}\
  }\textbf {\bibinfo {volume} {19}},\ \bibinfo {pages} {113042} (\bibinfo
  {year} {2017})}\BibitemShut {NoStop}%
\bibitem [{\citenamefont {Pezzutto}\ \emph {et~al.}(2018)\citenamefont
  {Pezzutto}, \citenamefont {Paternostro},\ and\ \citenamefont
  {Omar}}]{pezzutto2018out}%
  \BibitemOpen
  \bibfield  {author} {\bibinfo {author} {\bibfnamefont {M.}~\bibnamefont
  {Pezzutto}}, \bibinfo {author} {\bibfnamefont {M.}~\bibnamefont
  {Paternostro}}, \ and\ \bibinfo {author} {\bibfnamefont {Y.}~\bibnamefont
  {Omar}},\ }\href@noop {} {\bibfield  {journal} {\bibinfo  {journal} {Quantum
  Science and Technology}\ } (\bibinfo {year} {2018})}\BibitemShut {NoStop}%
\bibitem [{\citenamefont {Quan}\ \emph {et~al.}(2007)\citenamefont {Quan},
  \citenamefont {Liu}, \citenamefont {Sun},\ and\ \citenamefont
  {Nori}}]{quan2007quantum}%
  \BibitemOpen
  \bibfield  {author} {\bibinfo {author} {\bibfnamefont {H.}~\bibnamefont
  {Quan}}, \bibinfo {author} {\bibfnamefont {Y.}~\bibnamefont {Liu}}, \bibinfo
  {author} {\bibfnamefont {C.}~\bibnamefont {Sun}}, \ and\ \bibinfo {author}
  {\bibfnamefont {F.}~\bibnamefont {Nori}},\ }\href@noop {} {\bibfield
  {journal} {\bibinfo  {journal} {Physical Review E}\ }\textbf {\bibinfo
  {volume} {76}},\ \bibinfo {pages} {031105} (\bibinfo {year}
  {2007})}\BibitemShut {NoStop}%
\bibitem [{\citenamefont {Karimi}\ and\ \citenamefont
  {Pekola}(2016)}]{karimi2016otto}%
  \BibitemOpen
  \bibfield  {author} {\bibinfo {author} {\bibfnamefont {B.}~\bibnamefont
  {Karimi}}\ and\ \bibinfo {author} {\bibfnamefont {J.}~\bibnamefont
  {Pekola}},\ }\href@noop {} {\bibfield  {journal} {\bibinfo  {journal}
  {Physical Review B}\ }\textbf {\bibinfo {volume} {94}},\ \bibinfo {pages}
  {184503} (\bibinfo {year} {2016})}\BibitemShut {NoStop}%
\bibitem [{\citenamefont {Kosloff}\ and\ \citenamefont
  {Rezek}(2017)}]{kosloff2017quantum}%
  \BibitemOpen
  \bibfield  {author} {\bibinfo {author} {\bibfnamefont {R.}~\bibnamefont
  {Kosloff}}\ and\ \bibinfo {author} {\bibfnamefont {Y.}~\bibnamefont
  {Rezek}},\ }\href@noop {} {\bibfield  {journal} {\bibinfo  {journal}
  {Entropy}\ }\textbf {\bibinfo {volume} {19}},\ \bibinfo {pages} {136}
  (\bibinfo {year} {2017})}\BibitemShut {NoStop}%
\bibitem [{\citenamefont {Cavina}\ \emph
  {et~al.}(2017{\natexlab{a}})\citenamefont {Cavina}, \citenamefont {Mari},\
  and\ \citenamefont {Giovannetti}}]{slowdriving}%
  \BibitemOpen
  \bibfield  {author} {\bibinfo {author} {\bibfnamefont {V.}~\bibnamefont
  {Cavina}}, \bibinfo {author} {\bibfnamefont {A.}~\bibnamefont {Mari}}, \ and\
  \bibinfo {author} {\bibfnamefont {V.}~\bibnamefont {Giovannetti}},\ }\href
  {\doibase 10.1103/PhysRevLett.119.050601} {\bibfield  {journal} {\bibinfo
  {journal} {Physical Review Letters}\ }\textbf {\bibinfo {volume} {119}},\
  \bibinfo {pages} {050601} (\bibinfo {year} {2017}{\natexlab{a}})}\BibitemShut
  {NoStop}%
\bibitem [{\citenamefont {Cavina}\ \emph
  {et~al.}(2017{\natexlab{b}})\citenamefont {Cavina}, \citenamefont {Mari},
  \citenamefont {Carlini},\ and\ \citenamefont {Giovannetti}}]{optimalcontrol}%
  \BibitemOpen
  \bibfield  {author} {\bibinfo {author} {\bibfnamefont {V.}~\bibnamefont
  {Cavina}}, \bibinfo {author} {\bibfnamefont {A.}~\bibnamefont {Mari}},
  \bibinfo {author} {\bibfnamefont {A.}~\bibnamefont {Carlini}}, \ and\
  \bibinfo {author} {\bibfnamefont {V.}~\bibnamefont {Giovannetti}},\
  }\href@noop {} {\bibfield  {journal} {\bibinfo  {journal} {arXiv preprint
  arXiv:1709.07400}\ } (\bibinfo {year} {2017}{\natexlab{b}})}\BibitemShut
  {NoStop}%
\bibitem [{\citenamefont {Gardiner}\ \emph {et~al.}(2004)\citenamefont
  {Gardiner}, \citenamefont {Zoller},\ and\ \citenamefont {Zoller}}]{gardiner}%
  \BibitemOpen
  \bibfield  {author} {\bibinfo {author} {\bibfnamefont {C.}~\bibnamefont
  {Gardiner}}, \bibinfo {author} {\bibfnamefont {P.}~\bibnamefont {Zoller}}, \
  and\ \bibinfo {author} {\bibfnamefont {P.}~\bibnamefont {Zoller}},\
  }\href@noop {} {\emph {\bibinfo {title} {Quantum noise: a handbook of
  Markovian and non-Markovian quantum stochastic methods with applications to
  quantum optics}}},\ Vol.~\bibinfo {volume} {56}\ (\bibinfo  {publisher}
  {Springer Science \& Business Media},\ \bibinfo {year} {2004})\BibitemShut
  {NoStop}%
\bibitem [{\citenamefont {Esposito}\ \emph {et~al.}(2010)\citenamefont
  {Esposito}, \citenamefont {Kawai}, \citenamefont {Lindenberg},\ and\
  \citenamefont {Van~den Broeck}}]{esposito-EMP-bounds}%
  \BibitemOpen
  \bibfield  {author} {\bibinfo {author} {\bibfnamefont {M.}~\bibnamefont
  {Esposito}}, \bibinfo {author} {\bibfnamefont {R.}~\bibnamefont {Kawai}},
  \bibinfo {author} {\bibfnamefont {K.}~\bibnamefont {Lindenberg}}, \ and\
  \bibinfo {author} {\bibfnamefont {C.}~\bibnamefont {Van~den Broeck}},\ }\href
  {\doibase 10.1103/PhysRevLett.105.150603} {\bibfield  {journal} {\bibinfo
  {journal} {Physical Review Letters}\ }\textbf {\bibinfo {volume} {105}},\
  \bibinfo {pages} {150603} (\bibinfo {year} {2010})}\BibitemShut {NoStop}%
\bibitem [{\citenamefont {Alicki}(1979)}]{alickiwork}%
  \BibitemOpen
  \bibfield  {author} {\bibinfo {author} {\bibfnamefont {R.}~\bibnamefont
  {Alicki}},\ }\href@noop {} {\bibfield  {journal} {\bibinfo  {journal}
  {Journal of Physics A: Mathematical and General}\ }\textbf {\bibinfo {volume}
  {12}},\ \bibinfo {pages} {L103} (\bibinfo {year} {1979})}\BibitemShut
  {NoStop}%
\bibitem [{\citenamefont {Anders}\ and\ \citenamefont
  {Giovannetti}(2013)}]{anders-giova}%
  \BibitemOpen
  \bibfield  {author} {\bibinfo {author} {\bibfnamefont {J.}~\bibnamefont
  {Anders}}\ and\ \bibinfo {author} {\bibfnamefont {V.}~\bibnamefont
  {Giovannetti}},\ }\href@noop {} {\bibfield  {journal} {\bibinfo  {journal}
  {New Journal of Physics}\ }\textbf {\bibinfo {volume} {15}},\ \bibinfo
  {pages} {033022} (\bibinfo {year} {2013})}\BibitemShut {NoStop}%
\bibitem [{\citenamefont {Kieu}(2004)}]{kieuwork}%
  \BibitemOpen
  \bibfield  {author} {\bibinfo {author} {\bibfnamefont {T.~D.}\ \bibnamefont
  {Kieu}},\ }\href {\doibase 10.1103/PhysRevLett.93.140403} {\bibfield
  {journal} {\bibinfo  {journal} {Phys. Rev. Lett.}\ }\textbf {\bibinfo
  {volume} {93}},\ \bibinfo {pages} {140403} (\bibinfo {year}
  {2004})}\BibitemShut {NoStop}%
\bibitem [{\citenamefont {Parrondo}\ \emph {et~al.}(2015)\citenamefont
  {Parrondo}, \citenamefont {Horowitz},\ and\ \citenamefont
  {Sagawa}}]{parrondo}%
  \BibitemOpen
  \bibfield  {author} {\bibinfo {author} {\bibfnamefont {J.~M.}\ \bibnamefont
  {Parrondo}}, \bibinfo {author} {\bibfnamefont {J.~M.}\ \bibnamefont
  {Horowitz}}, \ and\ \bibinfo {author} {\bibfnamefont {T.}~\bibnamefont
  {Sagawa}},\ }\href@noop {} {\bibfield  {journal} {\bibinfo  {journal} {Nature
  Physics}\ }\textbf {\bibinfo {volume} {11}},\ \bibinfo {pages} {131}
  (\bibinfo {year} {2015})}\BibitemShut {NoStop}%
\bibitem [{\citenamefont {Esposito}\ and\ \citenamefont {den
  Broeck}(2011)}]{esposito-2ndlaw}%
  \BibitemOpen
  \bibfield  {author} {\bibinfo {author} {\bibfnamefont {M.}~\bibnamefont
  {Esposito}}\ and\ \bibinfo {author} {\bibfnamefont {C.~V.}\ \bibnamefont {den
  Broeck}},\ }\href@noop {} {\bibfield  {journal} {\bibinfo  {journal} {EPL
  (Europhysics Letters)}\ }\textbf {\bibinfo {volume} {95}},\ \bibinfo {pages}
  {40004} (\bibinfo {year} {2011})}\BibitemShut {NoStop}%
\bibitem [{\citenamefont {Holevo}(2012)}]{holevo}%
  \BibitemOpen
  \bibfield  {author} {\bibinfo {author} {\bibfnamefont {A.~S.}\ \bibnamefont
  {Holevo}},\ }\href@noop {} {\emph {\bibinfo {title} {Quantum systems,
  channels, information: a mathematical introduction}}},\ Vol.~\bibinfo
  {volume} {16}\ (\bibinfo  {publisher} {Walter de Gruyter},\ \bibinfo {year}
  {2012})\BibitemShut {NoStop}%
\bibitem [{\citenamefont {Andresen}\ \emph {et~al.}(1984)\citenamefont
  {Andresen}, \citenamefont {Berry}, \citenamefont {Ondrechen},\ and\
  \citenamefont {Salamon}}]{andresenFTT}%
  \BibitemOpen
  \bibfield  {author} {\bibinfo {author} {\bibfnamefont {B.}~\bibnamefont
  {Andresen}}, \bibinfo {author} {\bibfnamefont {R.~S.}\ \bibnamefont {Berry}},
  \bibinfo {author} {\bibfnamefont {M.~J.}\ \bibnamefont {Ondrechen}}, \ and\
  \bibinfo {author} {\bibfnamefont {P.}~\bibnamefont {Salamon}},\ }\href@noop
  {} {\bibfield  {journal} {\bibinfo  {journal} {Accounts of Chemical
  Research}\ }\textbf {\bibinfo {volume} {17}},\ \bibinfo {pages} {266}
  (\bibinfo {year} {1984})}\BibitemShut {NoStop}%
\bibitem [{\citenamefont {Curzon}\ and\ \citenamefont
  {Ahlborn}(1975)}]{curzon-ahlborn}%
  \BibitemOpen
  \bibfield  {author} {\bibinfo {author} {\bibfnamefont {F.~L.}\ \bibnamefont
  {Curzon}}\ and\ \bibinfo {author} {\bibfnamefont {B.}~\bibnamefont
  {Ahlborn}},\ }\href {\doibase 10.1119/1.10023} {\bibfield  {journal}
  {\bibinfo  {journal} {American Journal of Physics}\ }\textbf {\bibinfo
  {volume} {43}},\ \bibinfo {pages} {22} (\bibinfo {year} {1975})}\BibitemShut
  {NoStop}%
\bibitem [{\citenamefont {Abiuso}\ and\ \citenamefont {Perarnau}()}]{t.b.p.}%
  \BibitemOpen
  \bibfield  {author} {\bibinfo {author} {\bibfnamefont {P.}~\bibnamefont
  {Abiuso}}\ and\ \bibinfo {author} {\bibfnamefont {M.}~\bibnamefont
  {Perarnau}},\ }\href@noop {} {}\bibinfo {note} {Work in
  preparation}\BibitemShut {NoStop}%
\bibitem [{\citenamefont {Erdman}\ \emph {et~al.}(2018)\citenamefont {Erdman},
  \citenamefont {Cavina}, \citenamefont {Fazio}, \citenamefont {Taddei},\ and\
  \citenamefont {Giovannetti}}]{erdman2018maximum}%
  \BibitemOpen
  \bibfield  {author} {\bibinfo {author} {\bibfnamefont {P.~A.}\ \bibnamefont
  {Erdman}}, \bibinfo {author} {\bibfnamefont {V.}~\bibnamefont {Cavina}},
  \bibinfo {author} {\bibfnamefont {R.}~\bibnamefont {Fazio}}, \bibinfo
  {author} {\bibfnamefont {F.}~\bibnamefont {Taddei}}, \ and\ \bibinfo {author}
  {\bibfnamefont {V.}~\bibnamefont {Giovannetti}},\ }\href@noop {} {\bibfield
  {journal} {\bibinfo  {journal} {arXiv preprint arXiv:1812.05089}\ } (\bibinfo
  {year} {2018})}\BibitemShut {NoStop}%
\bibitem [{\citenamefont {Wilming}\ \emph {et~al.}(2016)\citenamefont
  {Wilming}, \citenamefont {Gallego},\ and\ \citenamefont {Eisert}}]{eisert1}%
  \BibitemOpen
  \bibfield  {author} {\bibinfo {author} {\bibfnamefont {H.}~\bibnamefont
  {Wilming}}, \bibinfo {author} {\bibfnamefont {R.}~\bibnamefont {Gallego}}, \
  and\ \bibinfo {author} {\bibfnamefont {J.}~\bibnamefont {Eisert}},\
  }\href@noop {} {\bibfield  {journal} {\bibinfo  {journal} {Physical Review
  E}\ }\textbf {\bibinfo {volume} {93}},\ \bibinfo {pages} {042126} (\bibinfo
  {year} {2016})}\BibitemShut {NoStop}%
\bibitem [{\citenamefont {Lekscha}\ \emph {et~al.}(2018)\citenamefont
  {Lekscha}, \citenamefont {Wilming}, \citenamefont {Eisert},\ and\
  \citenamefont {Gallego}}]{eisert2}%
  \BibitemOpen
  \bibfield  {author} {\bibinfo {author} {\bibfnamefont {J.}~\bibnamefont
  {Lekscha}}, \bibinfo {author} {\bibfnamefont {H.}~\bibnamefont {Wilming}},
  \bibinfo {author} {\bibfnamefont {J.}~\bibnamefont {Eisert}}, \ and\ \bibinfo
  {author} {\bibfnamefont {R.}~\bibnamefont {Gallego}},\ }\href@noop {}
  {\bibfield  {journal} {\bibinfo  {journal} {Physical Review E}\ }\textbf
  {\bibinfo {volume} {97}},\ \bibinfo {pages} {022142} (\bibinfo {year}
  {2018})}\BibitemShut {NoStop}%
\bibitem [{\citenamefont {Perarnau-Llobet}\ \emph {et~al.}(2018)\citenamefont
  {Perarnau-Llobet}, \citenamefont {Wilming}, \citenamefont {Riera},
  \citenamefont {Gallego},\ and\ \citenamefont {Eisert}}]{eisert3}%
  \BibitemOpen
  \bibfield  {author} {\bibinfo {author} {\bibfnamefont {M.}~\bibnamefont
  {Perarnau-Llobet}}, \bibinfo {author} {\bibfnamefont {H.}~\bibnamefont
  {Wilming}}, \bibinfo {author} {\bibfnamefont {A.}~\bibnamefont {Riera}},
  \bibinfo {author} {\bibfnamefont {R.}~\bibnamefont {Gallego}}, \ and\
  \bibinfo {author} {\bibfnamefont {J.}~\bibnamefont {Eisert}},\ }\href@noop {}
  {\bibfield  {journal} {\bibinfo  {journal} {Physical Review Letters}\
  }\textbf {\bibinfo {volume} {120}},\ \bibinfo {pages} {120602} (\bibinfo
  {year} {2018})}\BibitemShut {NoStop}%
\bibitem [{rel()}]{relax_hyp}%
  \BibitemOpen
  \href@noop {} {}\bibinfo {note} {This condition can be always guaranteed also
  when considering systems strongly out-of equilibrium (e.g. for the Otto cycle
  case) assuming a sufficient number $N$ of ancillary systems ${\cal
  A}_j^{(n)}$ ($j={\rm H,C}\; n=1,\dots,N$) sequentially interacting with $\cal
  S$ and relaxing afterwards.}\BibitemShut {Stop}%
\bibitem [{\citenamefont {Scandi}\ and\ \citenamefont
  {Perarnau-Llobet}(2018)}]{scandi-perarnau}%
  \BibitemOpen
  \bibfield  {author} {\bibinfo {author} {\bibfnamefont {M.}~\bibnamefont
  {Scandi}}\ and\ \bibinfo {author} {\bibfnamefont {M.}~\bibnamefont
  {Perarnau-Llobet}},\ }\href@noop {} {\bibfield  {journal} {\bibinfo
  {journal} {arXiv preprint arXiv:1810.05583}\ } (\bibinfo {year}
  {2018})}\BibitemShut {NoStop}%
\bibitem [{\citenamefont {Ma}\ \emph {et~al.}(2018)\citenamefont {Ma},
  \citenamefont {Xu}, \citenamefont {Dong},\ and\ \citenamefont
  {Sun}}]{universalcostraint}%
  \BibitemOpen
  \bibfield  {author} {\bibinfo {author} {\bibfnamefont {Y.-H.}\ \bibnamefont
  {Ma}}, \bibinfo {author} {\bibfnamefont {D.-Z.}\ \bibnamefont {Xu}}, \bibinfo
  {author} {\bibfnamefont {H.}~\bibnamefont {Dong}}, \ and\ \bibinfo {author}
  {\bibfnamefont {C.-P.}\ \bibnamefont {Sun}},\ }\href@noop {} {\bibfield
  {journal} {\bibinfo  {journal} {arXiv preprint arXiv:1802.09806}\ } (\bibinfo
  {year} {2018})}\BibitemShut {NoStop}%
\bibitem [{\citenamefont {Cavina}\ \emph {et~al.}(2018)\citenamefont {Cavina},
  \citenamefont {Mari},\ and\ \citenamefont {Giovannetti}}]{slowdriving1}%
  \BibitemOpen
  \bibfield  {author} {\bibinfo {author} {\bibfnamefont {V.}~\bibnamefont
  {Cavina}}, \bibinfo {author} {\bibfnamefont {A.}~\bibnamefont {Mari}}, \ and\
  \bibinfo {author} {\bibfnamefont {V.}~\bibnamefont {Giovannetti}},\
  }\href@noop {} {\bibfield  {journal} {\bibinfo  {journal} {Proceedings of
  IQIS Conference 2018}\ } (\bibinfo {year} {2018})}\BibitemShut {NoStop}%
\end{thebibliography}%
\pagebreak \newpage
\onecolumngrid
\appendix
\section{The dissipators} 
\label{APP-DISS}
Here we review few examples of dissipators 
$\mathcal{D}_t^{(j)}$ that obey the constraints (\ref{FFD}) and (\ref{STAB}).

The first, and simplest of such models, is provided by the super-operator 
\cite{optimalcontrol}
\begin{equation}\label{PRIMO} 
\mathcal{D}_t^{(j)} [\cdots]=\Gamma_j(\hat{\Omega}^{(j)}_{\hat{H}_t}-\cdots)\;,\\
\end{equation}
with $\Gamma_j>0$ constant, which do not need any specification of the system Hamiltonian.

Assuming instead the Hamiltonian of ${\cal S}$ to be $\hat{H}_t={\epsilon}(t) \hat{\sig}^z /2$ 
(see Eq.~(\ref{energy})), another example is provided by the dissipator 
\begin{equation} \mathcal{D}_t^{(j)}[\cdots]  
:= \sum_{\ell=\pm} \Gamma^{(j)}_\ell(\epsilon(t))( \hat{\sigma}_\ell \cdots  \hat{\sigma}_\ell^\dagger -\frac{1}{2} [ \hat{\sigma}_\ell^\dagger\hat{\sigma}_\ell , \cdots ]_+)\;, 
\label{DIS1} 
\end{equation} 
where $\hat{\sigma}_+$ and  $\hat{\sigma}_-(=\hat{\sigma}_+^\dag)$ are, respectively, the raising and lowering operators of ${\cal S}$, $[\cdots, \cdots]_+$ is the anti-commutator, which exhibit the functional dependence~(\ref{FFD}) upon
$\hat{H}_t$ through the rates $\Gamma^{(j)}_\pm(\epsilon)$
 fulfilling the detailed balance equation condition 
 \begin{eqnarray} 
 {\Gamma^{(j)}_+(\epsilon )}/{\Gamma^{(j)}_-(\epsilon)} 
 = e^{-\beta_j  \epsilon}\;, \end{eqnarray} 
 which ensures~(\ref{STAB}).
 In particular taking 
 \begin{eqnarray} \Gamma^{(j)}_-(\epsilon) : =  (1 - N_{\bf F}(\beta_j \epsilon)) \; \Gamma \;, 
\nonumber \quad  \label{FERMIRATE} 
  \Gamma^{(j)}_+(\epsilon) : =  N_{\bf F}(\beta_j  \epsilon)\; \Gamma \;,  
 \end{eqnarray} 
 with $\Gamma \geq 0$ and
 \begin{equation}
N_{\bf F}(x)=\frac{1}{e^{x}+1}\;, 
\end{equation}
equation~(\ref{DIS1}) can be used to describe the interaction of ${\cal S}$ with a Fermionic bath.
Instead taking 
 \begin{eqnarray}
\nonumber  \Gamma^{(j)}_-(\epsilon) : =  (1 + N_{\bf B}(\beta_j \epsilon)) \; \Gamma \;, \quad
  \Gamma^{(j)}_+(\epsilon) : =  N_{\bf B}(\beta_j  \epsilon)\; \Gamma \;,
   \label{BOSONICRATE}
 \end{eqnarray} 
 with $\Gamma \geq 0$ and
 \begin{equation}
N_{\bf B}(x)=\frac{1}{e^{x}-1}\;, 
\end{equation}
it describes the interaction of ${\cal S}$ with a Bosonic bath.

\section{S-D approximation implies low dissipation} \label{appendixNEW}
A virtue of the S-D approximation  is that it  provides a formal justification of the  
 low-dissipation (L-D) assumption~\cite{esposito-EMP-bounds} which is typically introduced in FTT analysis as a phenomenological working hypothesis.
To see this let us start observing that in the 
S-D theory, at the lowest order of the pertubative expansion~(\ref{expansion})
the von Neumann entropy of the density matrix $\hat{\rho}(t)$ can be expressed as 
\begin{equation}\label{PRIMA1} 
S(t) := \mbox{Tr} [ \hat{\rho}(t) \ln \hat{\rho}(t)]\simeq 
\mbox{Tr} [ (\hat{\rho}^{(0)}(t)+\hat{\rho}^{(1)}(t)) \ln (\hat{\rho}^{(0)}(t)+\hat{\rho}^{(1)}(t))]
\simeq S^{(0)}(t) + \beta_j \Tr[\hat{\rho}^{(1)}(t) \hat{H}_t] \;, 
\end{equation}
where in the last step we used the fact that the term $\hat{\rho}^{(1)}(t)$ is traceless, i.e. 
$\Tr[\hat{\rho}^{(1)}(t)]=0$, and the fact that $\hat{\rho}^{(0)}(t)$ is the instantaneous Gibbs state~(\ref{def:gibbs_state}), i.e. 
$\hat{\rho}^{(0)}(t)=\hat{\Omega}^{(j)}_{\hat{H}_t}$. 
A close inspection reveals that the second contribution of $S(t)$ 
corresponds to the first order correction to the internal energy of the
system defined in Eq.~(\ref{DEFINTE}), i.e.  $E_1(t):=
 \Tr[\hat{\rho}^{(1)}(t) \hat{H}_t]$, allowing us to  cast (\ref{PRIMA1}) as
\begin{equation}
S(t)\simeq S^{(0)}(t)+\beta_j E^{(1)}(t)\ .
\end{equation}
The temporal increment of this quantity can hence be computed as 
\begin{equation}
d S(t)\simeq d S^{(0)}(t)+\beta_j d E^{(1)}(t)= \beta_j d Q^{(0)}(t) + 
\beta_j d Q^{(1)}(t) + \beta_j d W^{(1)}(t)\;, 
\end{equation}
where we used Eq.~(\ref{ORDzero}) and wrote $d E^{(1)}(t)$  in terms of a work
and heat contribution, i.e. $d E^{(1)}(t)= d Q^{(1)}(t) +d W^{(1)}(t)$ (first thermodynamics
principle) with $d Q^{(1)}(t)$ as in Eq.~(\ref{DDF1}) and 
\begin{eqnarray} 
d W^{(1)}(t) := \Tr[\hat{\rho}^{(1)}(t) d \hat{H}_t]\;. \label{DEFW1} 
\end{eqnarray} 
Grouping together all the heat contributions we can hence finally write
\begin{equation}
\label{IRR}
d S(t)\simeq \beta_j d Q(t)  + \beta_j d W^{(1)}(t)\, \Longrightarrow 
d S^{(irr)}_j(t) \simeq \beta_j d W^{(1)}(t)\;,  
\end{equation}
where $d S^{(irr)}_j(t) :=d S(t) - \beta_j d Q(t)$ is the 
irreversible entropy production increment which  quantifies the
differences between  information transfer rates and the heat transfer rate in the system. 
When integrated over a finite time interval $\tau_j$, 
Eq.~(\ref{IRR}) provides an estimation of 
the associated finite irreversible entropy production  $\Delta S^{(irr)}_j$.
In FTT under L-D assumption this term is postulated to be expressed as
 inversely proportional to $\tau_j$ via a constant
term $\Sigma_j$ which only depends on  the coupling constants to the bath, and the cycle endpoints, i.e.~\cite{esposito-EMP-bounds,universalcostraint}
\begin{eqnarray} \Delta S^{(irr)}_j\Big\rvert_{\rm L-D}=\Sigma_j /\tau_j \label{FTTscaling} \;.
\end{eqnarray} Now a $1/\tau_j$ scaling as in  Eq.~(\ref{FTTscaling}) is exactly what one naturally get by computing $ \Delta S^{(irr)}_j$ via direct integration of (\ref{IRR}) due to the fact  that in the S-D expansion the  $\hat{\rho}^{(1)}(t)$ term
has an explicit linear dependence upon $1/\tau_j$, whilst $\hat{\rho}^{(0)}(t)$ and
the associated instantaneous Hamiltonian $\hat{H}_t$ are independent from such parameter~\cite{slowdriving,slowdriving1}.
According to this observation, on one side  we can hence say that S-D provide a natural framework for discussing L-D assumption. 
On the other side instead we can conclude that 
the general  results derived under FTT assumption~\cite{esposito-EMP-bounds,universalcostraint} must apply in the 
characterization  of system  driven under S-D approximation, at least at the first order of the perturbative analysis.
 In particular it is not difficult to see that the condition $\alpha_{\rm C}=\alpha_{\rm H}$ we require in Sec.~\ref{sec:carnot} corresponds to set $\Sigma_{\rm C}=\Sigma_{\rm H}$, which in turn implies that the Curzon-Ahlborn efficiency is the EMP in this regime \cite{esposito-EMP-bounds}. Getting rid of this symmetry simply means to explore the different ratios $\Sigma_{\rm H}/\Sigma_{\rm C}$ and the relative results \cite{esposito-EMP-bounds,universalcostraint} are valid.

\section{Otto cycle beyond the ITT limit} \label{sec:perfOTTO} 
To evaluate the correction terms appearing in Eqs.~(\ref{EFFICotto}) and (\ref{def:powerotto}) for a generic engine
we assume the  dissipation model of Eq.~(\ref{PRIMO}). Let then indicate with 
 $\hat{\rho}^{(in)}_{j,k}$ and $\hat{\rho}^{(out)}_{j,k}$  the states of ${\cal S}$ at the beginning and at the end of the time intervals ${\cal I}_j$ of the $k$-th Otto cycle.
 Due to the presence of the quenches at the steps 2) and 4), 
they must be related as follows \begin{eqnarray}
\hat{\rho}^{(in)}_{{\rm H},k} = \hat{\rho}^{(out)}_{{\rm C},k}\;,\qquad 
\hat{\rho}^{(in)}_{{\rm C},k} = \hat{\rho}^{(out)}_{{\rm H},k-1}\;, \label{COND11} 
\end{eqnarray} 
meaning that  input state of the $k$-th interval ${\cal I}_{\rm H}$
coincides with the output state of the $k$-th interval ${\cal I}_{\rm C}$, while
 the input of the $k$-th interval ${\cal I}_{\rm C}$ with the 
 output of the $k$-th interval ${\cal I}_{\rm H}$.
By direct integration of the ME~(\ref{eq:general_QTmachineEq})  we get \begin{equation}
\hat{\rho}^{(out)}_{{\rm C},k}  =  \hat{\Omega}^{({\rm C})}_{\epsilon_1} + e^{-\Gamma_{\rm C} \tau_{\rm C}} (
\hat{\rho}^{(in)}_{{\rm C},k} -  \hat{\Omega}^{({\rm C})}_{\epsilon_1} ) \;, \qquad
\hat{\rho}^{(out)}_{{\rm H},k}  =  \hat{\Omega}^{({\rm H})}_{\epsilon_2} + e^{-\Gamma_{\rm H} \tau_{\rm H}} ( 
\hat{\rho}^{(in)}_{{\rm H},k} -  \hat{\Omega}^{({\rm H})}_{\epsilon_2} ) \;,
\end{equation} 
which, with the help of (\ref{COND11}) can be equivalently cast in the following recursive expressions 
\begin{eqnarray}
\hat{\rho}^{(out)}_{{\rm C},k}  =  \hat{\Omega}^{({\rm C})}_{\epsilon_1} &+& e^{-\Gamma_{\rm C} \tau_{\rm C}} ( 
\hat{\Omega}^{({\rm H})}_{\epsilon_2}-  \hat{\Omega}^{({\rm C})}_{\epsilon_1} )
+
e^{-(\Gamma_{\rm C}\tau_{\rm C}+\Gamma_{\rm H}\tau_{\rm H})} (\hat{\rho}^{(out)}_{{\rm C},k-1}- \hat{\Omega}^{({\rm C})}_{\epsilon_1})
 \;,
\\
\hat{\rho}^{(out)}_{{\rm H},k}  =  \hat{\Omega}^{({\rm H})}_{\epsilon_2}
&-& e^{-\Gamma_{\rm H}\tau_{\rm H}}  ( \hat{\Omega}^{({\rm H})}_{\epsilon_2} -  \hat{\Omega}^{({\rm C})}_{\epsilon_1} )
+
e^{-(\Gamma_{\rm C}\tau_{\rm C}+\Gamma_{\rm H}\tau_{\rm H})} (
\hat{\rho}^{(out)}_{{\rm H},k-1} -  \hat{\Omega}^{({\rm H})}_{\epsilon_2} ) \;.
\end{eqnarray} 
Now in the ITT limit $\tau_j\rightarrow \infty$ these yields 
$\hat{\rho}^{(out)}_{{\rm C},k}= \hat{\Omega}^{({\rm C})}_{\epsilon_1}$ and 
$\hat{\rho}^{(out)}_{{\rm H},k}= \hat{\Omega}^{({\rm H})}_{\epsilon_2}$ for all $k$, leading to (\ref{IMPOII})
via Eq.~(\ref{COND11}).
For finite $\tau_j$ instead, keeping only the most relevant order, we obtain  
\begin{eqnarray}
\hat{\rho}^{(out)}_{{\rm C},k}  &\simeq&  \hat{\Omega}^{({\rm C})}_{\epsilon_1} + e^{-\Gamma_{\rm C} \tau_{\rm C}} (
 \hat{\Omega}^{({\rm H})}_{\epsilon_2} -  \hat{\Omega}^{({\rm C})}_{\epsilon_1} ) \;,
\\
\hat{\rho}^{(out)}_{{\rm H},k}  &\simeq& \hat{\Omega}^{({\rm H})}_{\epsilon_2} -  e^{-\Gamma_{\rm H} \tau_{\rm H}}
 ( \hat{\Omega}^{({\rm H})}_{\epsilon_2} -  \hat{\Omega}^{({\rm C})}_{\epsilon_1} ) \;,
\end{eqnarray} 
for all $k$, which, exploiting once more~(\ref{COND11}), gives
\begin{eqnarray}  \label{IMPOII} 
\Delta\hat{\rho}_{\rm C} = - \Delta\hat{\rho}_{\rm H} \simeq
(1-e^{-\Gamma_{\rm C} \tau_{\rm C}} -e^{-\Gamma_{\rm H} \tau_{\rm H}})
(  \hat{\Omega}^{({\rm C})}_{\epsilon_1}-\hat{\Omega}^{({\rm H})}_{\epsilon_2})\;.
\end{eqnarray} 
Inserting this into Eq.~(\ref{DD2OTTO}) we can express 
the first order corrections $\Delta Q^{(1)}_{j}$s as 
\begin{eqnarray}
\Delta Q^{(1)}_{j} = - (e^{-\Gamma_{\rm C} \tau_{\rm C}} +e^{-\Gamma_{\rm H} \tau_{\rm H}})
\Delta Q^{(0)}_{j} \;, 
\end{eqnarray} 
hence obtaining 
\begin{eqnarray}
\alpha_j = - (e^{-\Gamma_{\rm C} \tau_{\rm C}} +e^{-\Gamma_{\rm H} \tau_{\rm H}})\;, 
\end{eqnarray} 
for $j={\rm H,C}$. 
From Eq.~(\ref{EFFICotto}) then follows that the efficiency 
 remains un-effected by the ITT corrections, i.e.  $\eta=\eta_{\text o}$, while 
 according to Eq.~(\ref{eq:quasi-static_ottopow}) the power becomes 
  \begin{equation} 
\label{def:powerotto1}
P\simeq{\Delta Q^{(0)}_{\rm H}} \frac{\eta_{\text{o}} -\eta_{\text{c}}  (e^{-\Gamma_{\rm C} \tau_{\rm C}} +e^{-\Gamma_{\rm H} \tau_{\rm H}}) }{\tau_{\rm C}+\tau_{\rm H}}  \;.
\end{equation}

\section{Optimal protocol shape for the Quantum Carnot cycle}
\label{app:optimalshape}
In this section we solve the minimization of 
 the functional ${\cal F}[q(x)]$ of \eqref{def:F} (hereby ${\cal F}[q]$ for short)  under the constraints 
\begin{equation}
\label{eq:constraints}
\{q(0)=q_{in}\;, \dot{q}(0)=0\;, q(1)=q_{fin} \;, \dot{q}(1)=0\},
\end{equation}
which, according to Eq.~(\ref{res:pmaxS-W})  allow us to optimize the power production on the Quantum Carnot cycle. 
First of all we notice that it can be equivalently expressed as 
\begin{equation}
{\cal F}[q]=-\int_0^1 \text{d}x \  \ddot{q}\ln\big(\frac{q}{1-q}\big)\;,
\end{equation}
where the modulus has been replaced by a minus sign, 
due to the fact that integrand is guaranteed to be non-positive for all the allowed choices of the function $q(x)$ (same argument
we used in Eq.~(\ref{eq:heats1}) to establish the non positivity of  $\Delta Q_j^{(1)}$).

\begin{figure}
\centering
\includegraphics[scale=0.6]{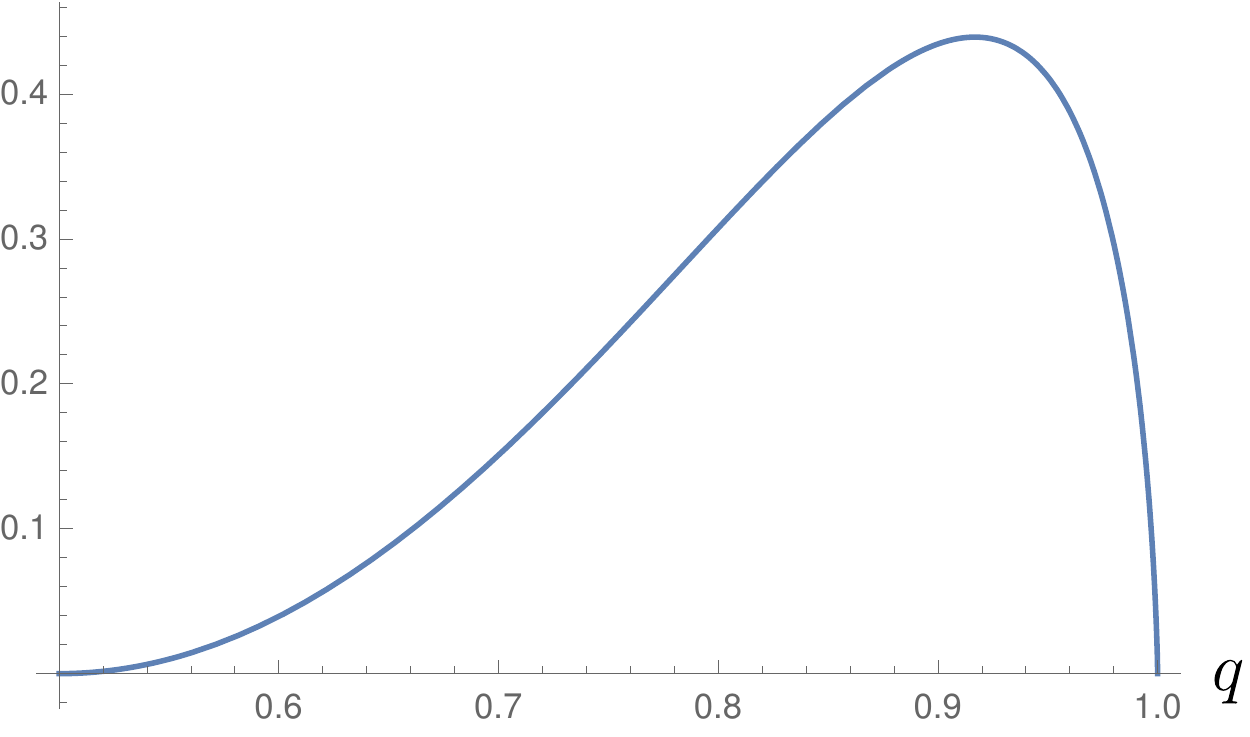}
\caption{Plot of $[\ln\big(\dfrac{q}{1-q}\big)]^2q(1-q)$ as a function of $q$.}
\label{bbb}
\end{figure}

For this purpose we consider the variation of the functional \eqref{def:F}  under a small variation of the control $q\rightarrow q +\delta q$,
\begin{equation}
\delta {\cal F}[q] = {\cal F}[q+\delta q] - {\cal F}[q] =- \int \delta\ddot{q}\ln\big(\frac{q}{1-q}\big)+\int \ddot{q}\ \delta\!\ln\big(\frac{q}{1-q}\big) 
=- \int \delta q \bigg[\frac{2 \ddot{q}}{q(1-q)}+\dot{q}^2\frac{2q-1}{q^2(1-q)^2}\bigg]\;, 
\end{equation}
where the last identity was obtained by integration by parts using 
the constraints \ref{eq:constraints}.
Imposing the latter to nullify under  arbitrary variation we can then obtain the differential equation
\begin{equation}
2\ddot{q}+\dot{q}^2\frac{2q-1}{q(1-q)}=0  \quad 
 \Longrightarrow \quad 2\ln\dot{q}-\ln(q(1-q))=constant\ ,
\end{equation}
which can be solved using separation of variables and the substitution $q'=q-\frac{1}{2}$, leading to optimal solutions of the form 
\begin{equation}
\label{res:optimal_sol}
\bar{q}(x)=\cos^2\bigg(\frac{\omega(x+\varphi)}{2}\bigg)=\frac{1+\cos(\omega(x+\varphi))}{2}.
\end{equation}
This class of solutions is parametrised by the two values $\{\omega,\varphi\}$ and is in general incompatible with the constraints \ref{eq:constraints} given at the extrema: this is a typical issue one meets in variational problems performed on given sets of functions that are not topologically closed; that is, it is possible to construct a sequence of functions $q_k(x)$ which decrease the functional toward an infimum which, however, is reached only for a function $\lim_{k\rightarrow\infty} q_k=\bar{q}$ that is outside the initial function space. We can build the sequence $q_k$ by simply stringing smoothly  $\bar{q}(0)$ to $\bar{q}(\varepsilon)$ and $\bar{q}(1-\varepsilon)$ to $\bar{q}(1)$ for small $\varepsilon=\frac{1}{k}$, 
\begin{equation}
\label{eq:smooth_approx}
q_k(x)=
\begin{cases}
s_k & 0\leq x \leq \frac{1}{k} \\
\bar{q}(t) & \frac{1}{k}\leq x \leq 1 -\frac{1}{k} \\
s_k & 1-\frac{1}{k} \leq x \leq 1 
\end{cases}
\end{equation}
with sufficiently smooth functions $s_k$ such that $s_k(0)=\bar{q}(0),\ s_k(\frac{1}{k})=\bar{q}(\frac{1}{k}),\ \dot{s}_k(0)=0,\ \dot{s}_k(\frac{1}{k})=\dot{\bar{q}}(\frac{1}{k})$ and similarly for $x=1-\frac{1}{k},1$.
The correction to the optimal functional $\delta_k {\cal F}={\cal F}[{q_k}]-{\cal F}[{\bar{q}}]$ will then be given from the contribution near the border $[0,\frac{1}{k}]$
\begin{equation}
\int_0^{\frac{1}{k}} d x \Big(\ddot{s}_k\ln\big(\frac{s_k}{1-s_k}\big)-\ddot{\bar{q}}\ln\big(\frac{\bar{q}}{1-\bar{q}}\big)\Big)
\sim \Big(\big(\dot{s}_k(\frac{1}{k})-\dot{s}_k(0)\big)\ln\big(\frac{s_k(0)}{1-s_k(0)}\big)-\big(\dot{\bar{q}}(\frac{1}{k})-\dot{\bar{q}}(0)\big)\ln\big(\frac{\bar{q}(0)}{1-\bar{q}(0)}\big)\Big)
\end{equation}
and the analogous term for $[1-\frac{1}{k},1]$. Using that $\dot{s}_k(\frac{1}{k})-\dot{s}_k(0)=\dot{\bar{q}}(\frac{1}{k})$ we can take the limit to obtain, adding the $[1-\frac{1}{k},1]$ contribution,
\begin{equation}
\label{eq:corr_limit}
\lim_{k\rightarrow\infty}\delta_k {\cal F}:=\bar{\delta}{\cal F}=\dot{\bar{q}}(0) \ln\bigg(\frac{\bar{q}(0)}{1-\bar{q}(0)}\bigg) -
\dot{\bar{q}}(1) \ln\bigg(\frac{\bar{q}(1)}{1-\bar{q}(1)}\bigg)\
\end{equation}
and thus the optimal value
\begin{equation}
\lim_{k\rightarrow\infty}{\cal F}[q_k]:={\cal F}_{min}={\cal F}[{\bar{q}}]+\bar{\delta} {\cal F}\ .
\end{equation}
It is however important to stress that the sequence $q_k$ will eventually break the S-D approximation, having a high second derivative near 0 and 1. Hence one should "stop" to a $k$ which is not too big to achieve these approximate results. Numerical plots support the achievability of the limit.
\paragraph{Quasi-Otto limit.}
 When allowing the control to vary the initial and final point $\{q(0),q(1)\}$, numerical plots show that the choice that maximizes the value of the power \eqref{res:pmaxS-W}, which is equal to
\begin{equation}
\label{eq:fracpow}
\dfrac{(\sqrt{T_H}-\sqrt{T_C})^2}{4A}\dfrac{\Big(\int \dot{q}\ln\big(\frac{q}{1-q}\big)\Big)^2}{-\int \ddot{q}\ln\big(\frac{q}{1-q}\big)}\; ,
\end{equation} 
is obtained in the limit of them being the same $q(1)-q(0)=\varepsilon\rightarrow 0$.
In this limit $\bar{q}(x)$ \eqref{res:optimal_sol} is essentially a line with
$
\dot{\bar{q}}\sim \frac{\bar{q}(1)-\bar{q}(0)}{1}-0=\varepsilon\;,
$
hence the numerator of \eqref{eq:fracpow} is
\begin{equation}
\Big(\int \dot{\bar{q}}\ln(\bar{q}/(1-\bar{q})) \text{d}t\Big)^2 \sim \ln\big(\frac{\bar{q}}{1-\bar{q}}\big)^2 \varepsilon^2\ .
\end{equation}
For the denominator ${\cal F}_{min}$ the contribution ${\cal F}[{\bar{q}}]$ nullifies while $\bar{\delta} {\cal F}$ can be estimated from \eqref{eq:corr_limit} as
\begin{equation}
 \dot{\bar{q}}\Big(\ln\big(\frac{\bar{q}(0)}{1-\bar{q}(0)}\big)-\ln\big(\frac{\bar{q}(1)}{1-\bar{q}(1)}\big)\Big)\sim \frac{\varepsilon^2}{\bar{q}(1-\bar{q})}\ ,
\end{equation}
where we use $\dot{q}=\varepsilon$ and the derivative of $\ln(q/(1-q))$, which is $1/(q(1-q))$. We can then write \eqref{eq:fracpow} in this limit
\begin{equation}
P_{Quasi-Otto}=\dfrac{(\sqrt{T_H}-\sqrt{T_C})^2}{4A}[\ln\big(\dfrac{\bar{q}}{1-\bar{q}}\big)]^2\bar{q}(1-\bar{q}) \ .
\end{equation}
It is possible plot this function to find the optimal value of $\bar{q}$ around which to perform the optimal control, as in Figure~\ref{bbb}, where is evident that $q_{optimal}\sim 0.92$ and $P_{max} \sim 0.11 \frac{(\sqrt{T_H}-\sqrt{T_C})^2}{A}$.

\section{Dynamical solution of the non-Markovian model}
In this Appendix we show exact and approximate (Slow-Driving) solutions of the non-Markovian model of Section \ref{sec:model3}. We remind that we are interested in the analysis of the dynamics of the qubit \s when coupled to one of the two baths, that without loss of generality can be considered to be the cold one, so that the state of the system and the ancillary qubit ${\cal{A}}_{\rm C}$ of the bath can be described by the density matrix $\hat{\bold{R}}_{\rm{C}}$, as well al the local states of \s ($\hat{\rho}:=\Tr_{\cal{A}_{\rm C}}[\hat{\bold{R}}_{\rm{C}}]$) and $\cal{A}_{\rm C}$ ($\hat{\rho}_{\rm C}:=\Tr_{\cal S}[\hat{\bold{R}}_{\rm{C}}]$). As pictured in Fig.\ref{fig:singlebath} we remind the local Hamiltonians and coupling interaction
\begin{equation}
\hat{H}_{tot}(t)=\hat{H}_t+\hat{H}_{{\cal A}_{\rm C}}+\hat{V}_{\rm C}=\frac{\epsilon(t)}{2}\sigma^z+\frac{E_C}{2}\sigma^z_{\mathcal{A}_{\rm C}}+\gamma_{\rm C}(\sigma^+\ten\sigma^-_{\mathcal{A}_{\rm C}}+\sigma^-\ten\sigma^+_{\mathcal{A}_{\rm C}})\ ,
\end{equation}
as well as the thermalizing dissipators
\begin{equation}
\mathcal{D}[\hat{\bold{R}}_{\rm{C}}]=\Gc(\hat{\Omega}^{(C)}_{\hat{H}_t}\ten\hat{\rho}_{\rm C}-\hat{\bold{R}}_{\rm{C}})+\Gamma_\Ac(\hat{\rho}\ten\hat{\Omega}^{(C)}_{\hat{H}_{{\cal A}_{\rm C}}}-\hat{\bold{R}}_{\rm{C}})
\end{equation}
so that in the interaction picture the dynamical equation is
\begin{equation}
\label{eq:singlebathgeneral}
\dot{\hat{\bold{R}}}_{\rm{C}}=-i\gamma_{\rm C}[e^{-i\delta t}\sig^+\ten\sig_{{\cal A}_{\rm C}}^-+e^{i\delta t}\sig^-\ten\sig_\Ac^+,\Rc]+\Gamma_\Ac(\hat{\rho}\ten\hat{\Omega}^{(C)}_{\hat{H}_{{\cal A}_{\rm C}}}-\Rc)+\Gc(\hat{\rho}_{\rm C}\ten\hat{\Omega}^{(C)}_{\hat{H}_t}-\Rc)\ ,
\end{equation}
with $\delta=\epsilon-E_C$.\\
Introducing the thermal ground state probabilities and a time dependent phase
\begin{equation}
p_s(t)=\frac{1}{e^{-\beta_{\rm C} \epsilon(t)}+1}\ , \quad \quad p_{\rm C}=\frac{1}{e^{-\beta_{\rm C} E_{\rm C}}+1}\ , \quad \quad
\phi(t)=e^{i\delta t}\ ,
\end{equation}
we can solve equation (\ref{eq:singlebathgeneral}) by writing it in the computational basis $
\Rc\equiv\sum_{\alpha,\beta,\mu,\nu=0,1} {\rho}_{\alpha\mu\beta\nu} |\alpha\rangle \!\langle\beta|_{_\Ac}\ten|\mu\rangle\! \langle\nu|_{_{\cal S}} \;.
$
We get, in matrix form,
\begin{multline}
\label{eq:generalsinglebathcomputationalbasis}
\bigg(\frac{d}{dt}-(\Gc+\Gamma_{{\cal A}_C})\eye\bigg)
\begin{pmatrix}
{\rho}_{0000} & {\rho}_{0001} & {\rho}_{0010} & {\rho}_{0011}  \\
{\rho}_{0100} & {\rho}_{0101} & {\rho}_{0110} & {\rho}_{0111}  \\
{\rho}_{1000} & {\rho}_{1001} & {\rho}_{1010} & {\rho}_{1011}  \\
{\rho}_{1100} & {\rho}_{1101} & {\rho}_{1110} & {\rho}_{1111}  \\
\end{pmatrix} 
= \\ \\
\left(
\begin{array}{c|c|c|c}
\Gamma_{{\cal A}_C} p_{\rm C}({\rho}_{_{0000}}\!\!+\!{\rho}_{_{1010}}) & \Gamma_{{\cal A}_C} p_{\rm C}({\rho}_{_{0001}}\!\!+\!{\rho}_{_{1011}})\!\!&  &   \\ 
+\Gc p_s({\rho}_{_{0000}}\!\!+\!{\rho}_{_{0101}})& & \Gc p_s({\rho}_{_{0010}}\!\!+\!{\rho}_{_{0111}})\!\! & 0\\
& +\! i\gamma_{\rm C} \phi {\rho}_{_{0010}}  & +i\gamma_{\rm C} \phi^*{\rho}_{_{0001}} & \\ \hline
 & \Gamma_{{\cal A}_C} p_{\rm C}({\rho}_{_{1111}}\!\!+\!{\rho}_{_{0101}})\!\! & &  \\
\cdots & +\Gc (1-p_s)({\rho}_{_{0000}}\!\!+\!{\rho}_{_{0101}}) & & \Gc (1-p_s)({\rho}_{_{0010}}\!\!+\!{\rho}_{_{0111}})\\ 
  &+\!i\gamma_{\rm C}(\phi{\rho}_{_{0110}}\!\!-\phi^*\!{\rho}_{_{0110}}^*) & +i\gamma_{\rm C}\phi^*({\rho}_{_{0101}}\!\!-\!{\rho}_{_{1010}}) & -i\gamma_{\rm C}\phi^*{\rho}_{_{1011}}  \\ \hline
 & & \Gamma_{{\cal A}_C} (1-p_{\rm C})({\rho}_{_{0000}}\!\!+\!{\rho}_{_{1010}})\!\!& \Gamma_{{\cal A}_C} (1-p_{\rm C})({\rho}_{_{0001}}\!\!+\!{\rho}_{_{1011}})\!  \\
\cdots & \cdots & +\Gc p_s({\rho}_{_{1111}}\!\!+\!{\rho}_{_{1010}})\!\! & \\
& & -\! i\gamma_{\rm C}(\phi{\rho}_{_{0110}}\!\!-\phi^*\!{\rho}_{_{0110}}^*) & -\! i\gamma_{\rm C}\phi {\rho}_{_{0111}}\\ \hline
 & & & \Gamma_{{\cal A}_C} (1-p_{\rm C}) ({\rho}_{_{1111}}\!\!+\!{\rho}_{_{0101}})  \\
\cdots & \cdots & \cdots & +\Gc (1-p_s) ({\rho}_{_{1111}}\!\!+\!{\rho}_{_{1010}}) \\
 & & & \\
\end{array}
\right)
\end{multline}
where the inferior triangular part has been omitted to improve readability and can be filled by just noting $\Rc$ is hermitian. For each matrix element the 3 different lines represent the contributions from the $\Ac$ dissipator
 ($\propto \Gamma_{{\cal A}_C}$), the \s dissipator ($\propto \Gc$),
 and the Hamiltonian exchange ($\propto \gamma_{\rm C}$). Looking at the equation we can note that the time evolution generator is a sparse super-operator, which couples separately different subsets of components, namely the ones highlighted here with different colors
\[
\begin{pmatrix}
\colorbox{yellow}{$\textcolor{blue}{{\rho}_{0000}}$}  & \textcolor{orange}{{\rho}_{0001}} & \textcolor{orange}{{\rho}_{0010}} & \textcolor{red}{{\rho}_{0011}}  \\
 & \colorbox{yellow}{$\textcolor{blue}{{\rho}_{0101}}$} & \colorbox{yellow}{$\textcolor{blue}{{\rho}_{0110}}$} & \textcolor{orange}{{\rho}_{0111}}  \\
 &  & \colorbox{yellow}{$\textcolor{blue}{{\rho}_{1010}}$} & \textcolor{orange}{{\rho}_{1011}}  \\
 &  &  & \colorbox{yellow}{$\textcolor{blue}{{\rho}_{1111}}$}  \\
\end{pmatrix}\ . 
\]
Three different sets of equations can be then solved separately, but we will be interested in the Thermodynamics of the system, hence mainly the highlighted \colorbox{yellow}{\textcolor{blue}{blue subset}}, because it contains the populations which determine thermodynamic variables (namely, the eigenstates of $\hat{H}_t+\hat{H}_\Ac$).
We can represent it as the vector
\begin{equation}
\label{eq:rho0}
\vec{\bold{R}}(t)=
\begin{pmatrix}
q_{00} \\ q_{10} \\ q_{01} \\ q_{00} \\ k
\end{pmatrix}
\end{equation}
where $q_{ab}={\rho}_{abab}$ is the population of \s in the state $b$ and $\Ac$ in $a$, while $k\equiv {\rho}_{0110}$ is the coherence between the states $\ket{01}$ and $\ket{10}$ which are the ones interacting by the exchange Hamiltonian $V_{\rm C}$.

\subsection{Resonant case $(\epsilon=E_{\rm C})$}
\label{sec:analytical_resonant_sol}
Consider the instance in which the gaps are fixed equal $\epsilon=E_{\rm C}=E$; in this case the algebra has some simplifications; indeed 
\begin{equation}
[\sigma^+\ten\sigma^-_\Ac+\sigma^-\ten\sigma^+_\Ac\;,\;\sigma^z+\sigma^z_\Ac]_-=0
\end{equation} 
so that the value of the interaction Hamiltonian is conserved in absence of the dissipative dynamics (or decreases exponentially, see below).
It is easy to check that the (only) stationary state is $\Rc=\hat{\Omega}^{(C)}_{\hat{H}_t}\ten\hat{\Omega}^{(C)}_{\hat{H}_{{\cal A}_{\rm C}}}$ (or $\hat{\Omega}_{\rm C}\ten \hat{\Omega}_{\rm C}$ for simplicity).Having the gap equal we can call  
\begin{equation}
p_s\equiv p_{\rm C}\equiv p_0 = \frac{1}{e^{- \beta_{\rm C} E_{\rm C}}+1}\ .
\end{equation}
We will also write $\mathcal{L}$, with a small abuse of notation, to indicate the Lindblad generator of the dynamics restricted to the different subsets of components.
The equation \eqref{eq:generalsinglebathcomputationalbasis} for the vector (\ref{eq:rho0}) can be written in this special case as
\begin{equation}
\label{eq:bluesubsetsystem}
\left\{
\begin{aligned}
\dot{q}_{00} +(\Gamma_{\Ac}+\Gamma_{\rm C}) q_{00}&=\Gamma_{\Ac} p_0(q_{00}+q_{10})+\Gamma_{\rm C} p_0(q_{00}+q_{01}) \\
\dot{q}_{01} +(\Gamma_{\Ac}+\Gamma_{\rm C}) q_{01}&=\Gamma_{\Ac} p_0(q_{11}+q_{01})+\Gamma_{\rm C} (1-p_0)(q_{00}+q_{01})+i\gamma_{\rm C}(k-k^*) \\
\dot{q}_{10} +(\Gamma_{\Ac} +\Gamma_{\rm C}) q_{10}&=\Gamma_{\Ac} (1-p_0)(q_{00}+q_{10})+\Gamma_{\rm C} p_0(q_{11}+q_{10})-i\gamma_{\rm C}(k-k^*) \\
\dot{q}_{11} +(\Gamma_{\Ac}+\Gamma_{\rm C}) q_{00}&=\Gamma_{\Ac} (1-p_0)(q_{11}+q_{01})+\Gamma_{\rm C} (1-p_0)(q_{11}+q_{10}) \\
\dot{k} +\Gamma_{\Ac} k+\Gamma_{\rm C} k &=i\gamma_{\rm C} (q_{01}-q_{10}) \\
\end{aligned}
\right.\quad  .
\end{equation}
Note that the real part of the coherence $\Re(k)$ satisfies \ $\Re(\dot{k}) +\Gamma\Re( k) =0$ , hence it is decoupled from the rest and it just dies exponentially\footnote{\label{footnote_interaction}Note that $\langle V_{\rm C}\rangle=\Tr[\Rc \gamma_{\rm C}(\sigma^+\ten\sigma_\Ac^-+h.c.)]=\gamma_{\rm C} (\rho_{_{0110}}+\rho_{_{1001}}=\gamma_{\rm C} 2\Re(k)$ which is then always decreasing. This means for initial condition given by a product state $\rho\ten\rho_{\rm C}$, $\Re(k)$ is constantly null, which in turn implies that the switch-on/switch-off work done to attach the system \s to the baths is null and can be safely neglected in the performance analysis.} $\sim e^{-\Gamma t}$.
Calling $\Im(k)\equiv I$ the system can be thus be written
\begin{equation}
\frac{d}{dt}
\begin{pmatrix}
q_{00} \\ q_{10} \\ q_{01} \\ q_{00} \\ I
\end{pmatrix}
=
\tilde{\mathcal{L}}
\begin{pmatrix}
q_{00} \\ q_{10} \\ q_{01} \\ q_{00} \\ I
\end{pmatrix}
-(\Gamma_{\rm C}+\Gamma_{\Ac})
\begin{pmatrix}
q_{00} \\ q_{10} \\ q_{01} \\ q_{00} \\ I
\end{pmatrix}\ ,
\end{equation}
the Lindblad generator being $\mathcal{L}=\tilde{\mathcal{L}}-(\Gamma_{\rm C}+\Gamma_{\Ac})$ and
\begin{equation}
\tilde{\mathcal{L}}=
\begin{pmatrix}
(\Gc+\Gamma_{{\cal A}_C})p_0 & \Gamma_{{\cal A}_C} p_0 & \Gc p_0 & 0 & 0\\
\Gamma_{{\cal A}_C} (1-p_0) & \Gamma_{{\cal A}_C} (1-p_0)+ \Gc p_0& 0 & \Gc p_0 & +2\gamma_{\rm C}\\
\Gc (1-p_0) & 0 & \Gamma_{{\cal A}_C} p_0+\Gc (1-p_0) & \Gamma_{{\cal A}_C} p_0 & -2\gamma_{\rm C}\\
0 & \Gc (1-p_0) & \Gamma_{{\cal A}_C} (1-p_0) & (\Gamma_{{\cal A}_C}+\Gc)(1-p_0) & 0\\
0 & -\gamma_{\rm C} & +\gamma_{\rm C} & 0 & 0\\
\end{pmatrix}\
\end{equation}
which in the $\Gamma_{{\cal A}_C}=\Gc=\Gamma$ case becomes ($\gamma_{\rm C}'=\gamma_{\rm C}/\Gamma$)
\begin{equation}
\tilde{\mathcal{L}}=\Gamma
\begin{pmatrix}
2p_0 & p_0 & p_0 & 0 & 0\\
(1-p_0) & 1 & 0 & p_0 & +2\gamma_{\rm C}'\\
(1-p_0) & 0 & 1 & p_0 & -2\gamma_{\rm C}'\\
0 & (1-p_0) & (1-p_0) & 2(1-p_0) & 0\\
0 & -\gamma_{\rm C}' & +\gamma_{\rm C}' & 0 & 0\\
\end{pmatrix}\ .
\end{equation}
To solve the dynamics one can find eigenvalues and eigenvectors of such a matrix.
In the $\Gc=\Gamma_{{\cal A}_C}$ case the particular symmetry of the problem is reflected in the tractable form of the eigensystem of $\mathcal{L}$, which is (subtracting already $-2\Gamma$ to all eigenvalues and expressing in units of $\Gamma$)
\begin{multline}
\label{res:eigensystemL}
\lambda_0=0 \rightarrow \vec{\rho}_0=
\begin{pmatrix}
 p_0^2 \\ p_0(1-p_0) \\ p_0(1-p_0) \\(1-p_0)^2 \\ 0
\end{pmatrix}\equiv\text{thermal state}\ \hat{\Omega}_{\rm C}\ten \hat{\Omega}_{\rm C} , \\
\lambda_1=-1 \rightarrow 
\vec{\rho}_1=\begin{pmatrix}
 p_0 \\ \frac{1-2p_0}{2} \\ \frac{1-2p_0}{2}\\ -(1-p_0)\\ 0 
\end{pmatrix}\ , \quad
\lambda_2=-2 \rightarrow \vec{\rho}_2= 
\begin{pmatrix}
1 \\ -1 \\ -1 \\ 1 \\ 0 
\end{pmatrix}\ , \\
\lambda_{3,4}=\frac{-3\pm\sqrt{1-16\gamma_{\rm C}'^2}}{2} \rightarrow 
\vec{\rho}_{3,4}=
\begin{pmatrix}
0 \\ \frac{1\pm\sqrt{1-16\gamma_{\rm C}'^2}}{4\gamma_{\rm C}'} \\ -\frac{1\pm\sqrt{1-16\gamma_{\rm C}'^2}}{4\gamma_{\rm C}'} \\ 0 \\ -1 
\end{pmatrix}\ .
\end{multline}
We can then solve completely the dynamics for an initial state of the form $\Rc (t_0)=\hat{\rho}\ten\hat{\Omega}_{\rm C}$, that is out of equilibrium on \s and thermal on $\Ac$, as requested by our model.\\
Suppose for the moment that also $\hat{\rho}$ is diagonal, that is
\begin{equation}
\label{eq:initialstate}
\hat{\rho}(t_0)=
\begin{pmatrix}
a & 0 \\ 0 & 1-a
\end{pmatrix}
\Rightarrow
\begin{pmatrix}
q_{00} \\ q_{10} \\ q_{01} \\ q_{00} \\ I
\end{pmatrix}
(t_0)
=
\begin{pmatrix}
ap_0 \\ a(1-p_0) \\ (1-a)p_0 \\ (1-a)(1-p_0) \\0
\end{pmatrix}\ .
\end{equation}
We write $a=p_0+\Delta$
to quantify how much $\hat{\rho}$ is out of equilibrium.
We decompose \eqref{eq:initialstate} as a combination of the eigenvectors \eqref{res:eigensystemL}, in order to write the solution which will be
\begin{equation}
\vec{q}(t)=\vec{\rho_0}+\Delta\vec{\rho}_1 e^{-\Gamma t}+\frac{\Delta\gamma_{\rm C}'}{\sqrt{1-16\gamma_{\rm C}'^2}}(\vec{\rho}_3 e^{\lambda_3 t}-\vec{\rho}_4 e^{\lambda_4 t})\ .
\end{equation}
Summing the first two components we can obtain the time-dependent ground state population of \s\! that is, calling
$\kappa_{\rm C}=\sqrt{1-16\gamma_{\rm C}'^2}\ ,$
\begin{equation}
\label{res:exactsolution_n-M0}
a(t)=p_0+\Delta\bigg(\frac{1}{2}e^{-\Gamma t}+\frac{1+\kappa_{\rm C}}{4\kappa_{\rm C}}e^{-\frac{3}{2}\Gamma t+\frac{\kappa_{\rm C}}{2}\Gamma t}-\frac{1-\kappa_{\rm C}}{4\kappa_{\rm C}}e^{-\frac{3}{2}\Gamma t-\frac{\kappa_{\rm C}}{2}\Gamma t} \bigg)=
p_0+\Delta f_{{\rm C}}(t)\ ,
\end{equation}
having defined
\begin{equation}
\label{def:f_kappa}
f_{{\rm C}}(t)\equiv \frac{e^{-\Gamma t}}{2}+e^{(-\frac{3}{2}+\frac{\kappa_{\rm C}}{2})\Gamma t}\bigg(\frac{1+\kappa_{\rm C}}{4\kappa_{\rm C}}\bigg)- e^{(-\frac{3}{2}-\frac{\kappa_{\rm C}}{2})\Gamma t}\bigg(\frac{1-\kappa_{\rm C}}{4\kappa_{\rm C}}\bigg)\ .
\end{equation}

\subsection{Non-resonant case $(\epsilon(t)\neq E_{\rm C})$ - Slow-Driving}
\label{sec:non-resonant_sol}
In case the two qubits are not resonant equation \eqref{eq:generalsinglebathcomputationalbasis} for the vector (\ref{eq:rho0}) takes the general form
\begin{equation}
\left\{
\begin{aligned}
\dot{q}_{00} +(\Gamma_{{\cal A}_C}+\Gc) q_{00}&=\Gamma_{{\cal A}_C} p_{\rm C}(q_{00}+q_{10})+\Gc p_s(q_{00}+q_{01})
\\
\dot{q}_{01} +(\Gamma_{{\cal A}_C}+\Gc) q_{01}&=\Gamma_{{\cal A}_C} p_{\rm C}(q_{11}+q_{01})+\Gc (1-p_s)(q_{00}+q_{01})+i\gamma_{\rm C}(\phi k- \phi^*k^*) 
\\
\dot{q}_{10} +(\Gamma_{{\cal A}_C} +\Gc) q_{10}&=\Gamma_{{\cal A}_C} (1-p_{\rm C})(q_{00}+q_{10})+\Gc p_s(q_{11}+q_{10})-i\gamma_{\rm C}(\phi k- \phi^*k^*)  
\\
\dot{q}_{11} +(\Gamma_{{\cal A}_C}+\Gc) q_{00}&=\Gamma_{{\cal A}_C} (1-p_{\rm C})(q_{11}+q_{01})+\Gc (1-p_s)(q_{11}+q_{10}) 
\\
\dot{k} +\Gamma_{{\cal A}_C} k+\Gc k &=i\gamma_{\rm C} \phi^* (q_{01}-q_{10}) 
\end{aligned}
\right.\quad .
\end{equation}
We note that the 2nd and 3rd equation here can be rewritten using\footnote{Remember $\phi(t)\equiv e^{i\delta t}$.} $\hat{k}=\phi k$, which satisfies
\begin{equation}
\dot{\hat{k}}=\dot{\phi}k+\phi\dot{k}=i\delta\phi k+\phi (i\gamma_{\rm C}\phi^*(q_{01}-q_{10})-k(\Gamma_{{\cal A}_C}+\Gc))
=i\delta\hat{k}+i\gamma_{\rm C}(q_{01}-q_{10})-(\Gamma_{{\cal A}_C}+\Gc)\hat{k}\ .
\end{equation}
In this way we can write,
calling $\Im(k)\equiv I$ and $\Re(k)\equiv R$,
\begin{equation}
\label{nonresonantvector}
\frac{d}{dt}
\begin{pmatrix}
q_{00} \\ q_{01} \\ q_{10} \\ q_{00} \\ I \\R
\end{pmatrix}
=
\tilde{\mathcal{L}}
\begin{pmatrix}
q_{00} \\ q_{01} \\ q_{10} \\ q_{00} \\ I \\ R
\end{pmatrix}
-(\Gamma_{\rm C}+\Gamma_{\Ac})
\begin{pmatrix}
q_{00} \\ q_{01} \\ q_{10} \\ q_{00} \\ I \\ R
\end{pmatrix}
\end{equation}
with 
$\mathcal{L}=\tilde{\mathcal{L}}-(\Gc+\Gamma_{{\cal A}_C})\eye\ ,$
\begin{equation}
\hat{\mathcal{L}}=
\begin{pmatrix}
\Gc p_s+\Gamma_{{\cal A}_C} p_{\rm C} & \Gc p_s & \Gamma_{{\cal A}_C} p_{\rm C} & 0 & 0 & 0\\
\Gc (1-p_s) & \Gamma_{{\cal A}_C} p_{\rm C}+\Gc (1-p_s) & 0 & \Gamma_{{\cal A}_C} p_{\rm C} & -2\gamma_{\rm C} & 0\\
\Gamma_{{\cal A}_C} (1-p_{\rm C}) & 0 & \Gamma_{{\cal A}_C} (1-p_{\rm C})+ \Gc p_s& \Gc p_s & +2\gamma_{\rm C} & 0\\
0 & \Gamma_{{\cal A}_C} (1-p_{\rm C}) & \Gc (1-p_s) & \Gamma_{{\cal A}_C}(1-p_{\rm C})+\Gc(1-p_s) & 0 & 0\\
0 & +\gamma_{\rm C} & -\gamma_{\rm C} & 0 & 0 & \delta\\
0 & 0 & 0 & 0 & -\delta & 0
\end{pmatrix} .
\end{equation}
The null eigenvector of $\mathcal{L}$ (i.e. the stationary state $\rho^{(0)}$) 
is not in general simply the thermal state $\Omega_{\rm C}\ten\Omega_{\rm C}$, but it reduces to it in the limit
\begin{equation}
\delta\rightarrow 0 \quad \quad(p_{\rm C}-p_s\rightarrow 0)
\Rightarrow
\vec{\rho}_0=
\begin{pmatrix}
p^2 \\ p(1-p) \\ p(1-p) \\ (1-p)^2 \\ 0 \\ 0
\end{pmatrix}\ .
\end{equation}
At first order\footnote{Note that $(p_s-p_{\rm C})\equiv \Delta_p$ is $\sim O(\delta)$.} in $\delta$ we find 
\begin{equation}
\label{res:quasi-staticnonresonant}
\vec{\rho}_0=
\begin{pmatrix}
p_{\rm C}p_s \\ p_{\rm C}(1-p_s) \\ p_s(1-p_{\rm C}) \\ (1-p_{\rm C})(1-p_s)\\ 0 \\ 0
\end{pmatrix}
-
\dfrac{\Delta_p}{N}
\begin{pmatrix}
2\gamma_{\rm C}^2(\Gamma_{{\cal A}_C} p_{\rm C}-\Gc p_s)
\\
2\gamma_{\rm C}^2((\Gc p_s-\Gamma_{{\cal A}_C} p_{\rm C})-\Gc)
\\
2\gamma_{\rm C}^2((\Gc p_s-\Gamma_{{\cal A}_C} p_{\rm C})+\Gamma_{{\cal A}_C})
\\
2\gamma_{\rm C}^2(\Gamma_{{\cal A}_C} (p_{\rm C}-1)-\Gc (p_s-1))
\\
-\Gc\Gamma_{{\cal A}_C} \gamma_{\rm C} 
\\
0
\end{pmatrix}\ ,
\end{equation}
the normalization being
$
N=(\Gc\Gamma_{{\cal A}_C}+2\gamma_{\rm C}^2)(\Gc+\Gamma_{{\cal A}_C})\ .
$
Note that in both approximations the real part of the coherence $\Re(\rho_{_{0110}})=[\vec{\rho}_0]_6=0$ is null; this allows us to neglect work contribution in the contacts and detachments from the baths, as in the resonant case\footnote{$\langle V_{\rm C}\rangle = \Tr[\Rc\gamma_{\rm C}(\sigma^+\ten\sigma_\Ac^-+h.c.)]=\gamma_{\rm C}(\rho_{_{0110}}+c.c.)=2\gamma_{\rm C}\Re(k)$ which is therefore null at the beginning and ending of each isothermal stroke, when $\Rc(t)=\hat{\rho}_{0}(t)$.}.

Following the approach described in Section \ref{sec:ii2} we can now compute the first order correction in Slow-Driving to the dynamics.
Looking at the formal solution of the S-D technique \eqref{eq:rho^1} we need for the computation:
\begin{itemize}
\item the quasi-static solution $\hat{\rho}^{(0)}$ found in (\ref{res:quasi-staticnonresonant}),
\item the dynamics generator $\mathcal{L}$ we wrote explicitly,
\item the projector on the null-trace subspace $\mathcal{P}$.
\end{itemize}
This last operator is easily found. The trace of the state is given from the sum of the 4 populations
\begin{equation}
\Tr[\Rc]=q_{00}+q_{01}+q_{10}+q_{11}\ .
\end{equation}
In order to project on the null-trace subspace we have to subtract to each population $\Tr[\rho]/4$, that is
$q_{ij} \rightarrow q_{ij}-\frac{1}{4}\sum_{a,b}q_{ab}$,
while the coherences stay unchanged. This can be written as 
\begin{equation}
\begin{pmatrix}
q_{00} \\ q_{01} \\ q_{10} \\ q_{11} \\ I \\ R
\end{pmatrix}
\rightarrow
\begin{pmatrix}
q_{00} \\ q_{01} \\ q_{10} \\ q_{11} \\ I \\ R
\end{pmatrix}
-\frac{1}{4}
\begin{pmatrix}
1 & 1 & 1 & 1 & 0 & 0 \\
1 & 1 & 1 & 1 & 0 & 0 \\
1 & 1 & 1 & 1 & 0 & 0 \\
1 & 1 & 1 & 1 & 0 & 0 \\
0 & 0 & 0 & 0 & 0 & 0 \\
0 & 0 & 0 & 0 & 0 & 0 \\
\end{pmatrix}
\begin{pmatrix}
q_{00} \\ q_{01} \\ q_{10} \\ q_{11} \\ I \\ R
\end{pmatrix}\ ,
\qquad \text{i.e.} \qquad
\mathcal{P}=
\begin{pmatrix}
\frac{3}{4} & -\frac{1}{4} & -\frac{1}{4} & -\frac{1}{4} & 0 & 0 \\
-\frac{1}{4} & \frac{3}{4}  & -\frac{1}{4} & -\frac{1}{4} & 0 & 0 \\
-\frac{1}{4} & -\frac{1}{4} & \frac{3}{4}  & -\frac{1}{4} & 0 & 0 \\
-\frac{1}{4} & -\frac{1}{4} & -\frac{1}{4} & \frac{3}{4}  & 0 & 0 \\
0 & 0 & 0 & 0 & 1 & 0 \\
0 & 0 & 0 & 0 & 0 & 1 \\
\end{pmatrix}\ .
\end{equation}
Now we have all the ingredients to compute (with the help of \emph{Wolfram Mathematica}) the first correction
$ \vec{\rho}_1=(\mathcal{LP})^{-1}\dot{\vec{\rho}}_0\ $.
The resulting expression is too complicated to be reported here, however, it is possible to compute the correction to the ground state population of \s\!, $q_{00}^{(1)}+q_{10}^{(1)}$, 
 which is in the form
\begin{equation}
q_{00}^{(1)}+q_{10}^{(1)}=[\rho_s^{(1)}]_{00}=-\dot{p}_s A_{\rm C}\ ,
\end{equation}
with an amplitude $A_{\rm C}$ that admits a closed but unfortunately still very convoluted general expression, which for the sake of readability we do not report here in its full extension.
Nevertheless, in the resonance ($\epsilon=E_{\rm C}$) limit we considered for our model we find 
\begin{equation}
\label{res:zerodeltaA}
\delta=0 \Rightarrow A_{\rm C}=\frac{\Gc((\Gamma_{{\cal A}_C}^2+\Gamma_{{\cal A}_C}\Gc)^2+2\gamma_{\rm C}^2(\Gamma_{{\cal A}_C}^2+2\Gamma_{{\cal A}_C}\Gc+4\gamma_{\rm C}^2)}{(\Gamma_{{\cal A}_C}+\Gc)^2(\Gamma_{{\cal A}_C}\Gc+2\gamma_{\rm C}^2)^2}\ .
\end{equation}

\section{Non-Markovian character of the dynamics}
\label{app:n-M_measure}
Here we show that the model  of Sec.~\ref{sec:n-M_model}  has an explicit non-Markovian character which depends on the non zero value of the parameters $\gamma_j$ that gauge
the coupling  between 
 \s and the ancillas ${\cal A}_j$.
 For this task, given  two input states $\hat{\rho}^{(1)}(0)$, $\hat{\rho}^{(2)}(0)$ of ${\cal S}$ and 
 $\hat{\rho}^{(1)}(t)$, $\hat{\rho}^{(2)}(t)$ their corresponding dynamical evolutions under the action of 
 the model,   we consider the  information-backflow BLP quantity~\citep{BLP} \begin{equation}
\mathcal{N}_{BLP}(\hat{\rho}^{(1)}(0), \hat{\rho}^{(2)}(0)):=\int dt  \; \dot{D}(\hat{\rho}^{(1)}(t),\hat{\rho}^{(2)}(t)) \; \Theta[\dot{D}(\hat{\rho}^{(1)}(t),\hat{\rho}^{(2)}(t))] \ , \label{BLP} 
\end{equation}
with $\Theta$ being the Heaviside function that restrict the domain of integration 
to the one where the integrand is positive, and where $D(\hat{\rho}^{(1)}(t),\hat{\rho}^{(2)}(t)):=\frac{1}{2}||\hat{\rho}^{(1)}(t)-\hat{\rho}^{(2)}(t)||_1$ is the trace distance~\cite{holevo}.
As discussed in Ref.~\citep{BLP}  value of $\mathcal{N}_{BLP}(\hat{\rho}^{(1)}(0), \hat{\rho}^{(2)}(0))$ 
greater then zero would imply non-Markovian character of the dynamics. 

\begin{figure}
\centering
\includegraphics[scale=0.6]{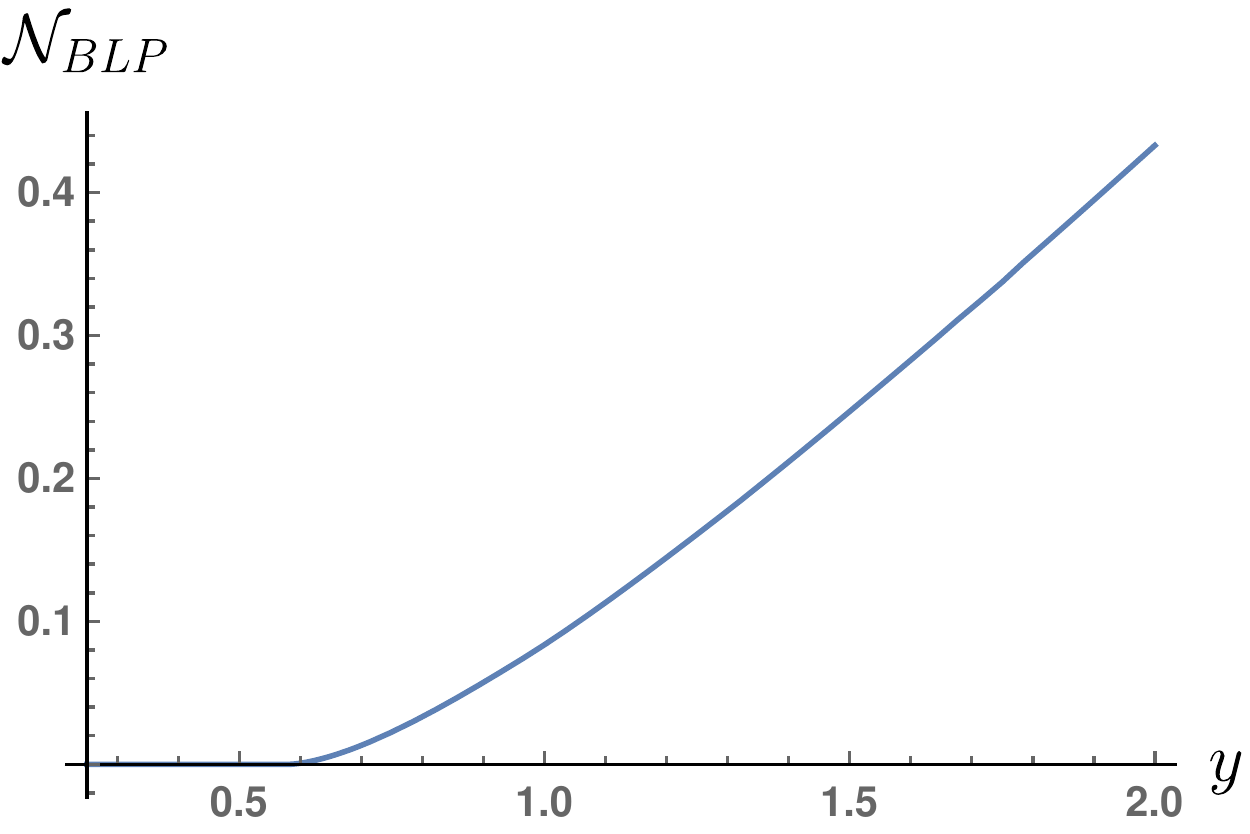}
\caption{The non-Markovian measure $\mathcal{N}_{BLP}$ as a function of $y_j$.
\label{fig:n-M_BLP}}
\end{figure}

In our case, focusing  only at $\hat{\rho}^{(1)}(0)$, $\hat{\rho}^{(2)}(0)$ having no coherence terms, we can simplify the analysis exploiting the fact that  
$D(\rho^{(1)}(t),\rho^{(2)}(t))
=|p^{(1)}(t)-p^{(2)}(t)|$, where for $j=1,2$,   $p^{(j)}(t)$ is the ground state population of the $\hat{\rho}^{(j)}(t)$. 
On-resonance ($\epsilon=E_j$) the solutions given in Sec. \ref{sec:analytical_resonant_sol} yields, in adimensional units ($\Gamma_j=1$),
\begin{equation}
\frac{p^{(1)}(t)-p^{(2)}(t)}{p^{(1)}(0)-p^{(2)}(0)}=
-\frac{e^{-t/2}}{2}-\frac{3}{8}e^{-\frac{3}{2}t}\bigg((1+\frac{1}{\kappa_j})e^{\kappa_j t}+(1-\frac{1}{\kappa_j})e^{-\kappa_j t}\bigg)
+\frac{1}{4}e^{-\frac{3}{2}t}\bigg((1+\kappa_j)e^{\kappa_j t}+(1-\kappa_j)e^{-\kappa_j t}\bigg)\;, 
\end{equation}
where $\kappa_j=\sqrt{1-16(\gamma_j/\Gamma_j)^2}$ .
Replacing this into (\ref{BLP}) and performing the integration numerically 
we obtain the results reported in Fig.~
\ref{fig:n-M_BLP} as function of $y_j=\gamma_j/\Gamma_j$.
As expected, the non-Markovianity is monotonously increasing with $\gamma_j$. Also, we find a threshold value under which this particular non-Markovianity witness is null.

\section{Symmetric Otto cycle has maximum power for $\tau_{\rm H}=\tau_{\rm C}$}
\label{app:sym_otto}
In this appendix we prove that the power expressed by Eq.~\eqref{eq:ottopower} is maximized, in case the coupling to the two baths is symmetric (i.e. $f_{\rm H}(t)=f_{\rm C}(t):=f(t)$ in Eq.~\eqref{eq:ottopower}), by choosing the time durations $\tau_{\rm H}=\tau_{\rm C}$ equal. In fact under this assumption the power can be written as
\begin{equation}  
\label{def:C}
P = (\epsilon_2-\epsilon_1)(p_{\rm C}-p_{\rm H}) 
\dfrac{\big(1-f(\tau_{\rm H})\big)\big(1-f(\tau_{\rm C})\big)}{\big(\tau_{\rm C}+\tau_{\rm H}\big)\big(1-f(\tau_{\rm C})f(\tau_{\rm H})\big)}:=(\epsilon_2-\epsilon_1)(p_{\rm C}-p_{\rm H}) C(\tau_{\rm C},\tau_{\rm H})\;.
\end{equation}
We show that when $\tau_{\rm H}\neq\tau_{\rm C}$ at least one between $C(\tau_{\rm C},\tau_{\rm C})$ and $C(\tau_{\rm H},\tau_{\rm H})$ is greater than $C(\tau_{\rm C},\tau_{\rm H})$, meaning that $\{\tau_{\rm H},\tau_{\rm C}\}$ would be outperformed by one of the two choices. To prove it we demonstrate that $C(\tau_{\rm C},\tau_{\rm H})\leq \sqrt{C(\tau_{\rm C},\tau_{\rm C})C(\tau_{\rm H},\tau_{\rm H})}$, which implies\footnote{Note that $C\geq 0$, by the definition \eqref{def:C} and $0\leq f\leq 1$.} the thesis.
This is equivalent to verify the following inequality holds
\begin{equation}
\label{eq:Ottopow_ineq}
\frac{(1-f(\tau_{\rm C}))(1-f(\tau_{\rm H}))}{(1-f(\tau_{\rm C})f(\tau_{\rm H}))(\tau_{\rm C}+\tau_{\rm H})}\leq\sqrt {\frac{(1-f(\tau_{\rm C}))^2}{(1-f^2(\tau_{\rm C}))2\tau_{\rm C}}\frac{(1-f(\tau_{\rm H}))^2}{(1-f^2(\tau_{\rm H}))2\tau_{\rm H}}}\ ,
\end{equation}
which is true by noting the numerator is the same and on the denominator by direct inspection
\begin{equation}
\tau_{\rm C}+\tau_{\rm H} \geq 2\sqrt{\tau_{\rm C}\tau_{\rm H}}\ ,
\end{equation}
\begin{equation}
(1-f(\tau_{\rm C})f(\tau_{\rm H}))\geq \sqrt{(1-f^2(\tau_{\rm C}))(1-f^2(\tau_{\rm H}))}\ .
\end{equation}

\end{document}